\documentclass[aps,twocolumn,prx,preprintnumbers, nofootinbib,10pt]{revtex4-2}
\usepackage{xpatch}
\makeatletter
\patchcmd{\@ssect@ltx}
    {\addcontentsline{toc}{#1}{\protect\numberline{}#8}}
    {}
    {}
    {}
\makeatother
\pdfoutput=1
\usepackage{color}
\usepackage{enumitem}
\usepackage[dvipsnames]{xcolor}
\usepackage{hyperref}
\definecolor{amaranth}{rgb}{0.9, 0.17, 0.31}
\definecolor{forestForestGreen(web)}{rgb}{0.13, 0.55, 0.13}
\definecolor{blue(munsell)}{HTML}{005567}
\definecolor{bblue}{rgb}{0.0, 0.58, 0.71}
\hypersetup{
pdfstartview={FitH}, % fits the width of the page to the window
pdftitle={}, % title
colorlinks=true, % false: boxed links; true: colored links
linkcolor=blue(munsell), % color of internal links (change box color with linkbordercolor)
citecolor=blue(munsell), % color of links to bibliography
filecolor=magenta, % color of file links
urlcolor=blue(munsell)% color of external links
}
\usepackage{pgfplots}
\usepackage{float}
\pgfplotsset{compat=1.18}
\usepackage{amssymb}
\usepackage{amsfonts}
\usepackage{graphicx}
\usepackage{epstopdf}
\usepackage{dcolumn}
\usepackage{amsmath}
\usepackage{latexsym,bm}
\usepackage{amsthm}
\usepackage{slashed}
\usepackage{color}
\usepackage{url}
\usepackage{longtable}
\usepackage{rotating}
\usepackage[normalem]{ulem}
\usepackage{faktor}
\usepackage{tikz-cd}
\usepackage{tensor}
\usepackage{array}
\usepackage{tabularx}
\usepackage{makecell}
\usepackage{changepage}

\usepackage{pgfplots}

\numberwithin{equation}{section}

\allowdisplaybreaks
\usepackage{tikz}
\usepackage{diagbox}
\usetikzlibrary{positioning}
\usetikzlibrary{calc}
\usetikzlibrary{decorations.pathreplacing,calligraphy}

\usepackage{xstring}
\usetikzlibrary{decorations.pathmorphing} 
\usetikzlibrary{decorations.markings} 
\usetikzlibrary{arrows} 
\usetikzlibrary{shapes} 
\usetikzlibrary{matrix} 
\usetikzlibrary{positioning} 
\usepackage[english]{babel} 
\usepackage[autostyle]{csquotes}
\usepackage{pifont}
\usetikzlibrary{shapes.multipart}

\tikzset{->-/.style={decoration={
  markings,
  mark=at position .5 with {\arrow{>}}},postaction={decorate}}}

\newcommand{\bea}{\begin{eqnarray}}
\newcommand{\eea}{\end{eqnarray}}
\newcommand{\be}{\begin{equation}}
\newcommand{\ee}{\end{equation}}
\newcommand{\ba}{\begin{aligned}}
\newcommand{\ea}{\end{aligned}}
\newcommand{\bit}{\begin{itemize}}
\newcommand{\eit}{\end{itemize}}
\newcommand{\ben}{\begin{enumerate}}
\newcommand{\een}{\end{enumerate}}
\newcommand{\nn}{\nonumber}

\newcommand{\id}{\text{id}}

\renewcommand{\ol}{\overline}

\newcommand{\SymTFT}{\text{SymTFT}}
\newcommand{\LL}{\langle\!\langle}
\newcommand{\RR}{\rangle\!\rangle}

\newcommand{\strong}{\text{strong}}
\newcommand{\weak}{\text{weak}}
\newcommand{\Tr}{\text{Tr}}

\newcommand{\sw}{\text{sw}}

\newcommand{\mbA}{\mathbb{A}}
\newcommand{\p}{L}
\renewcommand{\a}{R}

\newcommand{\diag}{\text{diag}}
\newcommand{\drangle}{ \rangle\!\rangle}
\newcommand{\dlangle}{ \langle\!\langle}

\newcommand{\Bsym}{\mathfrak{B}_{\text{sym}}}
\newcommand{\Bphys}{\mathfrak{B}_{\text{phys}}}
\newcommand{\Binp}{\mathfrak{B}_{\text{inp}}}

\newcommand{\lbb}{\left[}
\newcommand{\rbb}{\right]}

\newcommand{\bbI}{\mathbb{I}}

\newcommand{\wt}{\widetilde}

\newcommand{\Z}{{\mathbb Z}}

\newcommand{\bC}{{\mathbb C}}
\newcommand{\bI}{{\mathbb I}}

\newcommand{\cA}{\mathcal{A}}

\newcommand{\cc}{\mathcal{C}}
\newcommand{\cC}{\mathcal{C}}
\newcommand{\cD}{\mathcal{D}}

\newcommand{\cH}{\mathcal{H}}

\newcommand{\cL}{\mathcal{L}}
\newcommand{\cM}{\mathcal{M}}

\newcommand{\cO}{\mathcal{O}}
\newcommand{\cP}{\mathcal{P}}

\newcommand{\cS}{\mathcal{S}}

\newcommand{\cU}{\mathcal{U}}

\newcommand{\cW}{\mathcal{W}}

\newcommand{\cZ}{\mathcal{Z}}

\newcommand{\fB}{\mathfrak{B}}

\newcommand{\Hom}{\text{Hom}}

\renewcommand{\Vec}{\mathsf{Vec}}
\newcommand{\Rep}{\mathsf{Rep}}

\renewcommand{\dim}{\text{dim}}

\newcommand{\Ising}{\mathsf{Ising}}
\newcommand{\TY}{\mathsf{TY}}

\newcommand{\sym}{\text{sym}}
\newcommand{\phys}{\text{phys}}

\usepackage{comment}

\usepackage{physics}

\makeatother

\newcommand{\I}{\text{I}}
\newcommand{\II}{\text{II}}
\newcommand{\III}{\text{III}}

\usepackage{makecell}
\makeatletter
\newcommand\xlabel[2][]{\phantomsection\def\@currentlabelname{#1}\label{#2}}
\makeatother

\makeatletter
\def\l@subsubsection#1#2{}
\makeatother

\begin{document}

%\title{SymTFT for Mixed State Gapped Phases with Non-Invertible Symmetries}
% \title{SymTFT Approach for Non-invertible Symmetries of Mixed States} 

\title{SymTFT Approach for  Mixed States with Non-Invertible Symmetries} 

 \author{Sakura Sch\"afer-Nameki$^{1}$}
 \author{Apoorv Tiwari$^{2}$}
\author{Alison Warman$^{1}$}
 \author{Carolyn Zhang$^{3}\ $}

\thanks{Authors are listed alphabetically, reflecting equal contribution to this work.}

\affiliation{$^{1}$Mathematical Institute, University
of Oxford, Woodstock Road, Oxford, OX2 6GG, United Kingdom}
\affiliation{$^{2}$Center for Quantum Mathematics at IMADA, Southern Denmark University,
Campusvej 55, 5230 Odense, Denmark}
\affiliation{$^{3}$Department of Physics, Harvard University,
Cambridge, Massachusetts 02138, USA}

\begin{abstract}
\noindent
We develop a general framework for studying phases of mixed states with strong and weak symmetries, including non-invertible or categorical symmetries. The central idea is to consider a purification of the mixed state density matrix, which lives in a doubled Hilbert space. We propose a systematic classification of phases in this doubled Hilbert space, relying crucially on the Symmetry Topological Field Theory (SymTFT) approach. This framework applies not only to group symmetries but also, importantly, to non-invertible symmetries. We illustrate the approach in 1+1d to classify phases with strong (non-)invertible symmetries, which include strong-to-weak spontaneous symmetry breaking (SWSSB) phases and mixed strong/weak symmetry-protected topological phases (SPTs). We also develop an approach for studying symmetries that involve a combination of strong and weak symmetries. A noteworthy example of this has weak non-invertible Kramers-Wannier duality symmetry and  strong $\mathbb{Z}_2$ symmetry. 
The continuum description is complemented by a lattice model analysis informed by the SymTFT framework. 
\end{abstract}

\maketitle

\tableofcontents 

%%%%%%%%%%%%%%%%%%%%%%%%%%%%%%%%%%%%%%%

\section{Introduction}

The classification of pure state gapped phases with generalized symmetries -- including categorical or non-invertible ones -- has recently seen major progress. This includes the study of gapped phases, their order parameters, and second-order phase transitions in the presence of such symmetries. 
A key advance is the formulation of the problem and resulting classification that is completely systematic, relying solely on the information of the symmetry, and is applicable in any dimension. 
The central tool enabling these advances is the Symmetry Topological Field Theory (SymTFT) \cite{Ji:2019jhk,  Gaiotto:2020iye, Apruzzi:2021nmk, Freed:2022qnc} (also known as ``topological holography"), which is particularly powerful for finite symmetries, putting invertible and non-invertible symmetries on the same footing, and extends to higher dimensions. 
The formalism applies equally to standard symmetries described by groups, but also to more unusual, so-called categorical or non-invertible symmetries. For recent review of non-invertible symmetries see \cite{Schafer-Nameki:2023jdn,Shao:2023gho}.

 The core idea of the SymTFT is to associate a symmetry in $d+1$ dimensions with a $d+2$-dimensional topological quantum field theory, allowing for a characterization of symmetry properties largely independent of the system’s dynamics, i.e. a choice of symmetric Hamiltonian. This facilitates the classification of gapped phases purely on the grounds of the properties of the symmetry, including its anomalies. For groups this is of course well-known and part of the Landau paradigm. For non-invertible symmetries, the recent progress extending the Landau paradigm to categorical symmetries hinges very strongly on the SymTFT framework \cite{Bhardwaj:2023fca}, and can be carried out for gapped phases in 1+1d both in the continuum and in lattice models see \cite{Bhardwaj:2023idu, Bhardwaj:2024kvy, Chatterjee:2024ych, Bhardwaj:2024wlr, Warman:2024lir, Aksoy:2025rmg, Bottini:2025hri, Lu:2025rwd} and 2+1d \cite{Bhardwaj:2024qiv, Xu:2024pwd, Bullimore:2024khm,  Bhardwaj:2025piv,  Inamura:2025cum}, and for gapless phases, that are second order phase transitions between the gapped phases \cite{Chatterjee:2022tyg, Bhardwaj:2023bbf, Wen:2023otf, Bhardwaj:2024qrf,  Wen:2024qsg,  Bhardwaj:2025jtf, Wen:2025thg}.

In parallel development, much recent progress has been made on mixed state ``gapped" phases, in the presence of finite, group-like symmetries \cite{coser2019,degroot2022,ma2023average,lee2025,Ma:2023rji,chen2024,Ma:2024kma,sala2024,lessa2025swssb,lessa2025,wang2025open,xu2025}. Mixed states are described by density matrices $\rho$, and arise naturally in the study of open quantum systems. Symmetries of mixed states can be weak or strong, and the interplay between weak and strong symmetries can give rise to qualitatively new phenomena. A strong symmetry of a density matrix satisfies 
\be
U_g \rho = \rho U_g = e^{i\theta_g}\rho \,,
\ee
whereas a weak symmetry requires action on both sides of the density matrix
\be
U_g \rho U_g^\dagger = \rho 
\ee
for all elements $g$ in a group $G$. 
Particularly interesting mixed state phases include strong-to-weak spontaneous symmetry breaking (SWSSB) \cite{sala2024,lessa2025swssb,Liu:2024mme,zhang2025swssb,kuno2025swssb}, where a strong symmetry is spontaneously broken but its corresponding weak symmetry is preserved. There are also SPT phases protected by combinations of strong and weak symmetries \cite{degroot2022,ma2023average,zhang2022strange,sala2024,lee2025}. 
Most of the existing studies consider invertible, i.e. group-like symmetries, with the exception of \cite{Sun:2025que} that studies MPOs in 1+1d, and topological order with non-invertible 1-form symmetries \cite{Zini:2021lte, Ellison:2024svg,sohal2025,sala20251,sala20252}.

The goal of this work is to combine these two developments, and propose a general framework to systematically classify mixed state phases in the presence of symmetries, including non-invertible (categorical) symmetries. Let us denote the symmetry (category) by $\cS$. The key tool to achieve this is to map the mixed state problem to a pure state formulation using the Choi-Jamiolkowski isomorphism (also known as thermofield double), and then to formulate the classification of phases in terms of the SymTFT. 
Concretely, 
for a density matrix $\rho = \sum_n p_n  \ket{\psi_n}\bra{\psi_n}$
the Choi state in the  doubled Hilbert space (left and right) $\cH_{\p}\otimes \cH_{\a}$, where $\cH_{\p}\cong \cH_\a \cong \cH $ is 
\be\label{chois}
    |\rho \drangle = \frac{1}{\mathrm{Tr}(\rho^2)}\sum_{n}{ p_{n}}|\psi_n\rangle|\overline{\psi}_n\rangle \in \cH_{\p} \otimes \cH_{\a}\,,
\ee
where the overline denotes complex conjugation. The strong symmetry $s\in \cS$ then acts as 
\be\label{choistrong}
\ba
   (D_{s,L}\otimes \mathbf{1}_R)|\rho\RR &=c_s|\rho\RR\\
 (\mathbf{1}_L\otimes \ol{D}_{s,R})|\rho\RR& =\ol{c}_s|\rho\RR\,.
\ea
\ee
These symmetries are permuted under $T$: 
\be
T D_{s,L}T^{-1} = \ol{D}_{s,R}\,.
\ee
Weak symmetries of $\rho$ act on both $\mathcal{H}_L$ and $\mathcal{H}_R$:
\be\label{choiweak}
    D_{s,L}\otimes \ol{D}_{s,R}|\rho\RR=|c_s|^2|\rho\RR\,.
\ee
In this doubled description we can now apply the SymTFT approach, suitably modified to account for the fact that the resulting Choi state phase has a corresponding consistent mixed state phase. 
The requirement of a positive, Hermitian density matrix imposes additional restrictions on the SymTFT construction. We provide precise formulations how to implement strong and weak symmetric mixed phases in the SymTFT. 

The core of this paper exemplifies this in terms of 1+1d mixed state phases. 
More precisely,  we will classify phases of mixed states with strong and weak 0-form symmetries, including symmetries that are non-invertible (i.e. fusion category symmetries). We will study ``gapped" mixed states, which we define in this work as mixed states whose canonical purification admits a gapped local parent Hamiltonian with a finite number of degenerate ground states. In 1+1d, for every gapped phase, we can find a fixed point state $\rho$ that is proportional to a product of local projectors. For these states, the canonical purification and the normalized Choi state coincide, so for the purpose of studying these fixed points we lose no generality in computing Choi state properties.

The plan of the paper is as follows: 
 in the rest of this section, we will summarize our main results and provide some illustrative examples of fixed point mixed state density matrices. In Sec.~\ref{sec:Mix} we review mixed states and the Choi state reformulation, as well as some basic symmetry concepts in this context, in particular the diagnostics for phases of mixed states in the presence of symmetries. 
 The SymTFT description is provided in Sec.~\ref{sec:SymStrong}, where we start with a review of the standard pure state SymTFT in Sec.~\ref{sec:PureStateSymTFT}. 
 We then explain the study of strong symmetry phases via the SymTFT in Sec.~\ref{sec:SymStrongPhases}, explaining constraints that come from hermiticity and positivity of density matrices. % and the 
 %classification of strong symmetric mixed phases for any 1+1d fusion category symmetry in Sec.~\ref{sec:StrongGapped}. 
 Lattice models that realize these mixed phases with strong symmetry are constructed in Sec.~\ref{Sec:Lattice Models}, for any fusion category symmetry using the anyon chain. We will focus on group $G$ and non-invertible $\Rep (G)$ symmetries which can be realized in Hilbert spaces with a tensor product decomposition. 
Examples for strong symmetric mixed phases are given in Sec.~\ref{Sec:Examples}, for invertible and non-invertible symmetries, showing both the continuum SymTFT derivation and a lattice model realization. 

In Sec.~\ref{sec:WeakStrong} we turn to mixed phases with weak symmetries or combined weak and strong symmetries. Again, we give a fairly systematic way to construct weak symmetries, and provide constraints on what symmetries can be made weak. The most interesting setups are when there are both weak and strong symmetries. For instance we show that there are non-invertible symmetries like the Ising or Tambara-Yamagami cagetories, where a non-invertible Kramers-Wannier duality can be made weak, and the invertible part of the symmetry remains strong. Again, we give a SymTFT derivation and a lattice model description for such phases. 
{Finally, we also construct a mixed strong-weak SPT phase  with weak non-invertible duality and strong $\Z_2\times \Z_2$ symmetry, where one of the $\Z_2$ symmetries forms an SPT with the weak duality, while the other $\Z_2$ is SWSSB'ed.}

\subsection{Summary of Results: SymTFT Proposal For Mixed State Phases}

The SymTFT formulation of mixed states can be summarized as follows (for the reader unfamiliar with the pure state SymTFT, we have included a review in Sec.~\ref{sec:PureStateSymTFT}). 
Let $\cS$ be a fusion category symmetry. E.g. a finite group $G $(with choice of 3-cocycle $H^3(G,U(1))$), or a representation category of a finite group $\Rep (G)$, $\Ising$, Fibonacci, etc. 

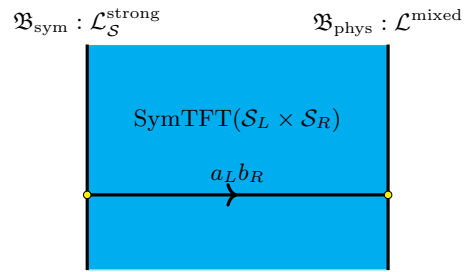
\begin{figure}
$
\begin{tikzpicture}
\begin{scope}[shift={(0,0)}]
\draw [cyan,  fill=cyan] 
(0,0) -- (0,3) --(4,3) -- (4,0) -- (0,0) ; 
\draw [white](0,0) -- (0,3) --(4,3) -- (4,0) -- (0,0) ;  ; 
\draw [very thick] (0,0) -- (0,3)  ;
\draw [very thick] (4,3) -- (4,0) ;
\node at (2,2) {$\text{SymTFT}(\cS_L\times \cS_R)$} ;
\node[above] at (0,3) {$\Bsym: \cL_{\cS}^\strong$}; 
\node[above] at (4,3) {$\Bphys: {\cL}^{\text{mixed}} $};
 \draw [very thick, ->-] (0,1)  -- (4,1) ;
 \node[above] at (2, 1.05) {$a_Lb_R$};
  \draw [black,fill=yellow] (0,1) ellipse (0.05 and 0.05);
  \draw [black,fill=yellow] (4,1) ellipse (0.05 and 0.05);
\end{scope}
% \begin{scope}[shift={(6,0)}]
% \node at (-1,1.5) {$=$} ; 
% \draw [very thick] (0,0) -- (0,3) ;
% \node[right] at (0, 1.5) {$\cO_a$};
%  \draw [black,fill=yellow] (0,1.5) ellipse (0.05 and 0.05);
% \end{scope}
\end{tikzpicture}
$
\caption{SymTFT for mixed states with strong symmetry $\cS_L\times \cS_R$ where $\cS_R= \ol{\cS_L}$. The symmetry boundary $\Bsym$ realizes the strong symmetry, which is given by a Lagrangian algebra $\cL_{\cS}^\strong= \cL_{\cS_L}\otimes \cL_{\cS_R}$, i.e. a product of the $L$ and $R$ Lagrangian algebras that give rise to the $\cS_{L/R}$ symmetry categories. For gapped mixed phases, the  physical boundary on the right hand side has to be a mixed state Lagrangian algebra satisfying ensuring positivity and hermiticity of the density matrix. The anyon $a_Lb_R$ ending on both boundaries yields a local order parameter, if it is also invariant under the $T$ operation \label{fig:Syms}}
\end{figure}

\smallskip
\noindent{\bf Strong symmetric mixed phases.}
To realize a strong $\cS$ symmetry, we double the category and consider instead of the SymTFT for $\cS$, the one for $\cS_L \boxtimes \cS_R$,\footnote{Here, the Deligne product $\boxtimes$ is roughly the generalization of the direct product operation on groups to categories: objects in $\mathcal{S}_L\boxtimes\mathcal{S}_R$ include all objects of the form $s_Ls_R$ for $s_L\in\mathcal{S}_L$ and $s_R\in\mathcal{S}_R$.} where 
\be
\cS_R = \ol{\cS_L}\,.
\ee
Most importantly, the starting point for strong $\cS$ symmetric phases is the Drinfeld center
\be\label{ZSLSR}
\cZ(\cS_L\boxtimes \cS_R)\,.
\ee
Denote by $\cL_{\cS}$ the Lagrangian algebra  of $\cZ(\cS)$, which gives rise to the $\cS$ symmetry for pure states $|\rho\RR$ in the doubled Hilbert space. Then the strong symmetry is realized by 
\be
\cL_\cS^\strong = \cL_{\cS_L} \otimes \cL_{\cS_R} \,.
\ee
To classify the strong symmetric phases, we need to specify the physical boundary condition to be a gapped boundary of (\ref{ZSLSR}). Not all Lagrangian algebras however give rise to consistent mixed states. We provide arguments, that in order for a Lagrangian algebra $\cL= \oplus_{a,b} n_{a_L, b_R} a_L \otimes b_R$ to give rise to a positive, Hermitian density matrix, it has to be a so-called {\bf mixed state Lagrangian algebra}, which has the following two properties: Hermiticity implies the anti-unitary $T$ invariance given by 
\be
n_{a_L, b_R} = n_{\ol{b_L}, \ol{a_R}}\,,
\ee
and positivity implies 
\be
n_{a_L,b_R}\leq n_{a_L,\overline{a}_R}n_{\overline{b}_L,b_R} \,.
\ee
This schematically means that the diagonal entries are bounded below by the off-diagonal entries, viewing $n_{a_L,b_R}$ as a matrix between $L$ and $R$ anyons. 
We show that this allows us to systematically determine the mixed state gapped phases with strong symmetry. 
The SymTFT setup is shown in Fig.~\ref{fig:Syms}.

Choi states for specific gapped phases are constructed from the interval compactification of the SymTFT with one boundary $\Bsym$ and the other ``physical" boundary condition $\mathfrak{B}_{\mathcal{L}}$ given by $\mathcal{L}$. Generally speaking, when there is spontaneous symmetry breaking, there are generally multiple locally indistinguishable density matrices (we will clarify the meaning of this in later sections) obtained from locally indistinguishable Choi states. In the SymTFT framework, different states corresponding to vacua in different symmetry sectors of the spontaneously broken symmetry are obtained by insertion of anyon lines that tunnel between both boundaries. For Choi states however, these anyon lines must be $T$ symmetric. Therefore, the number of vacua that are valid Choi states is generally smaller than the total number of vacua in the doubled Hilbert space.

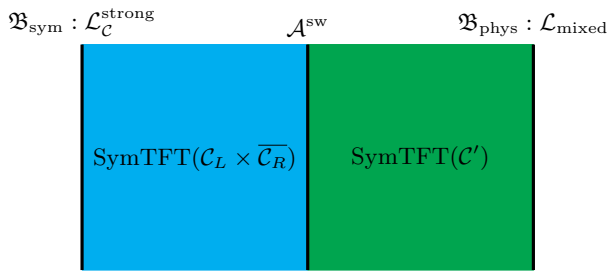
\begin{figure}
$
\begin{tikzpicture}
\begin{scope}[shift={(0,0)}]
\draw [cyan,  fill=cyan] 
(0,0) -- (0,3) --(3,3) -- (3,0) -- (0,0) ; 
\draw [Green, fill= Green](3,0) -- (3,3) --(6,3) -- (6,0) -- (3,0) ; 
\draw [very thick] (0,0) -- (0,3) ;
\draw [very thick] (3,3) -- (3,0) ;
\draw [very thick] (6,3) -- (6,0) ;
\node at (1.5,1.5) {$\SymTFT(\cC_L\times \overline{\cC_R})$} ;
\node at (4.5,1.5) {$\SymTFT(\cc')$} ;
\node[above] at (0,3) {$\Bsym: \cL_{\cC}^{\text{strong}}$}; 
\node[above] at (3,3) {$\cA^\sw$}; 
\node[above] at (6,3) {$\Bphys: {\cL}_{\text{mixed}} $}; 
\end{scope}
\end{tikzpicture}
$
\caption{SymTFT for Mixed States with weak and strong  symmetry. The starting point on the left is an enlarged SymTFT, which realizes a strong symmetry $\cC_L\times \cC_R$. The SymTFT that realizes a combination of weak and strong symmetries is obtained after partial anyon condensation along the interface. This is specified by a mixed state condensable algebra $\cA^{\text{mixed}}$, which yields a reduced SymTFT for a symmetry $\cC'$. The simplest example is when we have a purely weak symmetry $\cW$. In this case $\cC_{L/R}= \cW_{L/R}$ and $\cC'=\cW$. Then $\cA$ condense the diagonal charges.  
 \label{fig:Symws}}
\end{figure}

\smallskip

\noindent{\bf Weak and Strong Symmetric Mixed Phases.}
To construct mixed gapped phases with both weak and strong symmetries, we use an auxiliary, enlarged symmetry which is purely strong, and then partially condense to realize the mixed strong and weak symmetries. The setup is summarized in Fig.~\ref{fig:Symws}.

We start with a strong symmetry $\cC_L\times \cC_R$ and partially condense to a smaller symmetry, which can be interpreted as having weak and strong symmetries. 
The algebras that specify the anyon condensation have to again be consistent with the properties of density matrices and should be mixed state condensable algebras (again satisfying the positivity and $T$-invariance). Furthermore, in order to realize a weak symmetry, we need to condense partially a set of diagonal charges.  

Consider first a purely weak symmetry $\cW$. We start with the SymTFT for the enlarged symmetry, which is simply the strong symmetry TFT for $\cW$ and condense the diagonal charges 
\be
\cZ (\cW_L \boxtimes \cW_R) \to \cZ (\cW)\,.
\ee
Invertible symmetries can straight-forwardly be made weak in the SymTFT framework. However, interestingly, we show that 
a {\bf non-invertible symmetry cannot be entirely made weak.}  

However, one can make non-invertible symmetries partially weak, leading to mixed strong weak symmetries. In this case, we only condense a subset of diagonal charges, and part of the symmetry becomes weak after the condensation. The most interesting case arises for {\bf weak  non-invertible symmetries combined with strong invertible} ones. The main example to illustrate this is the case of Tambara-Yamagami categories for an abelian group $\mbA$
\be
\cC_L= \ol{\cC_R} = \TY (\mbA) \,,
\ee
e.g. $\mbA = \Z_2$ this is the Ising fusion category. 
We partially condense anyons so that only the diagonal duality symmetry survives and becomes weak, whereas the abelian symmetry $\mbA$ stays strong. The resulting SymTFT is 
\be
\cZ (\cC') = \cZ (\TY (\mbA_L \times \mbA_R)) \,.
\ee
The induced symmetry boundary condition for this reduced SymTFT is obtained by collapsing the first interval of Fig.~\ref{fig:Symws}. The gapped phases are again obtained by choosing mixed state Lagrangian algebras for $\cZ (\cC')$, which is invariant under the $T$ operation, that exchanges left and right. The advantage of this approach is that we can trace back the $L$ and $R$ assignments to the larger topological order (TO). 

Another reason to start in the larger symmetry $\mathcal{C}_L\boxtimes\mathcal{C}_R$ is this naturally allows for the construction of models for $|\rho\RR$ in the doubled Hilbert space. The condensation of diagonal charges is related to explicitly breaking the undesired strong symmetries to obtain the desired combination of strong and weak symmetries.
The same constraints on valid symmetry broken vacua as discussed previously also applies in this case.

\subsection{A Key Example}

To provide some intuition for the novel kinds of mixed state phases we will study, we begin with an illustrative example. We will encounter this example again and in more detail later in later sections.

We will first use the example to demonstrate what is known as \emph{strong-to-weak} spontaneous symmetry breaking. As we will review in more detail below, strong symmetries of density matrices is given by an operator $U$ for which $U\rho=s\rho$ for some complex scalar $s$. In the unitary case, $|s|=1$. A strong symmetry automatically implies a weak symmetry, i.e. invariance of $\rho$ up to a scalar under conjugation with $U$. Consider the Hilbert space of a 1+1d chain of $N$ qubits, with $\mathbb{Z}_2$ symmetry action $U=\prod_iX_i$. Then the following density matrix has a strong $\mathbb{Z}_2$ symmetry, that automatically comes with a weak $\mathbb{Z}_2$ symmetry:
\begin{equation}\label{z2swssb}
    \rho=\frac{1}{2^N}\Bigl(1+\prod_iX_i\Bigr) \,.
\end{equation}
This state is simply the projection onto the $\mathbb{Z}_2$ even subspace $\mathcal{H}_e$: $\rho=\sum_{|\psi_i\rangle\in\mathcal{H}_e}|\psi_i\rangle\langle\psi_i|$. One can check that $U\rho=\rho U=\rho$ and $U\rho U^\dagger = \rho$. The strong $\mathbb{Z}_2$ symmetry is spontaneously broken, as indicated by the long-range ``Renyi-2" 
correlator
\begin{equation}\label{renyi2zz}
    \mathrm{Tr}(\rho Z_iZ_j\rho Z_iZ_j)=1 \,.
\end{equation}
Note that this is actually a long-range connected correlation function because we can subtract $\mathrm{Tr}(\rho Z_i\rho Z_i)\mathrm{Tr}(\rho Z_j\rho Z_j)$, which is guaranteed to be zero by the strong symmetry. This correlator precludes localization of the strong symmetry, $U_A\rho=O_LO_R\rho$, where $U_A$ is the restriction of the operator $U$ to a region $A$ and $O_L,O_R$ act on the left and right endpoints of $A$ respectively. 
Localization of the strong symmetry is clearly true for the fixed point strongly symmetric state $|+\rangle\langle +|$ but cannot occur for any density matrix satisfying (\ref{renyi2zz}), as can be seen by choosing $A$ to contain $j$ but not $i$ and using the anti-commutation relations of $U_A$ and $Z_j$. The interesting aspect of (\ref{z2swssb}) is that its weak symmetry is \emph{not} spontaneously broken: 
the weak symmetry is not prevented from localizing by (\ref{renyi2zz}) and in fact does localize, since $U_A\rho U_A^\dagger =\rho$. In summary, the strong $\mathbb{Z}_2$ symmetry of (\ref{z2swssb}) is spontaneously broken but the weak $\mathbb{Z}_2$ symmetry is not.
Operationally, this means that (\ref{z2swssb}) is not two-way symmetric local channel connected to $|+\rangle\langle+|$ or $|\mathrm{GHZ}+\rangle\langle \mathrm{GHZ}+|$ (where $|\mathrm{GHZ}+\rangle$ is the GHZ state with eigenvalue $+1$ under the global symmetry); it lies in an intrinsically mixed phase. Notice that together with (\ref{z2swssb}), there is another density matrix $\frac{1}{2^N}(1-\prod_iX_i)$ and the two density matrices are indistinguishable by any local $\mathbb{Z}_2$ symmetric linear or Renyi-2 correlator. This two-fold ``degeneracy" reflects the spontaneous breaking of the strong symmetry.

The same density matrix (\ref{z2swssb}) also has a weak non-invertible duality symmetry $D\rho D^\dagger = 2\rho$. $D$ here is the lattice implementation of the Ising duality operator \cite{seiberg2024}. We will see later on that the weak duality symmetry is actually spontaneously broken in this state, despite the fact that there is no order parameter that is linear in $\rho$. 
{This may come as a surprise because a linear order parameter is required for SSB of weak invertible symmetries. (\ref{z2swssb}) also serves as a particularly interesting example where in the implementation of the symmetry on a chain of qubits as in \cite{seiberg2024}, where the fusion rules of $D$ mix with translation, there is no trivially symmetric density matrix with the strong $\mathbb{Z}_2$ symmetry together with the weak duality symmetry, even though there is a trivially symmetric state in the doubled Hilbert space. This subtle anomaly-like constraint comes from the fact that any trivially symmetric state in the doubled Hilbert space cannot be the Choi state of a Hermitian, positive semi-definite density matrix.}

\section{Mixed States, (Non-)invertible Symmetries and Phases} 
\label{sec:Mix}

In this section, we will first review the definitions of strong and weak symmetries of density matrices. We will begin with invertible symmetries, and then  give the natural generalization to non-invertible symmetries, focusing on symmetries in 1+1d. We will furthermore review how a density matrix can be mapped to a pure state in a doubled Hilbert space via the Choi-Jamiołkowski map and the canonical purification. For fixed-point density matrices, the normalized Choi state coincides with the canonical purification. Therefore, we will use the Choi state for much of the subsequent analysis. Phases of density matrices detected by linear and Renyi-2 order, disorder, and string order parameters can be understood from phases of the corresponding pure Choi states.

\subsection{Strong and Weak Symmetries}

Consider a density matrix $\rho = \rho^\dagger$, where $\langle A \psi | \phi \rangle = \langle  \psi | A^\dagger \phi \rangle $ for all $\phi, \psi \in \cH$, where $\cH$ is the Hilbert space. We can write 
\be\label{densitymatrix}
\rho = \sum_n p_n  \ket{\psi_n}\bra{\psi_n}  \,.
\ee
{with $0\leq p_n\leq 1$ and $\sum_np_n=1$.} The state is pure if and only if $\rho^2=\rho$, i.e. $\rho = |\psi\rangle\langle\psi|$. 

\vspace{2mm}
\noindent {\bf Invertible Symmetry Groups.}
Consider a finite symmetry group $G$ with a unitary representation on the Hilbert space given by $\{U_g\}_{g\in G}$. Then a density matrix $\rho$ has a strong $G$ symmetry if
\begin{equation}\label{strongsym}
    U_g\rho =e^{i\theta_g}\rho\qquad\forall g\in G\,,
\end{equation}
where $\{e^{i\theta_g}\}$ form a 1d representation of $G$. Notice that taking the conjugate transpose of both sides above gives $\rho U_g^\dagger =e^{-i\theta_g}\rho$. Therefore, $\rho$ has both a left strong symmetry $\{U_g\}$ and a right strong symmetry $\{U_g\}=\{U_g^\dagger\}$. It follows that the presence of a strong symmetry always implies the presence of a weak symmetry:
\begin{equation}\label{weaksym}
    U_g\rho U_g^\dagger = \rho\qquad \forall g\in G\,.
\end{equation}
In the invertible case, the above definition of a weak symmetry means that $[U_g,\rho]=0$. Note that while a strong symmetry implies a weak symmetry, just because $\rho$ has a weak symmetry satisfying (\ref{weaksym}) \emph{does not} mean it also has a strong symmetry. For example, if $\rho$ consists of a mixture of pure states in different symmetry sectors, it has a weak symmetry but not a strong symmetry.

{In addition to the symmetry group $G$, the unitary representation also encodes an $F$ symbol, or anomaly $[\omega]\in H^3(G,U(1))$. This function from three group elements to $U(1)$ describes fusion properties of symmetry defects.}

\vspace{2mm}
\noindent {\bf Non-Invertible Symmetries.}
Symmetries in 1+1d quantum systems can form more general fusion categories. A fusion category $\mathcal{S}$ has a finite set of simple objects $\{s\}$ and each object corresponds to an operator $D_s$ on the Hilbert space. However, for $s$ that is not invertible, $D_s$ is not unitary. Although we do not require every simple object to be invertible, for every $s$ we do require that there exists a $s^\dagger\in \cS$ such that 
\be
s \times s^\dagger = 1+ \sum_{i} m_i a_i \equiv 1+ a 
\ee
with non-negative integer coefficients $m_i \geq 0$ and $a_i \in \cS$.\footnote{In other papers, the dual of $s$ may be written as $\bar{s}$ rather than $s^\dagger$. 
Here we use $s^\dagger$ to refer to the dual in a fusion category (also called charge conjugate in a modular tensor category) because we will use $\bar{s}$ to refer to complex conjugation.} 
{Like for invertible symmetries, the same fusion rules may give rise to distinct fusion category symmetries distinguished by their $F$ symbols. For non-invertible symmetries, the $F$ symbol is not a 3-cocycle, and satisfies the more general constraint described by the pentagon equation.} Note that $D_{s^\dagger}= D_s^\dagger$ in known constructions of systems with non-invertible symmetries. 
All the models we consider in this work are constructible as anyon chain models where $D_{s^{\dagger}}$ is manifestly $D_{s}^{\dagger}$, since taking the dual of an object in the symmetry category reverses its orientation which is implemented as hermitian conjugation on the state space. 
We say that $\rho$ has a strong (non-invertible) symmetry if
\begin{equation}\label{NonInvonRhoStrong}
    D_s\rho=c_s\rho\qquad \forall s\in \mathcal{S} \,.
\end{equation}
Like in the invertible case, $c_s$ {can take different values for states in different symmetry sectors, and there is in general a ``singlet" state with $c_s=d_s$ where $d_s$ is the quantum dimension of $s\in\mathcal{S}$}. Taking the conjugate transpose of both sides above gives 
\be 
\rho D_s^\dagger = \overline{c}_s\rho\,,
\ee
(where $\overline{c}_s$ is the complex conjugate of $c_s$), so $\rho$ has both a left strong symmetry $\{D_s\}$ and a right strong symmetry given by the operators $\{D_s^\dagger\}$. For non-invertible symmetries, the presence of a strong symmetry implies the presence of the following weak symmetry:
\begin{equation}\label{weaksymnon}
    D_s\rho D_s^\dagger = |c_s|^2\rho \qquad \forall s\in\mathcal{S}\,.
\end{equation}
In the invertible case, the fact that $U_g^\dagger$ is a right strong symmetry of $\rho$ means that $U_g$ is also a right strong symmetry of $\rho$. Similarly, due to the requirement that $s$ comes with a dual $s^\dagger$, the fact that $D_s^\dagger$ is a right strong symmetry of $\rho$ means that $D_s$ is also a right strong symmetry of $\rho$, so $D_s$ commutes with $\rho$. 
Like in the invertible case, the presence of a weak symmetry satisfying (\ref{weaksymnon}) does not imply that there is a corresponding strong symmetry.

\subsection{Choi State and Canonical Purification}\label{schoiintro}
It will be useful to map the density matrix $\rho$ to a pure state in a doubled Hilbert space.
\begin{comment}
Consider a 1+1 dimensional quantum system with strong $\cS$ symmetry, where $\cS$ is a fusion category.
Such a system could be formulated within continuum quantum field theory or as a lattice model.
For now, we keep the discussion general, and will describe the corresponding lattice models in Sec.~\ref{Sec:Lattice Models}.
Let us denote the symmetry operator corresponding to a simple object $s\in \cS$ as $\cU_{s}$.
Symmetries of mixed states come in two broad families, so called weak (average) and strong (exact) symmetries, which satisfy the following properties:
\begin{equation}
\begin{alignedat}{3}
{\rm Strong:}& \quad && \cU_s \rho = \rho \,  \cU_s\propto \rho \,,
\\
{\rm Weak:}& \quad \quad &&\cU_s\rho\, \cU_s^\dagger  \propto \rho  \,.
\\
\end{alignedat}    
\end{equation}
In order to be able to characterize phases with such symmetries, we apply the Choi state operator map to a doubled Hilbert space.
\end{comment}
Specifically, given a density matrix of the form (\ref{densitymatrix}), we can obtain a state in a doubled Hilbert space $\cH_{\p}\otimes \cH_{\a}$, where $\cH_{\p}\cong \cH_\a \cong \cH $:
\begin{equation} \label{eq:rho_Choi}
    |\rho \drangle \equiv |\rho^{\text{Choi}}\drangle  = \frac{1}{\mathrm{Tr}(\rho^2)}\sum_{n}{ p_{n}}|\psi_n\rangle|\overline{\psi}_n\rangle \in \cH_{\p} \otimes \cH_{\a}\,.
\end{equation}
where $|\overline{\psi}_n\rangle$ denotes complex conjugation. Sometimes, the unnormalized state without the $\frac{1}{\mathrm{Tr}(\rho^2)}$ is referred to as the Choi state. We will use the normalized state. One can also construct a state in the same Hilbert space that is automatically normalized using the Choi state of $\sqrt{\rho}$. This state is also called the canonical purification:
\be
   |\rho^{\text{c}}\drangle = \sum_{n}\sqrt{ p_{n}}|\psi_n\rangle|\overline{\psi}_n\rangle\,.
\ee
The canonical purification is a purification of the state $\rho$ because tracing out the $\cH_\a$ degrees of freedom returns $\rho$:
\begin{equation}
    \Tr_{\cH_{\a}}|\rho^{\text{c}} \drangle \dlangle \rho^{\text{c}} | =\rho\,.
\end{equation}
Both the Choi state and the canonical purification automatically have a $\mathrm{SWAP}^*$ symmetry implemented by the anti-unitary operator $T$ that swaps states in $\mathcal{H}_L$ with those in $\mathcal{H}_R$ and performs complex conjugation (denoted by bar). This $\mathrm{SWAP}^*$ symmetry will be very important because while every density matrix maps onto a state in the doubled Hilbert space, not every state comes from a density matrix, and at the very least any state in $\mathcal{H}_L\otimes\mathcal{H}_R$ that comes from a valid density matrix must satisfy 
\be
T|\rho\RR=|\rho\RR\,.
\ee

In much of what follows, we will be working  with fixed-point state density matrices within each mixed-state gapped phase. Since the pure state density matrices are proportional to projectors, they satisfy $\rho^2 \propto \rho\propto\sqrt{\rho}$. In this case, $|\rho^{\text{c}}\drangle$ and $|\rho\drangle$ are often on equal footing. We will mostly work with $|\rho\drangle$ for simplicity.

For more general density matrices, $\rho$ and $\sqrt{\rho}$ are known to behave differently. Therefore, when interpolating a density matrix through a mixed state phase transition, the corresponding Choi state and canonical purification generically demonstrate transitions at different points in parameter space and also demonstrate different universality classes of critical behavior. For these more general states, behavior of the canonical purification is more faithful to the properties of $\rho$, i.e. its transition coincides with the actual mixed-state transition point.

Strong and weak symmetries of $\rho$ map to symmetries of the pure state $|\rho\RR$. The left and right strong symmetries map to symmetries of $|\rho\RR$ that act on only $\mathcal{H}_L$ or $\mathcal{H}_R$: 
\begin{align}
\begin{split}
    D_s\rho&=c_s\rho\ \to\  (D_{s,L}\otimes \mathbf{1}_R)|\rho\RR=c_s|\rho\RR\\
\rho D_s^\dagger&=\ol{c}_s\rho\ \to\  (\mathbf{1}_L\otimes \ol{D}_{s,R})|\rho\RR=\ol{c}_s|\rho\RR\,.
\end{split}
\end{align}
These symmetries are permuted under $T$: 
\be
T D_{s,L}T^{-1} = \ol{D}_{s,R} \,.
\ee
Weak symmetries of $\rho$ act on both $\mathcal{H}_L$ and $\mathcal{H}_R$:
\begin{equation}
    D_s\rho D_s^\dagger = |c_s|^2\rho\ \to\ 
    D_{s,L}\otimes \ol{D}_{s,R}|\rho\RR=|c_s|^2|\rho\RR\,.
\end{equation}
Weak symmetry operators are invariant under $T$:
\be
T(D_{s,L}\otimes \ol{D}_{s,R})T^{-1} = D_{s,L}\otimes \ol{D}_{s,R}\,.
\ee

\subsection{Phases with Strong and Weak Symmetry}

Here we summarize some of the standard diagnostics for various symmetry breaking and symmetric phases in mixed states, and connect them to the usual diagnostics for pure state phases of $|\rho\RR$ \cite{Ma:2023rji}. First we formulate this for finite group symmetries $G$. Explicit strong symmetry breaking is detected by a non-zero linear expectation value of a charged operator
\begin{equation}\label{linearO}
    \mathrm{Tr}(\rho \mathcal{O}(x))\propto\mathrm{Tr}(\rho^2 \mathcal{O}(x))=\LL\rho|\mathcal{O}_L(x)|\rho\RR\neq 0\,, 
\end{equation}
where the proportionality constant is given by $\frac{1}{\mathrm{Tr}(\rho^2)}$. Explicit strong symmetry breaking is also detected by nonzero Renyi-2 expectation value of a charged operator
\begin{equation}\label{nonlinearO}
    \mathrm{Tr}(\rho \mathcal{O}(x)\rho \mathcal{O}^\dagger(x))=\LL\rho|\mathcal{O}_L(x)\ol{\mathcal{O}}_R(x)|\rho\RR\neq 0\,.
\end{equation}

Note that (\ref{linearO}) also detects explicit breaking of the corresponding weak symmetry, but (\ref{nonlinearO}) does not. If the above correlation functions are zero, then $\rho$ can satisfy (\ref{strongsym}). Spontaneous strong symmetry breaking is detected by two-point correlation functions of oppositely charged (net neutral) operators:
\begin{equation}\label{linearcorr}
    \mathrm{Tr}(\rho\mathcal{O}(x)\mathcal{O}(y))\propto\mathrm{Tr}(\rho^2\mathcal{O}(x)\mathcal{O}(y))\neq 0 \,.
\end{equation}
Note that for generic states, we must take the limit as $|x-y|\to\infty$ but in the fixed-point states considered in this work, the above quantity converges for $|x-y|>1$. We can also detect spontaneous symmetry breaking using {\bf Renyi-2 order parameters}:
\begin{align}
\begin{split}\label{renyicorr}
    \mathrm{Tr}&(\rho \mathcal{O}(x)\mathcal{O}(y)\rho\mathcal{O}(y)^\dagger\mathcal{O}(x)^\dagger) \\
    &= \LL\rho|\mathcal{O}_L(x)\mathcal{O}_L(y)
    \ol{\mathcal{O}}_R(x)\ol{\mathcal{O}}_R(y)|\rho\RR\neq 0\,.
\end{split}
\end{align}
While the above quantity being nonzero indicates strong symmetry breaking, it does not indicate spontaneous breaking of weak symmetry. Therefore, the Renyi-2 correlator is useful for distinguishing complete symmetry breaking from strong-to-weak symmetry breaking, where a strong symmetry is spontaneously broken but its corresponding weak symmetry is not. In this case, the linear correlator (\ref{linearcorr}) would be zero but the Renyi-2 correlator (\ref{renyicorr}) would be nonzero.

Local order parameters do not distinguish between different symmetric phases. Disorder parameters, which include {\bf string order parameters}, are useful for this purpose. These are expectation values of truncations of the global symmetry operators $\{U_g\}$ with appropriate end-point dressings. Strong disorder parameters are given by expectation values of the form
\begin{equation}
    \mathrm{Tr}(\rho^2\mathcal{O}(x)U_{g}(x,y)\mathcal{O}(y))=\LL\rho|\mathcal{O}_L(x)U_{g,L}(x,y)\mathcal{O}_L(y)|\rho\RR\,,
\end{equation}
where $\mathcal{O}_L(x)\mathcal{O}_L(y)$ is net neutral under the strong symmetries, and $U_{g,L}(x,y)$ acts like the strong symmetry in the interval $[x,y]$ and acts trivially elsewhere. When $\mathcal{O}_L(x),\mathcal{O}_L(y)$ are individually charged under $U_{h,L}$ for some other strong symmetry element $h\in G$, $\mathcal{O}_L(x)U_{g,L}(x,y)\mathcal{O}_L(y)$ is known as a string order operator and its expectation value a string order parameter. Non-vanishing string order parameters detect SPT phases in 1+1d. 

We can also consider more general quantities of the form
\begin{equation}
    \LL\rho|\mathcal{O}(x)U_g(x,y)\mathcal{O}(y)|\rho\RR\,,
\end{equation}
where $\mathcal{O}(x)\mathcal{O}(y)$ act in $\mathcal{H}_L$, $\mathcal{H}_R$, or both $\mathcal{H}_L\otimes\mathcal{H}_R$. Similarly, $U_{g}(x,y)$ may be a restricted strong symmetry acting in $\mathcal{H}_L$ or $\mathcal{H}_R$, or it may be a restricted weak symmetry operator acting in $\mathcal{H}_L\otimes\mathcal{H}_R$. All of these expectation values can be translated to Renyi-2 type correlators of the density matrix. Different kinds of such operators detect SPTs involving both strong and weak symmetries. 

In all of the correlation functions above, we could have replaced $\rho$ with $\sqrt{\rho}$ and expectation values in $|\rho\RR$ by expectation values in $|\rho^c\RR$. Correlation functions in $|\rho^c\RR$ are also known as Wightman correlators, and the corresponding traces over products of operators with $\sqrt{\rho}$ are known as Holevo fidelities. As mentioned in the previous section, the density matrices we consider satisfy $\rho\propto\sqrt{\rho}$, so in this case Wightman correlators behave in the same way as Renyi-2 correlators.

Phases with non-invertible symmetries can also be detected by order parameters and disorder parameters, which include string order parameters. To understand these quantities for non-invertible symmetries, it will be useful to introduce the concept of symmetry topological field theory (SymTFT). We will see that in both the invertible and the non-invertible cases, there is a correspondence between the above order and disorder parameters to anyons that end at boundary conditions of the 2+1d SymTFT.

\section{SymTFT for Strong Categorical Symmetries}
\label{sec:SymStrong}

In this section we describe a SymTFT based framework for classifying and characterizing mixed state  phases with general finite strong symmetries. These finite strong symmetries can be either invertible (group-like) or non-invertible. In this section we will restrict to only strong symmetries. We will not consider states that have weak symmetries without corresponding strong ones. In Sec.~\ref{sec:WeakStrong} we will show how to use the SymTFT construction to understand more general symmetries that include such weak symmetries.

The standard SymTFT paradigm for 1+1d  closed quantum systems associates to a fusion category of a 1+1d system, a 2+1d topological field theory. More precisely it associates the 
2+1d Turaev-Viro TQFT (topological order) $\mathfrak{Z}(\cS)$ with topological defects (anyons) associated to the fusion category $\cS$ to the space of 1+1d $\cS$-symmetric pure states. 
The topological defects of this TQFT are central to the SymTFT paradigm, and form the Drinfeld center of the fusion category, $\cZ(\cS)$. We will abuse notation a bit and in the following not distinguish between $\mathfrak{Z}$ and $\cZ$. 
The SymTFT can be used to systematically classify and characterize pure state gapped and gapless phases given an input symmetry $\cS$. Furthermore, one can construct anyon-chain type lattice models exhibiting all universal symmetry-based features. To extend this framework to weak and strong symmetries of mixed states, we propose the following setup. 

Consider a density matrix with a strong $\cS$ symmetry describing a (possibly mixed) state in a 1+1d system, where $\cS$ is a fusion category.
Such a system could be formulated within continuum quantum field theory or as a lattice model.
For now, we keep the discussion general, and will describe the corresponding lattice models in Sec.~\ref{Sec:Lattice Models}.

In the context of the SymTFT, objects in $\cS$ become topological lines on a certain topological boundary of $\cZ(\cS)$. 
If we consider a density matrix with $\cS$ strong symmetry (without any additional independent weak symmetries), then the Choi state has an $\cS_L\boxtimes\cS_R$ symmetry coming from the left and right symmetry operators. Note that $\cS_R$ is constrained to be 
$\cS_R=\overline{\cS}_L$,
because $D_c^\dagger$ acting on the right side of $\rho$ gets transposed in the operator-state Choi map: $\rho D_c^\dagger $ gets mapped to $\overline{D}_{c,R}|\rho\RR$.

To study symmetries of the density matrix, we will use the SymTFT of the Choi state symmetry $\cS_\p  \boxtimes \cS_{R} $ where $\cS_L$ and 
\be
\cS_R=\ol{\cS_L}
\ee
act on the physical (left) and ancilla (right) Hilbert space of the Choi state, respectively.  
 The SymTFT for such a system has topological defects given by the Drinfeld center 
\be\label{DoubleCenter}
\cZ (\cS_\p  \boxtimes \cS_R ) \cong \cZ (\cS_\p ) \boxtimes \cZ (\cS_R )
\cong \cZ (\cS_\p ) \boxtimes \cZ (\ol{\cS_L} )
\,.
\ee
In general, $\cZ(\cS_L)\ncong \cZ(\cS_R)$. For example, the action of $\ol{\phantom{a}}$ on an invertible symmetry $\mathrm{Vec}_G^\omega$ with anomaly $[\omega]\in H^3(G,U(1))$ is to complex conjugate $\omega$. This means that for $G=\mathbb{Z}_N$ with $N>2$, $\cZ(\cS_L)$ differs from $\cZ(\cS_R)$ if its anomaly is not order two. 

Importantly, the Choi state automatically comes with the {\bf anti-unitary symmetry $T$} that swaps states in $\mathcal{H}_L$ with those in $\mathcal{H}_R$ and performs complex conjugation, as discussed in Sec.~\ref{schoiintro}. One way to incorporate this anti-unitary symmetry is to consider a symmetry-enriched topological order, where the SymTFT with the topological defects above is further enriched by a 0-form anti-unitary symmetry. In the following, we refer to the SymTFT, we actually refer to the symmetry enriched topological order, and the action of the anti-unitary symmetry will affect the kinds of boundary conditions we can allow. 

\subsection{Review of the SymTFT Approach for Pure State Phases}
\label{sec:PureStateSymTFT}

The approach proposed in the present work generalizes the well-established procedure for classifying pure state gapped phases via the SymTFT to classifying mixed state phases. Therefore, we briefly recall the pure state analysis  \cite{Bhardwaj:2023fca, Bhardwaj:2023idu}.

Any gapped phase with symmetry $\cS$ 
is realized in the SymTFT, which describes a topological order with gapped boundary conditions. The gapped boundary conditions are characterized by Lagrangian algebras, which take the form
\begin{equation}
    \cL=\bigoplus_{a}{n}_{a} a\,, \quad n_a\in \Z_{\geq 0}\,,
\end{equation}
where $a$ are objects in $\cZ(\cS)$, corresponding to anyons of the SymTFT.  
The set of objects that have non-vanishing coefficients $n_{a}$ in a Lagrangian algebra have to be are bosonic and  must satisfy certain consistency conditions involving braiding and fusion \cite{davydov2013witt,Kong:2013aya,EGNO,cong2017hamiltonian}. 
Physically the Lagrangian algebra $\cL$ describes how the bulk SymTFT lines can end on the boundary $\fB_{\cL}$.
Any line $a$ has $n_a$ different ways to end on $\fB_{\cL}$, meaning the junction of the line with the boundary hosts a $n_a$ dimensional vector space. There is further a maximality condition on the Lagrangian algebra 
\begin{equation}
\label{eq:Lagrangian condition}
    \cD(\cZ(\cS))=\dim(\cL)\,, \quad \dim(\cL)\equiv \sum_{a}n_a d_a\,.
\end{equation}
Here the $\cD(\cZ(\cS)) = \sqrt{\sum_a d_a^2}$ is the total quantum dimension of $\mathcal{Z}(\mathcal{S})$, with $d_a$ the quantum dimension of the anyon $a$.

In the SymTFT approach, we put the TQFT on an interval with two gapped boundary conditions. The symmetry $\cS$ specifies a particular gapped boundary condition, $\Bsym$, on which there are topological defects that realize the symmetry $\cS$. $\Bsym$ is characterized by a particular Lagrangian algebra of the SymTFT, denoted by $\mathcal{L}_{\mathcal{S}}$.
Gapped phases with the symmetry are in 1-1 correspondence with choices of gapped ``physical" boundary conditions $\{\fB_{\cL}\}$ of the SymTFT. This boundary condition can be given by any possible Lagrangian algebra $\cL$ of $\cZ(\cS)$. Upon compactifying the interval occupied by the SymTFT, this produces an $\cS$-equivariant 2d TQFT, i.e. an $\cS$-symmetric 1+1d IR gapped phase. 
This can be represented as follows:
\be
\label{eq:pure state local OP}
\begin{split}
\begin{tikzpicture}
\begin{scope}[shift={(0,0)}]
\draw [cyan,  fill=cyan] 
(0,0) -- (0,3) --(4,3) -- (4,0) -- (0,0) ; 
\draw [white](0,0) -- (0,3) --(4,3) -- (4,0) -- (0,0) ;  ; 
\draw [very thick] (0,0) -- (0,3)  ;
\draw [very thick] (4,3) -- (4,0) ;
\node at (2,2.3) {$\text{SymTFT}(\cS)$} ;
\node[above] at (0,3) {$\Bsym$}; 
\node[above] at (4,3) {$\Bphys= \mathfrak{B}_{\cL} $};
\draw [very thick, ->-] (0,1.5)  -- (4,1.5) ;
\node[above] at (2, 1.55) {$a$};
 \draw [black,fill=yellow] (0,1.5) ellipse (0.05 and 0.05);
 \draw [black,fill=yellow] (4,1.5) ellipse (0.05 and 0.05);
\end{scope}
\begin{scope}[shift={(6,0)}]
\node at (-1,1.5) {$=$} ; 
\draw [very thick] (0,0) -- (0,3) ;
\node[right] at (0, 1.5) {$\cO_a$};
 \draw [black,fill=yellow] (0,1.5) ellipse (0.05 and 0.05);
\end{scope}
\end{tikzpicture}
\end{split}
\ee

The SymTFT lines ending on $\fB_{\cL}$ furnish order and disorder parameters that characterize the symmetry properties of the phase. 
If a line in $\mathfrak{B}_{\mathcal{L}}$ also ends on $\Bsym$ (see \eqref{eq:pure state local OP}), part of $\mathcal{S}$ is spontaneously broken. Such an anyon tunneling operator can be used to toggle into other symmetry broken vacua, and its two-point correlation function in the basis of symmetric states is an order parameter detecting spontaneous symmetry breaking:
\begin{equation}\label{orderparamdef}
    \langle \psi|\mathcal{O}_{a}(x)\mathcal{O}_{a}^\dagger (y)|\psi\rangle\neq 0\,.
\end{equation}

In general, there are $n_a$ many different ways the line $a$ can end on the boundary specified by $\mathcal{L}$, so there are $n_{a,\mathrm{sym}}n_{a}$ local operators $\cO_a^i$, $i=1, \cdots, n_{a,\mathrm{sym}}n_a$. Here, $n_{a,\mathrm{sym}}$ is the coefficient of $a$ in $\mathcal{L}_{\mathcal{S}}$ (associated with $\mathfrak{B}_{\mathrm{sym}}$) and $n_a$ is the coefficient of $a$ in $\mathcal{L}$ (associated with $\mathfrak{B}_{\mathcal{L}}$). In total, such a pure state phase would have $\sum_an_{a,\mathrm{sym}}n_a$ degenerate ground states.

Lines in $\mathcal{L}$ that do not end on $\mathfrak{B}_{\mathrm{sym}}$ furnish disorder parameters, including string order parameters, which are useful for distinguishing between SPT phases of the preserved symmetry. These are expectation values of more general patch operators\cite{Ji:2019jhk,chaterjeepatch} (also known as elements of the string operator algebra in \cite{Moradi:2022lqp}), that are not completely local and include symmetry lines of $\mathfrak{B}_{\mathrm{sym}}$. Specifically, they arise by taking a finite length anyon line $a$ with $n_{a}\neq 0$ with both ends on the physical boundary. Because the anyon line cannot be absorbed into $\mathfrak{B}_{\mathrm{sym}}$, the endpoints cannot form local operators. 
The setup is as follows:
\be
\label{Patchy}
\begin{split}
\begin{tikzpicture}
\begin{scope}[shift={(0,0)}]
\draw [cyan,  fill=cyan] 
(0,0) -- (0,3) --(4,3) -- (4,0) -- (0,0) ; 
\draw [white](0,0) -- (0,3) --(4,3) -- (4,0) -- (0,0) ;  ; 
\draw [very thick] (0,0) -- (0,3)  ;
\draw [very thick] (4,3) -- (4,0) ;
\draw [orange, very thick] (0,1) -- (0,2) ;
\draw [orange,  fill=orange] 
(-0.05,1) -- (-0.05,2) --(0.05,2) -- (0.05,1) -- (-0.05,1) ; 
\node at (2,2.7) {$\text{SymTFT}(\cS)$} ;
\node[above] at (0,3) {$\Bsym$}; 
\node[above] at (4,3) {$\Bphys= \mathfrak{B}_{\cL} $};
\draw [very thick, ->-] (4,2)  -- (0,2) ;
\node[above] at (2, 2.05) {$\ol{a}$};
 \draw [black,fill=yellow] (0,2) ellipse (0.05 and 0.05);
 \draw [black,fill=yellow] (4,2) ellipse (0.05 and 0.05);
\draw [very thick, ->-] (0,1)  -- (4,1) ;
\node[above] at (2, 1.05) {$a$};
 \draw [black,fill=yellow] (0,1) ellipse (0.05 and 0.05);
 \draw [black,fill=yellow] (4,1) ellipse (0.05 and 0.05);
\end{scope}
\begin{scope}[shift={(6,0)}]
\node at (-1,1.5) {$=$} ; 
\draw [very thick] (0,0) -- (0,3) ;
\draw [orange,  fill=orange] 
(-0.05,1) -- (-0.05,2) --(0.05,2) -- (0.05,1) -- (-0.05,1) ;
\node[right] at (0, 1.5) {$\cP_a(\ell )$};
 \draw [black,fill=yellow] (0,2) ellipse (0.05 and 0.05);
 \draw [black,fill=yellow] (0,1) ellipse (0.05 and 0.05);
\end{scope}
\end{tikzpicture}
\end{split}
\ee
Physically, the anyon line along the interval gives a symmetry operator along the interval, decorated by local operators at its endpoints corresponding to splitting of this anyon line into lines that end on $\mathfrak{B}_{\mathrm{sym}}$ and/or $\mathfrak{B}_{\mathcal{L}}$. Such an operator thereby furnishing a nonzero disorder parameter:
\begin{equation}
    \langle\psi|\mathcal{P}_a(x,y)|\psi\rangle\neq 0\,.
\end{equation}

Conjugate pairs of lines tunneling between $\mathcal{L}_{\mathcal{S}}$ and $\mathcal{L}$, such as $\mathcal{O}_a(x)\mathcal{O}_a^\dagger(y)$ written in (\ref{orderparamdef}), are patch operators with the trivial symmetry line. 

For example when the symmetry is anomaly-free $\mathbb{Z}_2\times\mathbb{Z}_2$ symmetry, the $\mathrm{SymTFT}$ is (untwisted) $\mathbb{Z}_2\times\mathbb{Z}_2$ gauge theory generated by anyons $e_1,e_2,m_1,m_2$. The symmetry boundary is $\mathcal{L}_{\mathcal{S}}=1\oplus e_1\oplus e_2\oplus e_1e_2$ and the $m_1,m_2$ anyon worldlines can be interpreted as the global $\mathbb{Z}_2\times\mathbb{Z}_2$ symmetry generators at $\mathfrak{B}_{\mathrm{sym}}$. If we choose $\mathfrak{B}_{\mathcal{L}}$ given by $\mathcal{L}=1\oplus e_1e_2\oplus m_1m_2\oplus e_1e_2m_1m_2$, then there is a charged local operator given by the tunneling $e_1e_2$ line resulting in two degenerate ground states. There is a disorder parameter given by the $m_1m_2$ line along a finite interval and both ends on $\mathfrak{B}_{\mathcal{L}}$. The operator is the diagonal $\mathbb{Z}_2$ symmetry restricted to an interval. $e_1e_2m_1m_2$ also furnishes a disorder parameter, with the same restricted diagonal $\mathbb{Z}_2$ symmetry together with endpoint operators that end on both $\mathfrak{B}_{\mathrm{sym}}$ and $\mathfrak{B}_{\mathcal{L}}$.

In a semi-infinite system, if we extend a patch operator that acts nontrivially along its length, we get an operator that takes us to a twisted sector: 
\be
\label{fig:pureSstringOP}
\begin{split}
\begin{tikzpicture}
\begin{scope}[shift={(0,0)}]
\draw [cyan,  fill=cyan] 
(0,0) -- (0,3) --(4,3) -- (4,0) -- (0,0) ; 
\draw [white](0,0) -- (0,3) --(4,3) -- (4,0) -- (0,0) ;  ; 
\draw [very thick] (0,0) -- (0,3)  ;
\draw [very thick, orange] (0,0) -- (0,1.5)  ;
\draw [very thick] (4,3) -- (4,0) ;
\node at (2,2.3) {$\text{SymTFT}(\cS)$} ;
\node[above] at (0,3) {$\Bsym$}; 
\node[above] at (4,3) {$\Bphys= \mathfrak{B}_{\cL} $};
\draw [very thick, ->-, orange] (0,1.5)  -- (4,1.5) ;
\node[above] at (2, 1.55) {$a$};
 \draw [black,fill=yellow] (0,1.5) ellipse (0.05 and 0.05);
 \draw [black,fill=yellow] (4,1.5) ellipse (0.05 and 0.05);
\end{scope}
\begin{scope}[shift={(6,0)}]
\node at (-1,1.5) {$=$} ; 
\draw [very thick] (0,0) -- (0,3) ;
\draw [very thick, orange] (0,0) -- (0,1.5)  ;
\node[right] at (0, 1.5) {$\cO_a$};
 \draw [black,fill=yellow] (0,1.5) ellipse (0.05 and 0.05);
\end{scope}
\end{tikzpicture}
\end{split}
\ee
{The corresponding patch operator for every $a$ is straightforward to define for invertible symmetries and will be studied concretely in subsequent examples. When a patch operator contains a non-invertible symmetry operator, it would naively insert extra degrees of freedom with Hilbert space dimension $d_a$, mapping a state to a state in a larger Hilbert space. Intuitively, this comes from the fact that the two endpoints of the patch operator create two defects, each of which can be associated with a vector space of dimension $d_a$. To obtain an operator acting on the physical Hilbert space, one has to either trace over this ``virtual space" \cite{Fechisin:2023dkj} or pick a particular vector in this Hilbert space \cite{Lu:2025rwd}. Note that here we have assumed that $d_a$ is integer. If this is not the case, then the state cannot be symmetric under the operator anyways \cite{Chang:2018iay}, so the disorder parameter is zero.\footnote{{Patch operators with non-integer $d_a$ can be written in Hilbert spaces that do not have a tensor product decomposition. We will not study these in detail in this work, although the construction in Sec.~\ref{Sec:Lattice Models} is flexible enough to describe such models.}} In this work, $\mathcal{P}_a(x,y)$ will always refer to operators acting on the Hilbert space.} 

\subsection{SymTFT Approach for Mixed State Phases}
\label{sec:SymStrongPhases}

We now move onto the SymTFT description of gapped mixed state phases.
In this work, we will define a gapped mixed state as those states whose corresponding Choi states admit gapped parent Hamiltonians i.e., demonstrate clustering of correlations.\footnote{We will consider states for which the correlation length defined by decay of $\langle O_iO_j\rangle - \langle O_i\rangle\langle O_j\rangle$ may not vanish, but for any gapped Hamiltonian, correlations cluster in the sense that $\langle O_iO_j\rangle - \langle O_iP_{\mathrm{GS}} O_j\rangle$ decays exponentially in $|i-j|$. Here, $P_{\mathrm{GS}}$ is the projector onto the ground state subspace.
}
Therefore it is natural to characterize mixed-state gapped phases via the classification of pure states in the doubled Hilbert space.
This strategy was employed in \cite{Ma:2024kma} to study mixed-state SPTs of invertible symmetries and will be lifted to the SymTFT, which in turn opens up the generalization to non-invertible categorical symmetries. 

We start with the SymTFT for $\cS_\p\boxtimes \cS_\a$ whose lines form $\cZ(\cS_\p\boxtimes \cS_\a)$.
The symmetry boundary for a strong symmetry $\cS$ is a special Lagrangian algebra which has the form 
\be\label{Lstrong}
\cL_{\cS}^{\text{strong}} = (\cL^\cS)_\p  \otimes {(\cL^{\overline{\cS}})}_\a\,, 
\ee
where $(\cL^\cS)_\p$ and ${(\cL^{\overline{\cS}})}_\a$ are canonical Dirichlet boundaries of $\cZ(\cS_L)$ and $\cZ(\cS_{\a})$, that carry the $\cS_L$ and $\cS_{R}=\overline{\cS}_L$ symmetry respectively.
Then picking a certain gapped boundary specified by a Lagrangian algebra $\mathcal{L}$ of $\mathcal{Z}(\mathcal{S}_L\boxtimes\mathcal{S}_R)$ as the physical boundary $\mathfrak{B}_{\mathrm{phys}}$ and compactifying the SymTFT produces a $\cS_{L}\boxtimes \cS_R$ symmetric gapped phase. 

This is not the end of the story, however. Below, we will show that to apply the SymTFT approach to Choi states of mixed states, we must include additional constraints on allowed condensable algebras, related to hermiticity and positivity of density matrices. We also show that after one specifies the physical boundary by a choice of $\mathcal{L}$, the multiplet of degenerate ``ground states" forms only a subset of those given by all of the possible tunneling operators.

\subsubsection{Mixed State Lagrangian Algebras}
An important observation is that, unlike the case of pure states, not all gapped boundaries of the SymTFT furnish consistent mixed state gapped phases.

Every density matrix has a corresponding Choi state, but not every state in $\mathcal{H}_L\otimes\mathcal{H}_R$ is the Choi state of a positive semi-definite, Hermitian density matrix. For example, any state in the doubled Hilbert space that is the Choi state of a density matrix must at the very least be invariant under the anti-unitary operator $T$, but this is not the only constraint. Hermiticity and positivity of the density matrix impose constraints on possible $\cS_L\boxtimes \cS_R$ gapped phases and corresponding possible gapped boundaries of $\cZ(\cS_L\boxtimes \cS_R)$. The purpose of this section is to describe necessary conditions on $\mathcal{L}$ of $\cZ(\cS_L) \boxtimes \cZ(\cS_{R})$ for it to describe a gapped mixed state phase. {We conjecture that these are also sufficient, and give a route to showing that they are sufficient.}

Any gapped boundary $\fB_{\cL}$ of $\cZ(\cS_L) \boxtimes \cZ(\cS_{R})$ is 
characterized by a Lagrangian algebra
\be\label{Langis}
\cL = \bigoplus_{a_L, b_R} n_{a_L,b_R} a_L  b_R  \,,
\ee
where $a_L\in\cZ(\cS_L)$ and $b_R\in\cZ(\cS_R)$, with the usual consistency conditions of Lagrangian algebras \cite{davydov2013witt,Kong:2013aya,EGNO,cong2017hamiltonian}. $\mathcal{L}$ must furthermore satisfy the maximality condition \eqref{eq:Lagrangian condition}.  
The Lagrangian algebra corresponding to the symmetry boundary has the special form (\ref{Lstrong}) and can be written more explicitly as 
\begin{equation}
\label{Langis_strong}
\cL^{\rm strong}_{\cS}    = \bigoplus_{a_L,b_R} n^{\cS}_{a_L} n^{\overline{\cS}}_{b_R} \
a_L b_R  \,.
\end{equation}

Like in the pure state case, every anyon $a_Lb_R\in\mathcal{Z}(\mathcal{S}_L)\boxtimes\mathcal{Z}(\mathcal{S}_R)$ gives $n_{a_L,b_R}$ patch operators $\mathcal{P}_{a_Lb_R}^i(x,y)$ that have nonzero expectation value in $|\rho_{\mathcal{L}}\RR${. For fixed-point states, we can normalize the operator to have expectation value 1}:
\begin{equation}\label{patchexp}
\LL\rho_{\mathcal{L}}|\mathcal{P}_{a_Lb_R}^i(x,y)|\rho_{\mathcal{L}}\RR=1 \,.
\end{equation}

{Note that we get expectation value 1 even for patch operators $\mathcal{P}_{a_Lb_R}^i(x,y)$ with non-invertible truncated symmetry lines. This is because the endpoints of the patch operators at the boundaries are topological, and can be moved to squeeze away the symmetry line. Therefore, the expectation value of a patch operator is equal to the expectation value of the product of its endpoint operators, which is equal to 1 like order parameters. Writing $\mathcal{P}_{a_Lb_R}^i(x,y)=\mathcal{R}_{a_L}^i(x,y)\otimes\mathcal{R}_{b_R}^i(x,y)$ where $\mathcal{R}_{a_{L/R}}^i(x,y)$ is the part of $\mathcal{P}_{a_Lb_R}^i(x,y)$ acting on $\mathcal{H}_{L/R}$ (note that the patch operator must factorize over $\mathcal{H}_L\otimes\mathcal{H}_R$ because the anyon $a_Lb_R$ factorizes), we get }
\begin{equation}
    \mathrm{Tr}(\rho_{\mathcal{L}}\mathcal{R}_{a}^i(x,y)\rho_{\mathcal{L}}\mathcal{R}_{b}^{iT}(x,y))=1
\end{equation}

We {\it conjecture} that any gapped boundary $\fB_{\cL}$ of $\cZ(\cS_L) \boxtimes \cZ(\cS_{R})$ 
that satisfies the following two properties corresponds to a distinct mixed state phase with strong $\cS$ symmetry. In other words, we conjecture that the following two conditions are both necessary and sufficient for $\mathcal{L}$ to correspond to a Hermitian, positive semi-definite density matrix $\rho_{\mathcal{L}}$. In principle one should make the dependence of $\rho$ on the the algebra $\cL$ notational explicit, but we will often refrain from doing so, if it is clear from the context. 
We define the Lagrangian algebras that satisfy these constraints to be {\bf mixed-state Lagrangian algebras}:
\begin{enumerate}
\item {\bf $T$ Invariance:} This condition follows from the Hermiticity of $\rho$ and implies  
\be
T(\cL) = \mathcal{L} \,, \quad T(a_\p  b_\a ) = \overline{b}_\p \overline{a}_\a \,,
\ee
which for any anyons $a_L$, $b_R$ swaps the two copies and performs complex conjugation (i.e. $\theta_{a_L}=\overline{\theta}_{\overline{a}_R}$). Recall that in general $\mathcal{Z}(\mathcal{S}_L)\ncong\mathcal{Z}(\mathcal{S}_R)$ but rather $\mathcal{Z}(\mathcal{S}_L)\cong\overline{\mathcal{Z}(\mathcal{S}_R)}$, where here the overline can be taken as time-reversal.
This translates to the following condition on the mixed state Lagrangian algebras:
\begin{equation}\label{hermcond}
    \boxed{n_{a_L,b_R} = n_{\overline{b}_L, \overline{a}_R}}
\end{equation}

\item {\bf Positivity:}
Another constraint arises from the positivity of $\rho$. 
As mentioned earlier in this section, every anyon $a_Lb_R$ corresponds to $n_{a_L,b_R}$ patch operators $\mathcal{P}_{a_L}^i(x,y)\otimes\mathcal{P}_{b_R}^i(x,y)$ (with $ i=1,2,\cdots n_{a_L,b_R}$) that have non-zero expectation value in $|\rho_{\mathcal{L}}\RR$. These operators can give rise to charged local operators or to disorder/string order parameters. Positivity leads to the constraint
\begin{equation}\label{poscond}
   \boxed{ n_{a_L,b_R}\leq n_{a_L,\overline{a}_R}n_{\overline{b}_L,b_R}}
\end{equation}
It is straightforward to see that the above inequality is satisfied for $a_L=\overline{b}_L$ since in this case $n_{a_L,b_R}=n_{a_L,\overline{a}_R}=n_{\overline{b}_L,b_R}$ and all of these are either 0 or $\geq 1$. It is also not hard to see that it is satisfied for Lagrangian algebras of the form $\mathcal{L}_L\otimes\mathcal{L}_R$. In fact these two cases saturate the stronger inequality $n_{a_L,b_R}^2\leq n_{a_L,\overline{a}_R}n_{\overline{b}_L,b_R}$.\footnote{Constraints on general Lagrangian algebras also give $n_{a_L,b_R}n_{\overline{b}_L,\overline{a}_r}=n_{a_L,b_R}^2\leq\sum_{c_L,c_R'}N_{a_L,\overline{b}_L}^{c_L}N_{b_R,\overline{a}_R}^{c_R'}n_{c_L,c_R'}$ where we used factorization of fusion in the two layers since the SymTFT takes the form $\mathcal{Z}(\mathcal{S}_L)\boxtimes \mathcal{Z}(\mathcal{S}_R)$. We also have $n_{a_L,\overline{a}_R}n_{\overline{b}_L,b_R}\leq\sum_{c_L,c_R'}N_{a_L,\overline{b}_L}^{c_L}N_{b_R,\overline{a}_R}^{c_R'}n_{c_L,c_R'}$. This seems to suggest the stronger inequality $n_{a_L,b_R}^2\leq n_{a_L,\overline{a}_R}n_{\overline{b}_L,b_R}$, since they are both upper bounded by the same quantity.}
\end{enumerate}

To prove (\ref{poscond}), we follow the approach in \cite{Ma:2024kma} (leaving the argument $(x,y)$ implicit to minimize notation):
\be\label{Kautschi}
\ba
0 \ &<  | \LL \rho | \mathcal{R}_{a_L}^i \otimes \mathcal{R}_{b_R}^i |\rho\RR|^2 \cr 
&=  | \Tr( \rho\mathcal{R}_{a}^i \rho  \mathcal{R}_{b}^{iT}  )|^2  \cr 
&= | \Tr( \rho^{1/2} \mathcal{R}_{a}^i  \rho^{1/2}  \rho^{1/2}  \mathcal{R}_{b}^{iT} \rho^{1/2} )|^2  \cr 
& \leq \Tr( \rho^{1/2} \mathcal{R}_{a}^i  \rho^{1/2} ( \rho^{1/2}  \mathcal{R}_{a}^{i} \rho^{1/2})^\dagger )\times \cr 
&\phantom{leq} \times\Tr( \rho^{1/2} \mathcal{R}_{b}^{iT}  \rho^{1/2} ( \rho^{1/2}  \mathcal{R}_{b}^{iT} \rho^{1/2})^\dagger )\cr 
& =   \LL\rho | \mathcal{R}_{a_L}^i \otimes \overline{\mathcal{R}}_{a_R}^{i} |\rho\RR  \times \LL\rho | \overline{\mathcal{R}}_{{b_L}}^i \otimes \mathcal{R}_{b_R}^i |\rho\RR\,.
\ea
\ee
The fourth line follows from the Cauchy-Schwarz inequality. This shows that if $n_{a_L,b_R}\neq 0$, then $n_{a_L,\overline{a}_R}\neq 0$ and $n_{\overline{b}_L,b_R}\neq 0$. Since $\LL\rho_{\mathcal{L}}|\mathcal{R}_{a_L}^i\otimes\mathcal{R}_{b_R}^i|\rho_{\mathcal{L}}\RR>0$ for every $i\in 1,2\cdots n_{a_L,b_R}$, this lower bounds the number of distinct combinations of $\mathcal{R}_{a_L}^i\otimes\overline{\mathcal{R}}_{a_R}^i$ and $\overline{\mathcal{R}}_{b_L}^i\otimes\mathcal{R}_{b_R}^i$ operators that also have expectation value $>0$ in $|\rho_{\mathcal{L}}\RR$. (\ref{poscond}) follows directly from this observation.

In fact, because we are considering fixed-point density matrices, we can strengthen the inequality \eqref{Kautschi} to an equality. In the following, for brevity, we will drop the index $i$. For fixed-point density matrices,  $\mathcal{R}_{a_L}\otimes\mathcal{R}_{b_R}|\rho\RR=|\rho\RR$,
so $\mathcal{R}_{a}\rho \mathcal{R}_{b}^{T} = \rho$. The above condition means that
\begin{equation}\label{o2o2star}
    \mathcal{R}_{a}\rho\mathcal{R}_{b}^T\overline{\mathcal{R}}_{b}=\rho\overline{\mathcal{R}}_{b}=\mathcal{R}_{a}\rho\,,
\end{equation}
where we used the invertible composition of patch operators on the state justified below eq.~\ref{patchexp}.
Multiplying by $\rho$ on both sides and using $\rho\propto\rho^{1/2}$, we have
\begin{equation}\label{lindep}
\rho^{1/2}\mathcal{R}_{a}\rho^{1/2}=\rho^{1/2}\overline{\mathcal{R}}_{b}\rho^{1/2} \,.
\end{equation}
Then 
\begin{align}
\begin{split}
    \mathrm{Tr}&(\rho^{1/2}\mathcal{R}_{a}\rho^{1/2}\rho^{1/2}\mathcal{R}_{b}^T\rho^{1/2})\\
    &=\mathrm{Tr}(\rho^{1/2}\mathcal{R}_{a}\rho^{1/2}(\rho^{1/2}\mathcal{R}_{b}^*\rho^{1/2})^\dagger)\\
    &=\LL\rho^{1/2}\overline{\mathcal{R}}_{b}\rho^{1/2}|\rho^{1/2}\mathcal{R}_{a}\rho^{1/2}\RR\,,
\end{split}
\end{align}
where (\ref{lindep}) ensures that the vectorized operators are linearly dependent. As a result, the Cauchy-Schwarz inequality is saturated:
\begin{equation}
|\LL\rho|\mathcal{R}_{a_L}\otimes\mathcal{R}_{b_R}|\rho\RR|^2=\LL\rho|\mathcal{R}_{a_L}\otimes\overline{\mathcal{R}}_{a_R}|\rho\RR\LL\rho|\overline{\mathcal{R}}_{b_R}\otimes\mathcal{R}_{b_R}|\rho\RR
\end{equation}
as long as $\mathcal{P}_{a_Lb_R}^i|\rho\RR=\mathcal{R}_{a_L}^i\otimes\mathcal{R}_{b_R}^i|\rho\RR= |\rho\RR$ {(in fact, this result also holds if the two vectors differ by a scalar)}.

Clearly these are necessary conditions, and we believe that they are also sufficient. We outline a possible way to proving this more rigously in section~\ref{sec:Complete}.

\subsubsection{Constraints on States in the Multiplet}\label{multiplet}

{We will now discuss the constraints on states that arise in the SymTFT after interval compactification. These correspond to OPs, but not all of these will give rise to valid Choi states. Only those anyons that can end on both symmetry and physical boundary, that are also swap-star, i.e. $T$ symmetric give rise to Choi states, that are associated to density matrices.}

As in the previous section, there are in general 
\begin{equation}
N_{\mathcal{L}}=\sum_{a_L,b_R}n_{a_L,b_R}^{\mathrm{strong}}n_{a_L,b_R}^{\mathcal{L}}
\end{equation}
states associated with a choice of $\mathcal{L}$. These different states are labeled by different anyons that can connect $\mathfrak{B}_{\mathrm{sym}}^\strong$ and $\mathfrak{B}_{\mathrm{phys}}=\mathfrak{B}_{\mathcal{L}}$: 
\be
\begin{tikzpicture}
\begin{scope}[shift={(0,0)}]
\draw [thick] (0,-1) -- (0,1) ;
\draw [thick] (3,-1) -- (3,1) ;
\draw [thick] (0, 0) -- (3, 0) ;
\node[above] at (1.5,0.1) {$a_L b_R$} ;
\node[above] at (0,1) {$\Bsym^\strong$}; 
\node[above] at (3,1) {$\Bphys=\mathfrak{B}_{\cL}$}; 
\draw [black,fill=black] (3,0) ellipse (0.05 and 0.05);
\draw [black,fill=black] (0,0) ellipse (0.05 and 0.05);
\end{scope}
\end{tikzpicture} 
\ee
These give rise to states in different symmetry sectors of the strong and weak symmetries. In the context of pure states, these different states form a degenerate ground state subspace from spontaneous symmetry breaking.  

We will focus on the particular state $|\rho_{\mathcal{L}}^0\RR$ that has a real, positive eigenvalue $d_s$ for every $s\in\mathcal{S}_L\boxtimes\mathcal{S}_R$:
\begin{equation}
    D_s|\rho_{\mathcal{L}}^0\RR=d_s|\rho_{\mathcal{L}}^0\RR\qquad\forall s\in\mathcal{S}_L\boxtimes\mathcal{S}_R
\end{equation}

This state is given by the vacuum line tunnel between the two boundaries,. The corresponding density matrix is given by
\begin{equation}
    \rho_{\mathcal{L}}^0=\mathrm{Tr}_{R}(|\rho_{\mathcal{L}}^0\RR\LL\rho_{\mathcal{L}}^0|)
\end{equation}
Since $\LL\rho_{\mathcal{L}}^0|D_{s_L}\otimes\mathbf{1}_R|\rho_{\mathcal{L}}^0\RR=d_{s_L}=\mathrm{Tr}(D_{s}(\rho_{\mathcal{L}}^0)^2)$ for any strong symmetry operator, this means that the singlet state satisfies
\begin{equation}    D_s\rho_{\mathcal{L}}^0=d_s\rho_{\mathcal{L}}^0=\rho_{\mathcal{L}}^0 D_s\qquad\forall s\in\mathcal{S}\,.
\end{equation}

We can get states in other symmetry sectors $|\rho_{\mathcal{L}}^i\RR$ for $i=1,\cdots N_{\mathcal{L}}$, but not all of these states are valid Choi states. In particular,{\bf  we can only  tunnel operators that are symmetric under $T$}. All such states are singlets under all of the weak symmetries. This means that size of the multiplet of valid Choi states, $N_{\mathcal{L}}^T$ does not need to be equal to $N_{\mathcal{L}}$ and in fact can be smaller than $N_{\mathcal{L}}$:
\begin{equation}
    N_{\mathcal{L}}^T\leq N_{\mathcal{L}} \,.
\end{equation}
For example, if we choose $\mathcal{L}=\mathcal{L}^{\mathrm{strong}}_{\mathcal{S}}$, then $N_{\mathcal{L}}^T=\sqrt{N_{\mathcal{L}}}$. 

This set of $N_{\mathcal{L}}^T$ states $|\rho_{\mathcal{L}}^i\RR$ are locally indistinguishable under any local, symmetric operator. {\bf The only operators that take different expectation values in different states are global symmetry operators.} It follows that the corresponding density matrices $\rho_{\mathcal{L}}^i$ are also locally indistinguishable via any linear or Renyi-2 correlator of symmetric operators:
\begin{equation}
    \mathrm{Tr}(\mathcal{O}\rho_{\mathcal{L}}^i)=\mathrm{Tr}(\mathcal{O}\rho_{\mathcal{L}}^j)\qquad \forall i,j\in 0,\cdots N_{\mathcal{L}}^T
\end{equation}
\begin{equation}
    \mathrm{Tr}(\mathcal{O}\rho_{\mathcal{L}}^i\mathcal{O}'\rho_{\mathcal{L}}^i)=\mathrm{Tr}(\mathcal{O}\rho_{\mathcal{L}}^j\mathcal{O}'\rho_{\mathcal{L}}^j)\qquad \forall i,j\in 0,\cdots N_{\mathcal{L}}^T
\end{equation}
where $\mathcal{O},\mathcal{O}'$ are local and strongly symmetric.\footnote{Recall that we are working with fixed point density matrices. More generally, there may be some discrepancies between the expectation values that are exponentially small in system size}

A physical implication of this local indistinguishability is that these density matrices form a degenerate subspace similar to ground states. The analogue of parent Hamiltonian for mixed states is parent channel. The local indistinguishability condition means that if one were to construct a gapped, symmetric {\bf parent channel} stabilizing the singlet state, $E[\rho_{\mathcal{L}}^0]=\rho_{\mathcal{L}}^0$, it would also stabilize the other density matrices $\rho_{\mathcal{L}}^{i}$. By gapped, symmetric channel, we mean a quantum channel built out of local, symmetric Krauss operators, i.e. a finite number of layers of mutually commuting local quantum channels implementing $\rho\to\sum_{i,x}K_{i,x}\rho K_{i,x}^\dagger$ with $\sum_iK_{i,x}^\dagger K_{i,x}=1$ to preserve trace. $\{K_{i,x}\}$ here are the Krauss operators of the local channel near site $x$. This means if the eigenvalues of $E$ take the form $e^{i\epsilon_n}$, there are (at least) $N_{\mathcal{L}}^T$ $\epsilon_n$'s with zero imaginary part (and a gap in the imaginary part to other eigenvalues).\footnote{We can also consider a gapped parent Lindbladian for $\rho$. The same lower bound on steady state degeneracy applies in this context.}

\subsection{A Route to a Complete Classification}\label{sec:Complete}

In this section, we will outline a procedure for proving that the two conditions on the Lagrangian algebra (\ref{hermcond}) and (\ref{poscond}) are both necessary and sufficient for a fixed point Choi state to correspond to a fixed point density matrix. The basic idea is to take the Choi state given by the gapped boundary $\mathcal{L}$, $|\rho_{\mathcal{L}}^0\RR$, and trace out the states in $\mathcal{H}_R$ to get $\rho_{\mathcal{L}}^0$ (note that we can do this because for fixed point density matrices $\rho\propto\sqrt{\rho}$). $|\rho_{\mathcal{L}}^0\RR$ then corresponds to a valid Choi state if all of the patch operators given by $\mathcal{P}_{a_L}^i\otimes\mathcal{P}_{b_R}^i$ (again, dropping the $(x,y)$ indices for ease of notation), $i=1,\cdots n_{a_L,b_R}$, satisfy
\begin{equation}\label{choicond}
    \LL\rho_{\mathcal{L}}^0|\mathcal{P}_{a_L}^i\otimes\mathcal{P}_{b_R}^i|\rho_{\mathcal{L}}^0\RR=\mathrm{Tr}(\mathcal{P}_a^i\rho_{\mathcal{L}}^0\mathcal{P}_{b}^{iT}\rho_{\mathcal{L}}^0)
\end{equation}
If this is the case, then $\mathrm{Tr}_{R}(|\rho_{\mathcal{L}}^0\RR\LL\rho_{\mathcal{L}}^0|)$ explicitly constructs a state with the desired symmetry and symmetry breaking patterns. If (\ref{choicond}) does not hold, then this indicates that $|\rho_{\mathcal{L}}^0\RR$ is not the Choi state of a fixed-point density matrix.

As a simple example, let us consider a strong $\mathbb{Z}_2$ symmetry and $\mathcal{L}=1\oplus e_Lm_R\oplus m_Le_R\oplus e_Le_Rm_Lm_R$. Clearly this does not satisfy (\ref{poscond}) so it does not give a valid Choi state. We can also deduce this result by showing that in this case, $|\rho_{\mathcal{L}}^0\RR$ does not satisfy (\ref{choicond}).

Let the strong symmetry be $U=\prod_iX_i$ on a 1+1D chain of qubits. Then it is not hard to see that
\begin{equation}
    \mathrm{Tr}_R(|\rho_{\mathcal{L}}^0\RR\LL\rho_{\mathcal{L}}^0|)=\rho_{\mathcal{L}}^0=\frac{1}{2^N}(1+\prod_iX_i)
\end{equation}
This is in fact the same density matrix as the one obtained from $\mathcal{L}=1\oplus e_Le_R\oplus m_Lm_R\oplus e_Le_Rm_Lm_R$ in Sec.~\ref{sec:SymTFT_Z2}. Clearly, it does not satisfy (\ref{choicond}):
\begin{equation}
    \LL\rho_{\mathcal{L}}^0|Z_{i,L}\left(\prod_{k=i}^jX_{k,R}\right)Z_{j,L}|\rho_{\mathcal{L}}^0\RR=1
\end{equation}
However
\begin{equation}
    \mathrm{Tr}(Z_iZ_j\rho_{\mathcal{L}}^0\prod_{k=i}^jX_k\rho_{\mathcal{L}}^0)=0
\end{equation}

One can try to prove that assuming (\ref{hermcond}) and (\ref{poscond}) are satisfied, (\ref{choicond}) holds for all $\mathcal{P}_{a_L}^i\otimes\mathcal{P}_{b_R}^i$. This would show that these conditions are not only necessary, but also sufficient. We leave this for future work. 

\subsection{Classification of Mixed State Gapped Phases}
\label{sec:StrongGapped}

Starting with a strong symmetry specified by $\cL_{\cS}^\strong$, we can determine all the gapped phases by inserting as the physical boundary conditions all mixed-state Lagrangians $\cL_{\phys}$ satisfying (\ref{hermcond}) and (\ref{poscond}). In this section, we will give some features of broad classes of mixed state phases.

\vspace{2mm}
\noindent
{\bf Pure-state Phases.} Some choices of $\mathcal{L}_{\mathrm{phys}}$ give pure states. These Lagrangian algebras are those that factorize:
\begin{equation}
    \mathcal{L}^{\mathrm{strong}}_{\mathrm{phys}}=\mathcal{L}_L\otimes\mathcal{L}_R\,,
\end{equation}
where $\mathcal{L}_L$ describes a gapped boundary of $\mathcal{Z}(\mathcal{S}_L)$ while $\mathcal{L}_R$ describes a gapped boundary of $\mathcal{Z}(\mathcal{S}_R)$. The Choi state then takes the form
\begin{equation}
|\rho_{\mathcal{L}}\RR=|\psi_{\mathcal{L}_L}\rangle\otimes|\psi_{\mathcal{L}_R}\rangle\,.
\end{equation}
It follows that
\begin{equation}
\rho_{\mathcal{L}}=\mathrm{Tr}_{R}(|\rho_{\mathcal{L}}\RR\LL\rho_{\mathcal{L}}|)=|\psi_L\rangle\langle\psi_L|
\end{equation}
describes a pure state corresponding to the boundary condition $\mathcal{L}_L$ of $\mathcal{Z}(\mathcal{S}_L)$. For example, if $\mathcal{L}_L=\left(\mathcal{L}^S\right)_L$ then the strong symmetry $\cS$ is completely spontaneously broken. This is also referred to as ``strong-to-nothing" SSB phase. On the other hand, if $\left(\mathcal{L}^\mathcal{S}\right)_L$ has trivial overlap with $\mathcal{L}_L$ in the sense that the only anyon included in both Lagrangian algebras is the vacuum, then we get a pure state that respects the symmetry. 

\vspace{2mm}
\noindent
{\bf Mixed-state Phases.} If $\Bphys$ is given by a Lagrangian algebra $\cL_\phys$ that satisfies (\ref{hermcond}) and (\ref{poscond}) and does not factorize into $\mathcal{L}_L\otimes\mathcal{L}_R$, then $\rho_{\mathcal{L}}=\mathrm{Tr}_R(|\rho\RR\LL\rho|)$ is mixed. Lack of factorization means that
\begin{equation}
    n_{a_L,b_R}\neq n_{a_L,1}n_{1,b_R}
\end{equation}
for some $a_L$ and $b_R$. In particular, suppose that $b_R=T(a_L)$ and $n_{a_L}^{\mathcal{S}}n_{b_R}^{\ol{\mathcal{S}}},n_{a_L,b_R}\neq 0$, meaning $\mathcal{L}$ contains a diagonal ``gauge charge" of $\mathcal{Z}(\mathcal{S}_L)\boxtimes\mathcal{Z}(\mathcal{S}_R)$. If $n_{a_L,1}=n_{1,b_R}=0$, then a strong symmetry operator $D_{s,L}$ for which $a_L$ gives an order parameter has a corresponding weak symmetry that may not spontaneously broken, because $a_Lb_R$ may not give an order parameter for $D_{s,L}\otimes D_{s,R}$. The phenomenon where a strong symmetry that is spontaneously broken but its corresponding weak symmetry that is not is called {\bf strong-to-weak SSB (SWSSB)}. Note that for factorized Lagrangian algebras, $n_{a_L,b_R}\neq 0$ implies that $n_{a_L,1}n_{1,b_R}\neq 0$, and the latter results in a nonzero order parameter for the weak symmetry, so SSB of strong symmetry implies SSB of its corresponding weak symmetry.

In addition to the above spontaneous symmetry breaking patterns, there can also be SPTs that involve both strong and weak symmetries. We will give some examples of these in section  \ref{sec:NonInvSPT}.

\section{Lattice Models for Mixed State Phases}
\label{Sec:Lattice Models}

In this section, we describe a class of one dimensional lattice models that can be used to realize the mixed state gapped phases that we discussed previously. In particular, we show how to construct microscopic fixed-point models for each of the gapped phases. 

The most general such models are known as anyon chain models \cite{Feiguin:2006ydp, 2009arXiv0902.3275T, Aasen:2016dop, Buican:2017rxc, Lootens:2021tet, Inamura:2021szw, Bhardwaj:2024kvy} which can realize general fusion category symmetries.
These models can also be naturally situated within the SymTFT, more precisely on a codimension-2 topological interface between the symmetry boundary and a reference (topological) boundary \cite{Bhardwaj:2024kvy}.

In Sec.~\ref{SubSec:Anyon Chain}, we outline this construction for any fusion category symmetry, and summarize the general theory of these models. We will use this construction to obtain models for mixed state gapped phases with strong categorical symmetries in Sec.~\ref{Sec:Examples}.

After outlining the general construction, we will focus on two special cases of fusion category symmetries which give models on tensor product Hilbert spaces:   (1) invertible (but possibly non-abelian) finite group symmetries $G$ (Sec.~ \ref{SubSec:G Chain}) and (2) closely-related (non)-invertible $\Rep(G)$ symmetries, of the representations of finite groups (Sec.~\ref{SubSec:Rep(G) Chain}).  Both symmetries give models that are realized in Hilbert spaces built out of $G$-qudits.

\subsection{Anyon Chains from SymTFT}
\label{SubSec:Anyon Chain}

We explained the SymTFT as a means of constructing gapped phases. This continuum analysis has a concrete corresponding lattice model implementation, which we now describe. The advantage of this setup is that it applies to any 
$\cS_{L}\boxtimes \cS_{R}$  strong fusion category symmetry, and takes in information directly from the classification of phases from the SymTFT. 

\begin{figure}
\centering
$
\includegraphics[scale=.75]{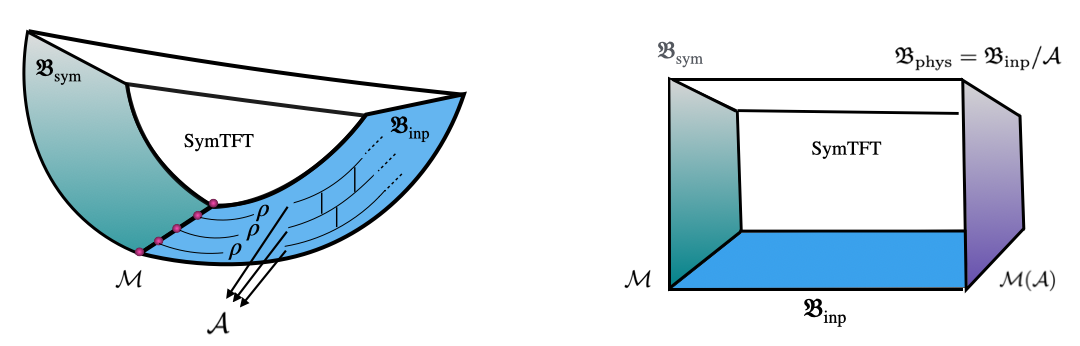}
$
\caption{The relation between the anyon chain (realized in a SymTFT picture on the LHS) and the standard SymTFT (RHS). The SymTFT is the standard TQFT setup, which is simply identical after interval compactification to the  gapped phase. The anyon chain (shown at the bottom of the well on the LHS, in terms of red dots with $\rho$ lines attached) is a lattice model that  takes in data from the SymTFT, such as the symmetry, and the physical boundary. Instead of $\Bphys$, the starting point here is (an often simpler choice) $\Binp$. Denote by $\cC$ the fusion category that is formed by the topological defects on $\Binp$. Then 
$\Bphys$ is related to the boundary $\mathfrak{B}_{\text{inp}}$ by gauging a Frobenius algebra $\cA$ in $\cC$.  In the anyon chain this is encoded in the data that specifies the Hamiltonian.  \label{eq:lattice SymTFT}}
\end{figure}

The relationship between the anyon chain and the SymTFT is illustrated in Fig.~\ref{eq:lattice SymTFT}. Conceptually we can describe the setup as follows --  unlike the SymTFT, which is simply a reformulation of a gapped 1+1d model in terms of a TQFT in one higher dimension, the anyon chain construction gives a representative microscopic lattice model for said gapped phase. The main advantage of this construction  is that the anyon chain model can be completely systematically constructed from the data of the SymTFT realization of the gapped phase and is more straightforward to write down than the actual microscopic SymTFT representation of the 1+1d system.

\onecolumngrid

\twocolumngrid

 Concretely, in the SymTFT construction, we start with a symmetry boundary $\Bsym$ fixed by the symmetry of interest $\mathcal{S}$ and a physical boundary $\Bphys$ of choice. What goes into the SymTFT is however not necessarily the physical boundary, but what we will denote $\fB_{\rm inp}$. After gauging a Frobenius  algebra $\cA$ on the boundary (the non-invertible generalization of gauging a subgroup), $\fB_{\rm inp}$ gives rise to the physical boundary. The topological defects on $\Binp$ form a fusion category $\cC$, which is not necessarily the same as $\cS$. 

Within the SymTFT, $\cS_{L}\boxtimes \cS_R$ is the fusion category of lines on the symmetry boundary, denoted as $\fB_{\sym}$, while $\cC$ is the category of lines on  $\fB_{\rm inp}$, which is related to $\Bphys$ by gauging. 
The interface between $\fB_{\sym}$ and $\fB_{\rm inp} $ is mathematically described by a module category $\cM$.\footnote{The precise relation between $\cM$ and $\cC$ is 
$\cC=\left(\cS_{L}\boxtimes \cS_{R}\right)^{*}_{\cM}\equiv {\rm Fun}_{\cS_{L}\boxtimes \cS_R}(\cM\,,\cM)$.} Physically, this means that we can gauge the symmetry on $\Bsym$ to get the one on $\Binp$, in the present case, this gauging needs to be $T$-symmetric.

The lattice models  are then defined using the following input data:
\begin{enumerate}
    \item $\cS_{L}\boxtimes \cS_{R}$, which determines $\Bsym$.
    \item An $\cS_{L}\boxtimes \cS_{R}$-module category $\cM$, which in turn fixes $\cC$.
    \item A (generically) non-simple object $\rho$ in $\cC$. 
\end{enumerate}

In the application to mixed states,  $\fB_{\rm inp}$  always corresponds to a mixed state Lagrangian algebra. 
In order to construct the state space for a lattice model, one considers a periodic array of objects $\rho \in \cC$ attached to $\cM$ from the input boundary as in Fig.~\ref{eq:lattice SymTFT}.
The space of configurations on $\cM$, including various objects in $m_j\in \cM$ and morphisms $\mu_{j+1/2}\in \Hom(m_{j}\,, \rho \otimes m_{j+1})$ compatible with the module structure, forms a vector space which is the underlying Hilbert space of the model:
\begin{equation}
\label{eq:anyon chain HS}
\begin{split}
\text{\includegraphics[scale=.5]{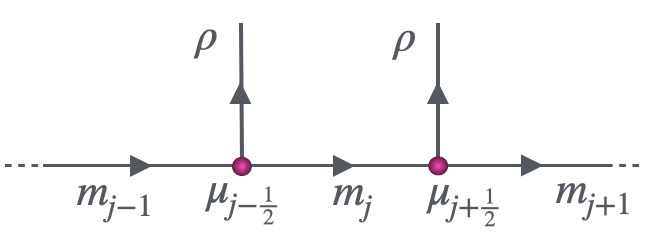}}
\end{split}
\end{equation}
For instance, for a strong (anomaly-free) $G$ group symmetry, we have
\be
 \cS_L\boxtimes \cS_R=  (\Vec_{G})_L\boxtimes (\Vec_{G})_R= \cM = \cC\,.
 \ee
Instead in order to get a $\Rep (G)_L\boxtimes \Rep (G)_R$ symmetry we again start with 
\be
\cC= (\Vec_{G})_L\boxtimes (\Vec_{G})_R\,,
\ee
but in order to get the symmetry to be $\Rep (G)_L\boxtimes \Rep (G)_R$, we need to gauge all of $\cC$, which corresponds to the choice 
\be
\cM= \Vec\,.
\ee
We will discuss these two cases in depth later on.
In summary, for a given $\mathcal{S}$ there are multiple choices for $\mathcal{C}$, and each has a corresponding $\mathcal{M}$. When possible, it is more convenient to work with a choice of $\cM$ which give rise to a category $\cC$ with all simple objects being invertible.

The state space decomposes into $\cH_L\otimes \cH_R$.
In many but not all cases, the state spaces $\cH_L\cong \cH_R$ also admit a tensor product decomposition into onsite state spaces. 
However, a tensor product decomposition may not be possible for symmetries with objects that have non-integer quantum dimension. A vast space of examples that admit a tensor product Hilbert space arise in the context of group theoretical categorical symmetries, i.e., those that are related via gauging to invertible (group-like) symmetries, possibly with 't Hooft anomalies. The theory for these examples will be described in further detail in the subsequent sections \ref{SubSec:G Chain} and \ref{SubSec:Rep(G) Chain}.

\smallskip \noindent {\bf Symmetry Action.} The symmetry category is the fusion category of topological defects on $\Bsym$ which for the case of strong symmetries is $\cS_L\boxtimes \cS_R$. 
These act on $\cM$ from the symmetry boundary or from ``below" as depicted in \eqref{eq:anyon chain HS}.

The state space by construction has the structure of a left $\cS_L\boxtimes \cS_R$ module category.
For some object $s\in \cS_\p\boxtimes \cS_\a$, the symmetry action on the state space can be read-off by evaluating the fusion diagram
\begin{equation}
\label{eq:anyon chain HS}
\begin{split}
\text{\includegraphics[scale=.55]{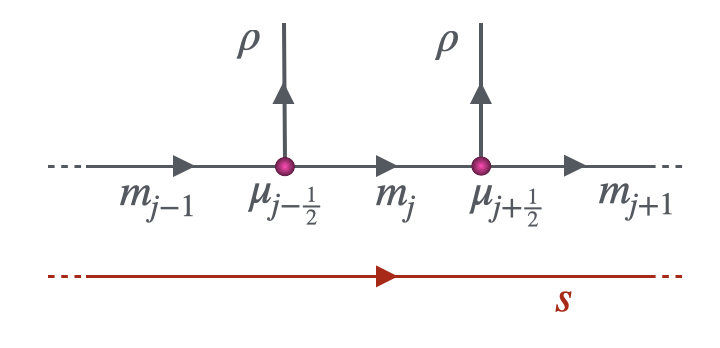}}
\end{split}
\end{equation}
{\bf Hamiltonian Operators.} In contrast the Hamiltonian operators in the anyon chain model act from the boundary $\fB_{\rm inp}$ as depicted in Fig.~\ref{eq:lattice SymTFT} or from ``above" in \eqref{eq:anyon chain HS}. 
Operators appearing in the Hamiltonian are constructed as endomorphisms of $\rho^{\otimes k}$ in $\cC$, where $k\in \mathbb Z_{+}$ sets the interaction range of the model.
For instance, choosing $k=2$, one obtains operators that act on up to five degrees of freedom
\begin{equation}
    \{m_{j-1},\mu_{j-1/2},m_j,\mu_{j+1/2},m_{j+1}\}\,,
\end{equation}
and are depicted as
\begin{equation}
\label{eq:anyon chain Ham Op}
\begin{split}
\text{\includegraphics[scale=.55]{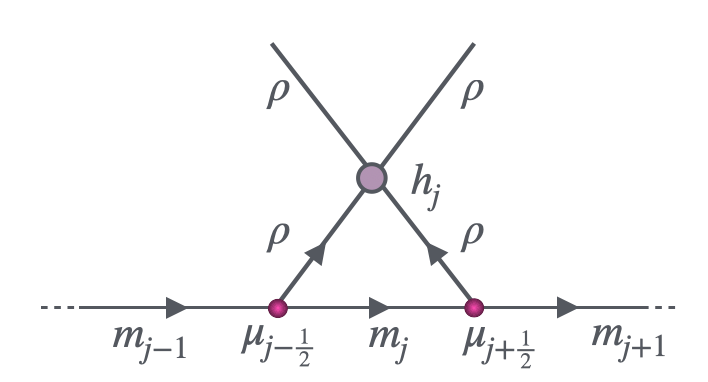}}
\end{split}
\end{equation}
The Hamiltonian constructed from such operators has the form
\begin{equation}
\label{eq:Ham}
    \mathcal H=-\sum_{j}h_{j}\,.
\end{equation}
The state space is a right $\cC$ module category with the module action compatible with symmetry action from $\Bsym$.
Physically this implies that the action of Hamiltonian operators commutes with the symmetry action.
Hence these models are naturally endowed with a $\cS_{L}\boxtimes \cS_{R}$ symmetry.

\smallskip \noindent {\bf Fixed-Point Hamiltonians.} The SymTFT provides a systematic construction of a fixed-point Hamiltonian in any symmetric gapped phase.
We recall that within the SymTFT a gapped phase is obtained by picking a topological boundary condition on the physical boundary $\Bphys$. 
By starting with the reference boundary $\fB_{\rm inp}$, one can obtain any other topological boundary by gauging a Frobenius algebra $\cA$ in $\cC$:
\be
\Bphys= \fB_{\rm inp}/\cA \,.
\ee
Concretely, for the lattice model, this entails requiring that the Hamiltonian is defined using 
$\cA$ in $\cC$ such the $h_{j}$ in \eqref{eq:Ham} is of the form
\begin{equation}
\label{eq:anyon chain Ham Op}
\begin{split}
\text{\includegraphics[scale=.45]{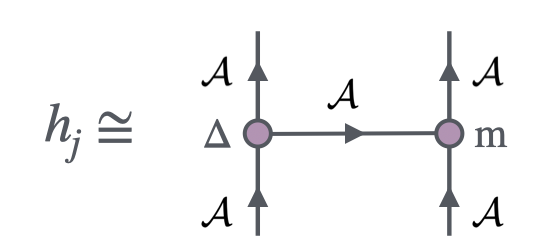}}
\end{split}
\end{equation}
where $\Delta$ and $\rm m$ denote the co-product and product in the algebra. {This projects $\rho$ onto $\mathcal{A}$.}
This can be understood as dynamically inserting a network of Frobenius algebra objects on the input boundary implementing a generalized gauging.
We emphasize that this gauging does not alter the symmetry of the model but is simply a way to obtain a new topological physical boundary condition with the SymTFT and therefore a new gapped phase.
The Frobenius condition also ensures that the Hamiltonian terms are mutually commuting and therefore described a zero-correlation length fixed point limit.
See \cite{Bhardwaj:2024kvy} for details.

\smallskip \noindent {\bf Fixed-point Density Operator.} Given a commuting projector Hamiltonian, it is straightforward to construct a projector onto its ground state space.
The ground state space is greater than one dimensional when the symmetry is spontaneously broken.
However the projector onto each symmetry eigenspace within the ground state subspace has a one-dimensional image.
By tracing over $\cH_{R}$ one obtains a fixed-point density matrix $\rho_{\mathcal{L}}$.

\smallskip \noindent {\bf Lattice Order Parameters.} Consider a chosen gapped phase whose corresponding SymTFT topological boundary condition is given by the Lagrangian algebra $\cL$.
The order parameters for this phase are labeled by anyons in $a_Lb_R\in\cL$ (some anyon might give multiple distinct order parameters).
The construction of such lattice order parameters was developed in \cite{Bhardwaj:2024kvy}.
It involves taking a SymTFT line $a_Lb_R\in \cZ(\cS_L\boxtimes\cS_R)$  in a configuration with one end each on the symmetry and input boundaries.
Then squeezing this line onto the interface where $\cM$ is located produces an action on the state space from which the concrete form of the operator can be read-off.
Naturally, the form of this operator depends on the choice of ends of $a_Lb_R$  on $\fB_{\rm inp}$ and $\Bsym$.
For the present work, we will adapt such lattice order parameters by taking a product of the operator and its Hermitian conjugate separated by a finite segment on the lattice as illustrated in \ref{Patchy}. The total operator is the uncharged, and can take a nonzero expectation value in symmetric states.

\subsection{$G$-Symmetric Lattice Models from SymTFT}
\label{SubSec:G Chain}

To study lattice systems with strong finite group $G$ symmetry, we will first construct lattice systems for the doubled symmetry $\wt G=G_L\times G_{R}$ with $G_L=G_R=G$.
We briefly describe the SymTFT based construction of such models.
The SymTFT corresponding to a non-anomalous finite group $\wt G$ is the $\wt G$ Dijkgraaf-Witten theory or the $\wt G$ Quantum Double model with a trivial topological action/twist.
The topological lines of $\cZ(\wt G)$ form a modular tensor category.
Each simple object of $\cZ(\wt G)$ is a topological line labeled by two pieces of data as
\begin{equation}
    ([g]\,, \Gamma_g)\,,
\end{equation}
where $[g]$ is a conjugacy class in $G$ and $\Gamma_{g}$ is a representation of the centralizer group of a chosen element in $[g]$. 
The SymTFT for such non-anomalous symmetries always has at least two topological boundary conditions. 
One of these boundaries is where all the pure charge lines (i.e., lines for which $[g]=[1]$) can end. 
The Lagrangian algebra corresponding to this boundary condition is
\begin{equation}\label{eq:canonical symmetry boundary}
    \cL_{\wt G}=\bigoplus_{\Gamma}{\rm dim}_\Gamma\,({[1]},\Gamma)\,,
\end{equation}
where the sum is over irreducible representations of $\wt G$ and ${\rm dim}_\Gamma\in \Z_{>0}$ is the dimension of the $\Gamma$ representation.
The (uncondensed) lines on the charge condensed boundary form a $\Vec_{\wt G}$ fusion category, i.e. invertible, (possibly non-abelian) $\wt{G}$ symmetry.
There is another boundary where all of the flux lines can end.
The Lagrangian algebra corresponding to this is 
\begin{equation}\label{eq:canonical symmetry boundary}
    \cL_{\Rep(\wt G)}=\bigoplus_{[g]}\,([g]\,,1)\,,
\end{equation}
where the sum is over conjugacy classes of $\wt G$.

To describe the $\wt G$ symmetric model, we consider the setup as illustrated in Fig.~\ref{eq:lattice SymTFT}.
We pick the symmetry boundary to be defined via the Lagrangian algebra 
\be
\cL_{\sym} = \cL_{\wt G}\ ,,
\ee
such that we have a $\wt G$ symmetry.
We pick the $\Vec_{\wt G}$ module category $\cM$ to be the regular module category such that $\fB_{\rm inp}=\Bsym$. 
and $\cC=\Vec_{\wt G}=\Vec_{G_L}\boxtimes \Vec_{G_R}$ whose simple objects are $g_Lg_R\in \wt G$.
We pick 
\begin{equation}    \rho=\bigoplus_{(g_L,g_R)\in \wt{G}} g_Lg_R\,.
\end{equation}% 
i.e. we sum over all group elements in $\wt{G}$.
With these choices, one obtains

We start with a lattice model whose Hilbert space decomposes into a tensor-product of onsite state spaces as  
\be
\wt \cH = \bigotimes_j\mathbb{C}[\wt G] \cong \bigotimes_j(\mathbb{C}[ G_\p] 
\otimes 
\mathbb{C}[G_\a]
) \,.
\ee
The onsite Hilbert space states are spanned by a group basis
\begin{equation}
    |g_L\,, g_{R}\rangle\,,   \quad g_\p\,, g_\a\in G\,.
\end{equation}
We define left and right group multiplication operators on $\bC[\wt G]$
\be
\begin{split}
L^{g_{\p} g_{\a}} | h_{L}\,,h_{R} \rangle =& | g_Lh_L\,,g_Rh_R \rangle \,, \\    
R^{g_{\p} g_{\a}} | h_{L}\,,h_{R} \rangle =& | h_Lg_L\,, h_Rg_R \rangle \,,
\end{split}
\ee
and diagonal operators 
\be
\begin{split}
    (Z^{\Gamma_L}_{I,J}) | h_L,h_R\rangle =& \cD_{I,J}^{\Gamma_L}  (h_L) | h_L,h_R \rangle \,, \\ 
    (Z^{\Gamma_R}_{I,J}) | h_L,h_R\rangle =& \cD_{I,J}^{\Gamma_R}  (h_L) | h_L,h_R \rangle \,, 
\end{split}
\ee
where $\Gamma_L\,, \Gamma_R \in \Rep(G)$, $I,J\in\{ 1, \cdots, \text{dim} (\Gamma)\}$ and $\cD_\Gamma (h)$ is the matrix representing $h\in G$ in the irrep $\Gamma$.

The strong symmetry operators that generate the $\wt{G}$ symmetry acting on the Choi state are 
\begin{equation}
    \prod_{i}R_{i}^{g}\otimes 1\,, \quad 1\otimes \prod_{i}\left(\overline{R_{i}^{g}}\right)\,.
\end{equation}

\medskip\noindent{\bf $G$-symmetric SWSSB phases.} 
Gapped phases with strong $G$ symmetry form  a subset of gapped phases with $G_L\times G_{R}$ symmetry.
As described in Sec.~\ref{sec:SymStrong}, this subset comprises those $\wt G$ symmetric  phases whose ground states furnish hermitian and positive strong $G$ symmetric density matrices upon tracing out $\cH_{R}$. 
A priori, gapped phases with $\wt G$ symmetry are classified  by tuples $(\wt F,\beta)$ where $\wt F\subseteq \wt G$ and $\beta \in H^2(\wt F,U(1))$. These label the physical boundary conditions $\Bphys$ of the SymTFT, i.e. Lagrangian algebras 
\be\label{GLags}
\cL_{(\wt{F}, \beta)} \,.
\ee
Physically the label $(\wt F,\beta)$ is associated to a phase that spontaneously breaks the $\wt G$ symmetry down to a subgroup $\wt F$, such that each ground state is an SPT, labeled by $\beta$, of the unbroken group $\wt F$. 
For the $\wt{G}$ symmetric phase to furnish a consistent strong $G$ symmetric gapped phase we firstly require $\wt F$ to be a swap symmetric subgroup of $\wt{G}$. 
There are further constraints that the 2-cocycle $\beta$ must satisfy.
To deduce these, we recall that the patch operators / order parameters for an SPT labeled by $\beta$ are 
\begin{equation}
    \mathcal{P}_{f,\beta}(i,j)=(\cO_{\iota_f\beta}^{i})\cU_{f}^{i\to j}(\cO_{\iota_f \beta}^j)^{\dagger}\,, \qquad f\in \wt F\,,
\end{equation}
where $\cU_{f}^{i\to j}$ is an $f\in \wt F$ symmetry string (disorder operator) extending from site $i$ to site $j$,\footnote{{Note that in this case the symmetry operator is invertible, so there is no need to trace over any virtual space as discussed at the end of Sec.~\ref{sec:PureStateSymTFT}.}} while $\iota_f \beta\in H^{1}(\wt F,U(1))$ is obtained via a slant product from $\beta$ and $\cO_{\iota_f \beta}$ is an operator that transforms under the representation $\iota_f \beta$. This has the concrete form
\begin{equation}
    \iota_f \beta(f')=\frac{\beta(f\,, f')}{\beta(f'\,, (f')^{-1}ff')}\,.
\end{equation}
Now, let us denote the decomposition of the patch operator onto operators acting on $\mathcal{H}_L\otimes\mathcal{H}_R$ as 
\begin{equation}
    \mathcal{P}_{f,\beta}(i,j)=\mathcal{R}_{f,\beta,L}(i,j)\otimes\mathcal{R}_{f,\beta,R}(i,j) \,.
\end{equation}
The fact that $\LL\rho|\mathcal{P}_{f,\beta}(i,j)|\rho\RR\neq 0$ for the gapped phase labeled $(\wt F,\beta)$ in conjunction with the condition \eqref{Kautschi} translates to the fact that the following operators have a nonzero expectation value on the states $|\rho\drangle$ 
\begin{align}
    \begin{split}
        \LL\rho|\mathcal{R}_{f,\beta,L}(i,j)\otimes\overline{\mathcal{R}_{f,\beta,L}(i,j)}|\rho\RR\neq 0\\
        \LL\rho|\overline{\mathcal{R}_{f,\beta,R}(i,j)}\otimes\mathcal{R}_{f,\beta,R}(i,j)|\rho\RR\neq 0\,.
    \end{split}
\end{align}
These patch operators need to mutually commute for all $f\in \wt F$ which imposes constraints on $\beta$ that lead to consistent mixed phases. 
Next, one may construct commuting projector Hamiltonians on $\wt{\cH}$ in the gapped phase $(\wt F,\beta)$ as
\be
    \wt{H}_{({\wt F}\,, \beta)}=- \sum_{j}h_{j}^{(\wt{F}, \beta)}\,.
\ee
where $h_{j}^{(\wt F,\beta)}$ acts on three $\wt G$ qudits as
\begin{equation}
\begin{split}
   \langle g_{j-1}\,, &hg_{j}\,, g_{j+1} |h_{j}^{(\wt F,\beta)} |g_{j-1}\,, g_{j}\,, g_{j+1}\rangle \\
   &= \,\delta^{\wt F}_{g_{j}g_{j-1}^{-1}}\delta^{\wt F}_{g_{j+1}g_{j}^{-1}}\frac{\beta(h,g_{j}g_{j+1}^{-1})}{\beta(g_{j-1}g_{j}^{-1}h^{-1}\,, h)} 
   \,,    
\end{split}
\end{equation}
where $\delta^{F}_{g} = 1$ if $g\in F $ and $0$ otherwise.

\subsection{$\Rep (G)$-symmetric Lattice Models from SymTFT}
\label{SubSec:Rep(G) Chain}
We now extend the lattice model and density matrices for the weak/strong symmetric phases for any $\Rep (G)$ symmetry. 
Note that $\Rep(G)$ symmetry is non-invertible for a non-Abelian $G$.
Following \cite{Bhardwaj:2024kvy,Bhardwaj:2024wlr,Warman:2024lir} we construct lattice models for strongly $\Rep (G)$ symmetric theories using the SymTFT based construction. 
Again the SymTFT is $\cZ(\Vec_G)$, i.e., the $G$ quantum double model.
The symmetry boundary for the strong symmetry $\Rep (G)$ coresponds to the flux condensed or Neumann boundary for which the Lagrangian algebra is  
\be
\cL_{\rm sym}= \bigoplus_{[g]\in G_\p  \times G_\a } ([g], 1) \,.
\ee
We choose again 
\be
\Binp= \cL_{\wt G}
\ee
in \eqref{eq:canonical symmetry boundary} as the input boundary.
in which case then 
\be
\cM= \Vec
\ee
so that the symmetry is $\Rep (G)$ string. 
With these choices, one obtains a state space that decomposes into a tensor product of local state spaces (each isomorphic to the group algebra $\bC[\wt G]$) associated to the morphism spaces of the anyon chain model.
\begin{equation}
%\label{eq:lattice SymTFT}
\begin{split}
\text{\includegraphics[scale=.4]{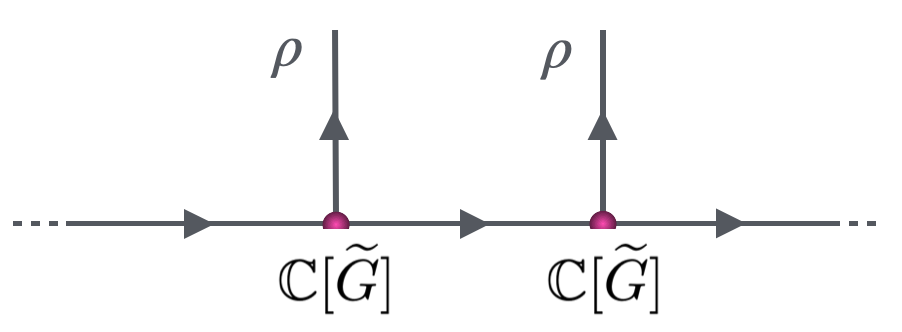}}
\end{split}
\end{equation}
The double Hilbert space is thus spanned by states
\be
|\vec{g}_L\,, \vec{g}_R \rangle\,,\,, 
\ee
where $\vec{g}_L=g_{L,1}\,, \cdots g_{L,N}$ etc.
There is $\Rep (G_L\times G_R)$ symmetry generated by $\Gamma_L\Gamma_R$ with $\Gamma_L\in \Rep(G_L)$ and $\Gamma_R\in \Rep(G_R)$.
These act as 
\be
\begin{split}
S_{\Gamma_L\Gamma_R}&  |\vec{g}_L\,, \vec{g}_R \rangle = \\ &\chi_{\Gamma_L}({\rm hol}(\vec{g}_L))\chi_{\Gamma_R}({\rm hol}(\vec{g}_R))|\vec{g}_L\,, \vec{g}_R \rangle\,.      
\end{split}
\ee
where ${\rm hol}(\vec{g}_L)\equiv \prod_j g_{L,j}$ etc.
Different gapped phases are again labeled by the Lagrangian algebras \ref{GLags}, specified by a pair $(\widetilde{F},\beta)$, which is subject to   the same constraints as those for the $G$ symmetric models described in Sec.~\ref{SubSec:G Chain}.
The commuting projector Hamiltonian for the gapped phase labeled $(\wt F,\beta)$ has the form
\be\label{DoubleHam}
\wt{H}_{(\widetilde{F}, \beta)}^{\Rep (G)} = - {1\over |F|} \sum_{j,f} (R^{\ol{f}}_{\beta})_{j}  (L^{f}_\beta)_{ j+1}  - \sum_{j} P_j^{(\widetilde{F})} \,.
\ee
Where the $\beta$-twisted left and right group multiplication operators are defined by the action
\begin{equation}\label{eq:beta_on_ket}
    \begin{split}
         R^{ \ol{f}}_\beta| f'\rangle &= \frac{1}{\beta( f' f^{-1}\,,  f)}| f' f^{-1}\rangle\,, \\
         L^{ f}_\beta| f'\rangle &= {\beta( f\,,  f')}| f f'\rangle\,, \\
    \end{split}
\end{equation}
The projector is defined as 
\be 
P^{(\widetilde{F})} |g \rangle = \delta_{g}^{\wt F}|g\rangle\,.
\ee
By taking the symmetric ground states for this doubled description we can extract the density matrix on the doubled Hilbert space, and then trace out half the degrees of freedom. 
The density matrix in this doubled Hilbert space can be written as the product of all the commuting projectors in the Hamiltonian (\ref{DoubleHam}) 
\be
\widetilde{\rho}_{(\widetilde{F},\beta)} \propto \prod_j \left(\sum_{f\in \widetilde{F}} \left(R^{\ol{f}}_{\beta}\right)_{j}  \left(L^{f}_\beta\right)_{ j+1}\right)P_j^{(\widetilde{F})} \,.
\ee
We note that the image of the projector $\widetilde{\rho}_{(\widetilde{F},\beta)}$ might be multi-dimensional.
Therefore we single out the state invariant under $\Rep(G)$ symmetry by using the projector
\begin{equation}
    P^{(\Rep(G))}=\sum_{\Gamma \in \Rep(\wt G)}\frac{\dim(R)}{|\wt{G}|} \cS_\Gamma\,.
\end{equation}
We define the fixed-point density matrix as the projector onto the $\Rep(\wt G)$ singlet in the ground state space of the $(\wt F,\beta)$ phase as
\begin{equation}
    \rho^{\Rep(G)}_{(\wt F,\beta)}= P^{(\Rep(G))}\times \widetilde{\rho}_{(\widetilde{F},\beta)}\,.
\end{equation}
{In addition to the singlet density matrix, one can also have density matrices with non-trivial eigenvalues under $\Rep(G)$ symmetry, as will be illustrated in the examples of Sec.~\ref{sec:strongRepS3}. These are simply obtained as projectors onto ground states  in other symmetry sectors.}
The density matrix acting on the physical Hilbert space $\cH$ is obtained by tracing out half the degrees of freedom. 
\begin{equation}
\rho_{(\wt F,\beta)}=    \Tr_{\cH_{\a}}\left[\rho^{\Rep(G)}_{(\wt F,\beta)} \right]\,.
\end{equation}
In this case the symmetry operators are non-invertible, so in order to obtain patch operators that act on the Hilbert space one needs to remove the extra ``virtual" Hilbert space introduced by the endpoints of the truncated symmetry operator. One example of how this can done is Ref.~\cite{Fechisin:2023dkj}, which constructed patch operators (i.e. string order parameter operators) for $G\times\mathrm{Rep}(G)$ SPTs. Similar patch operators would apply here.

\section{Examples with Strong Symmetries}
\label{Sec:Examples}

We will now illustrate this general framework with several examples. Each example will be discussed first in  the continuum SymTFT framework and subsequently in a lattice model realization. 
Note that the examples we consider involve {\bf no explicit symmetry breaking}, neither of strong nor weak symmetries, therefore they will always have 
\be\ba
\Tr(\rho \,\cO_i)&=0\,,\qquad \Tr(\rho\, \cO_i\rho \cO_i^\dagger)&=0\,,
\ea\ee
for $\cO_i$ a local operator charged under the symmetry.
However, the phases we discuss will include {\bf spontaneous} symmetry breaking: we will focus in particular on SWSSB phases.

To start off, we recap the well-known case of $\Z_2$ strong symmetries, then $\Z_2\times\Z_2$ strong symmetry which includes a mixed strong-weak SPT phase, then the lesser-known examples of non-abelian groups $S_3$ and then the non-invertible symmetries $\Rep (S_3)$ and $\Ising$. In all these we start with a purely strong symmetry. The mixed strong weak starting point will be discussed in the next section.

\subsection{$\Z_2$ Strong Symmetry Mixed Phases}

\subsubsection{SymTFT} \label{sec:SymTFT_Z2}

As a warmup consider the starting point to be a strong anomaly-free $\Z_2$ symmetry, sometimes written as $\mathrm{Vec}_{\mathbb{Z}_2}$. In the following, we will just write $\mathbb{Z}_2$ in place of $\mathrm{Vec}_{\mathbb{Z}_2}$. $\cZ (\mathbb{Z}_2\times\mathbb{Z}_2)$ is simply the $\mathbb{Z}_2\times\mathbb{Z}_2$ gauge theory without any Dijkgraaf-Witten twist, i.e. two copies of toric code. The anyons are generated by gauge charges $e_L,e_R$ and gauge fluxes $m_L,m_R$. Here, everything is real, so we do not explicitly write out the complex conjugates. 
The strong Lagrangian algebras are 
\begin{align}
     \cL_{\mathbb{Z}_2\times\mathbb{Z}_2} & = (1\oplus  e_\p ) \otimes (1\oplus e_\a ) \label{eq:LZ2_e} \\
     \cL_{\Rep(\mathbb{Z}_2\times\mathbb{Z}_2)} & = (1\oplus m_\p ) \otimes (1\oplus m_\a ) \,. \label{eq:LZ2_m}
\end{align}
We pick the symmetry boundary to be $\cL_{\Z_2\times\Z_2}$. Then immediately we find two pure-state phases: 
\be
\begin{split}
(\cL_{\mathbb{Z}_2\times\mathbb{Z}_2}, \cL_{\mathbb{Z}_2\times\mathbb{Z}_2}) &:\quad \Z_2^{\strong} \text{ to nothing SSB} \cr 
(\cL_{\mathbb{Z}_2\times\mathbb{Z}_2}, \cL_{\Rep(\mathbb{Z}_2\times\mathbb{Z}_2)}) &:\quad \Z_2^{\strong} \text{ preserving} \cr    
\end{split}
\ee
given by two Lagrangians that are simply products of Lagrangian algebras in the physical and ancilla sectors. 

Indeed, $(\cL_{\mathbb{Z}_2\times\mathbb{Z}_2}, \cL_{\mathbb{Z}_2\times\mathbb{Z}_2})$ gives the sandwich
\be\begin{split} 
\begin{tikzpicture}
\node at (1.5,-1.5) {$\Longrightarrow \ \Z_2$ SSB};
\draw [thick] (0,-1) -- (0,1) ;
\draw [thick] (3,-1) -- (3,1) ;
\draw [thick] (0, 0.25) -- (3, 0.25) ;
\draw [thick] (3, -0.5) -- (0, -0.5) ;
\node[above] at (1.5,0.25) {$\bm{e_\p  e_\a }$} ;
\node[above] at (1.5,-0.5) {$e_\p ,\; e_\a $} ;
\node[above] at (0,1) {$\cL_{\mathbb{Z}_2\times\mathbb{Z}_2}$}; 
\node[above] at (3,1) {$\cL_{\mathbb{Z}_2\times\mathbb{Z}_2}$}; 
\draw [black,fill=black] (0,0.25) ellipse (0.05 and 0.05);
\draw [black,fill=black] (3,0.25) ellipse (0.05 and 0.05);
\draw [black,fill=black] (3,-0.5) ellipse (0.05 and 0.05);
\draw [black,fill=black] (0,-0.5) ellipse (0.05 and 0.05);
\end{tikzpicture}
\end{split}\ee
which provides charged local operators for both the strong and weak symmetry, naively giving four states. Note however that only the $T$-symmetric states are the singlet state and the one given by insertion of $\cO_{e_Le_R}$,  which is in the odd sector of the strong $\mathbb{Z}_2$ symmetry. We will bold-face the charged operators that are $T$-symmetric in the following SymTFT configurations throughout. 

There is one additional Lagrangian algebra of the double that satisfies the hermiticity and positivity conditions:
\be
\cL_\text{weak} = 1\oplus e_\p  e_\a  \oplus m_\p  m_\a  
\oplus e_\p m_\p  e_\a  m_\a  \,.
\ee
$\cL_\text{weak}$ cannot be written as $\mathcal{L}_L\otimes\mathcal{L}_R$.
Considering the anyons in the SymTFT Sandwich 
\be
\cL_{\mathbb{Z}_2\times\mathbb{Z}_2} \cap \cL_\text{weak} = \{ 1, e_\p  e_\a \} \,,
\ee
which can be depicted as follows:
\vspace{-2mm}
\be\begin{split} 
\begin{tikzpicture}
\node at (1.5,-1) {$\Longrightarrow \ \Z_2$ SWSSB};
\draw [thick] (0,-0.5) -- (0,1) ;
\draw [thick] (3,-0.5) -- (3,1) ;
\draw [thick] (0, 0.25) -- (3, 0.25) ;
\node[above] at (1.5,0.25) {$\bm{e_\p  e_\a}$} ;
\node[above] at (0,1) {$\cL_{\mathbb{Z}_2\times\mathbb{Z}_2}$}; 
\node[above] at (3,1) {$\cL_{\weak}$}; 
\draw [black,fill=black] (0,0.25) ellipse (0.05 and 0.05);
\draw [black,fill=black] (3,0.25) ellipse (0.05 and 0.05);
\end{tikzpicture}
\end{split}\ee
There is a charged local operator $\cO_{e_\p  e_\a }$, which is $T$-symmetric,  that has non-trivial charge under both $m_\p $ and $m_\a $ and thus indicates strong SSB. However the weak symmetry, that originates from $m_\p m_\a $ links trivially with $e_\p  e_\a $ and remains unbroken. In summary: the $\Z_2$ strong symmetry is SSB'ed and the weak symmetry is preserved, i.e. this is a SWSSB:
\be
(\cL_{\mathbb{Z}_2\times\mathbb{Z}_2},\cL_\text{weak}) = \Z_2 \text{ SWSSB}  \,.
\ee

\subsubsection{Density Matrices}

We can derive lattice models for  phases with $\Z_2$ strong symmetry from the SymTFT discussion in Sec.~\ref{sec:SymTFT_Z2}, following the approach of \cite{Bhardwaj:2024kvy},  reproducing the well-known results for $\Z_2$. The method is however generalizable to non-abelian groups and non-invertible symmetries as explained in the preceding sections and as we will illustrate in subsequent examples.

For anomaly-free $\Z_2$ strong symmetry, we take the physical Hilbert space to be $\cH = \bigotimes_j \mathbb{C}_j^{2}$ on sites $j=1,...,N$ 
and $|\rho_{\mathcal{L}}\RR$ lives in $\mathcal{H}\otimes\mathcal{H}$.

We will now derive density matrices on $\cH$ strongly symmetric under the $\Z_2$ symmetry generated by $U=\prod_jX_j$ where $X_j$ is the Pauli $X$ matrix at site $j$. We will derive them from the tgrangian algebras, with the Lagrangian algebra \eqref{eq:LZ2_e} on the SymTFT symmetry boundary.

\vspace{2mm}
\noindent {\bf $\mathbb{Z}_2\times\mathbb{Z}_2$ preserving phase.}
Condensing the SymTFT Lagrangian algebra \eqref{eq:LZ2_m} on the physical boundary corresponds to taking $\wt{F}=\Z_2^\p\times\Z_2^\a$, which gives the Hamiltonian
\begin{align}
    \wt{H}_{\mathbb{Z}_2\times\mathbb{Z}_2}=-\sum_j\left(\frac{\bI+X_{j,L}}{2}\right)\left( \frac{\bbI+X_{j,R}}{2}\right).
\end{align}
We can write the density matrix in the doubled Hilbert space by taking products of the terms in $\wt{H}$ on each site,
which, upon tracing out $\cH_R$ and identifying $X_{j,L}$ with $X_j$, gives the pure state
\begin{align}
    {\rho}_{\mathbb{Z}_2\times\mathbb{Z}_2}=\prod_j\frac{\bI+X_j}{2}=\bigotimes_j(\ket{+}\bra{+})_j\,.
\end{align}
where $\ket{+}_j$ is the $+1$ eigenstate of $X_j$. From the SymTFT, no anyons end on both boundaries, indicating that the full $\Z_2$ strong symmetry is preserved. Indeed, we can check that (dropping the $\mathbb{Z}_2\times\mathbb{Z}_2$ subscript on $\rho$ to reduce cluttering of notation) 
\begin{align}
    \Tr(\rho Z_iZ_j)&=0, &  \Tr(\rho Z_iZ_j\rho Z_j Z_i)&=0\,.
\end{align}

\noindent {\bf Strong-to-nothing (SSB) phase.}
Choosing instead the physical boundary of the SymTFT to have \eqref{eq:LZ2_e} condensed corresponds to taking $\wt{F}=1$, which gives the Hamiltonian:
\begin{align}
    \wt{H}_{\Rep(\mathbb{Z}_2\times\mathbb{Z}_2)}=-\sum_j
    \left(\frac{\bI+Z_{j,L}Z_{j+1,L}}{2}\right)\left(\frac{\bI+Z_{j,R}Z_{j+1,R}}{2}\right)\,.
\end{align}
As mentioned earlier, only two of the four degenerate ground states of $\wt{H}_{\Rep(\mathbb{Z}_2)}$ correspond to valid Choi states. These are the two states in the $+1$ sector of the weak $\mathbb{Z}_2$ symmetry and opposite sectors of the strong $\mathbb{Z}_2$ symmetry.
Tracing out $\mathcal{H}_R$ gives 
\begin{align}
   {\rho}_{\Rep(\mathbb{Z}_2\times\mathbb{Z}_2)}^\pm=\frac{P^\pm}{2^{N}}\prod_j\lbb\bI+Z_jZ_{j+1}\rbb =\ket{\text{GHZ}\pm}\bra{\text{GHZ}\pm}\,.
\end{align}
where $P^\pm=\tfrac{1}{2}(1\pm \prod_jX_j)$ is the projector onto the $\mathbb{Z}_2$ even/odd subspace and $|\mathrm{GHZ}\pm\rangle$ is the $\mathbb{Z}_2$ even/odd GHZ state.
From the SymTFT, the non-trivial anyons $e_\p e_\a,\,e_\p,\,e_\a,$ end on both boundaries, indicating that the $\Z_2$ strong and weak symmetries are spontaneously broken. Indeed, we can check that 
\begin{align}
    \Tr(\rho Z_iZ_j)&=1, &  \Tr(\rho Z_iZ_j\rho Z_j Z_i)&=1\,,
\end{align}
detecting that the $\Z_2$ strong symmetry is SSB'ed to nothing. 

\vspace{2mm}
\noindent {\bf Strong-to-Weak SSB phase (SWSSB).}
Finally, taking the physical boundary of the SymTFT to have 
\be
\cL_\text{weak} = 1\oplus e_\p  e_\a  \oplus m_\p  m_\a  \oplus e_\p m_\p  e_\a  m_\a
\ee
condensed corresponds to setting $\wt{F}=\Z_2^\text{diag}$, which gives the Hamiltonian:
\be
\ba
    \wt{H}_{\mathrm{weak}}=
    &-\sum_j 
    \frac{\bbI+X_{j,L} X_{j,R}}{2}\\
    &\phantom{=}-\sum_j\frac{\bI+Z_{j,L}Z_{j+1,L}Z_{j,R}Z_{j+1,R}}{2}\,.
\ea
\ee
Following the same steps as above, we obtain the mixed states 
\be
    {\rho}_{\mathrm{weak}}^\pm=\frac{1}{2^{N-1}}P^\pm\,,
\ee
which demonstrate SWSSB:
\begin{align}
    \Tr(\rho Z_iZ_j)&=0\,, &  \Tr(\rho Z_iZ_j\rho Z_j Z_i)=1\,,
\end{align}
indicating that the strong $\Z_2$ symmetry is spontaneously broken but the weak $\mathbb{Z}_2$ symmetry is preserved.

\subsection{$\Z_2\times\Z_2$ Strong Symmetry Mixed Phases}
\label{sec:Z2Z2}
Let us now discuss $\Z_2^A\times\Z_2^B$ strong symmetry. The anyons in the doubled center $\cZ(\Z_2\times\Z_2)\boxtimes\cZ(\Z_2\times\Z_2)$ are spanned by $e_k^A,e_k^B,m_k^A,m_k^B$, where $k=\p,\a$ labels each center. 

The Lagrangian algebra for the strong symmetry is:
\be
\ba
    \cL_{\Z_2\times\Z_2}^{\strong}&=(1 \oplus e^A_\p \oplus e^B_\p  \oplus e^A_\p e^B_\p )(1 \oplus e^A_\a  \oplus e^B_\a  \oplus 
e^A_\a e^B_\a )\,.
\ea
\ee

\begin{widetext}
{\begin{center}
\begin{table}
    \centering
    $\begin{array}{|c|c|c|}
        \hline
        \cL_\phys & \text{Charged Local operators} & \text{Phase} \\
        \hline
        \cL_{1} & e^A_\p ,\;e^A_\a ,\; e^A_\p e^A_\a ,\;e^B_\p e^B_\a ,\;e^B_\p e^A_\a e^B_\a ,\;e^A_\p e^B_\p e^B_\a ,\;e^A_\p e^B_\p e^A_\a e^B_\a  & \Z_2^A \text{ SSB},\;\Z_2^{AB} \text{ SSB},\; \Z_2^B \text{ SWSSB} \\
        \cL_{2} & e_\p ^B,\;e^B_\a ,\;e^B_\p e^B_\a ,\;e^A_\p e^A_\a ,\;e^A_\p e^A_\a e^B_\a ,\;e^A_\p e^B_\p e^A_\a ,\;e^A_\p e^B_\p e^A_\a e^B_\a   &  \Z_2^B \text{ SSB},\;\Z_2^{AB} \text{ SSB},\; \Z_2^A \text{ SWSSB}\\
        \cL_{3} & e^A_\p e^B_\p ,\;e^A_\a e^B_\a ,\;e^A_\p e^A_\a ,\;e^B_\p e^B_\a ,\;e^A_\p e^B_\a ,\;e^B_\p e^A_\a ,\;e^A_\p e^B_\p e^A_\a e^B_\a  &  \Z_2^A \text{ SSB},\;\Z_2^B \text{ SSB},\; \Z_2^{AB} \text{ SWSSB}\\
        \cL_{4} & e^A_\p e^A_\a ,\;e^B_\p e^B_\a ,\;e^A_\p e^B_\p e^A_\a e^B_\a  & \Z_2^A\times\Z_2^B \text{ SWSSB}\\
        \cL_{5} & e^A_\p e^B_\p e^A_\a e^B_\a  & \Z_2^A \text{ SWSSB},\;\Z_2^B \text{ SWSSB}\phantom{,\;\text{with SPT}}\\
        \cL_{6} & e^A_\p e^B_\p e^A_\a e^B_\a  & \Z_2^A \text{ SWSSB},\;\Z_2^B \text{ SWSSB},\;\text{with SPT}\\
        \cL_{7} & e^A_\p e^A_\a  & \Z_2^A \text{ SWSSB}\phantom{,\;\text{with SPT}}\\
        \cL_{8} & e^A_\p e^A_\a  & \Z_2^A \text{ SWSSB},\;\text{with SPT}\\
        \cL_{9} & e^B_\p e^B_\a  & \Z_2^B \text{ SWSSB}\phantom{,\;\text{with SPT}}\\
        \cL_{10} & e^B_\p e^B_\a  & \Z_2^B \text{ SWSSB},\;\text{with SPT}\\
        \hline
    \end{array}$
    \caption{Strong to weak spontaneous symmetry breaking patters starting from strong $\Z_2^A\times\Z_2^B$ symmetry.}
    \label{tab:Z2Z2_SWSSB}
\end{table}
\end{center}}
\end{widetext}

There are ten Lagrangian algebras in $\cZ(\Z_2\times\Z_2)\boxtimes\cZ(\Z_2\times\Z_2)$ that satisfy the positivity and symmetry conditions and do not factorize into a product of Lagrangian algebras for each factor. These Lagrangian algebras describe various patterns of SWSSB, including possible non-trivial SPTs for the preserved symmetry. These are provided in appendix \ref{app:LagsZ2Z2} and the phases are summarized in table \ref{tab:Z2Z2_SWSSB}.

By taking the SymTFT sandwich with $\cL_\sym$ symmetry algebra and each of above Lagrangian algebras as a physical boundary, we can identify the charged local operators, given by the anyons ending on both boundaries, that reveal the patter of strong to weak spontaneous symmetry breaking. 

Note that we also have SPTs for the unbroken symmetries.
These are detected by string order parameters, i.e. anyons ending on the physical boundary that are not in $\mathcal{L}_{\mathbb{Z}_2\times\mathbb{Z}_2}^{\mathrm{strong}}$.
After compactification of the SymTFT sandwich, these give rise to a local operator attached to symmetry `strings'.

\vspace{2mm}
\noindent {\bf Analysis of SPTs.}
Let's consider the two Lagrangian algebras $\cL_7$ and $\cL_8$:
\be
\ba
       \cL_{7}=&1\oplus m^B_\a \oplus m^B_\p \oplus m^B_\p m^B_\a \oplus m^A_\p m^A_\a \\
       &\oplus m^A_\p m^A_\a m^B_\a \oplus m^A_\p m^B_\p m^A_\a \\
       &\oplus m^A_\p m^B_\p m^A_\a m^B_\a \oplus  \underline{e^A_\p e^A_\a }\oplus e^A_\p e^A_\a m^B_\a\\& \oplus e^A_\p m^B_\p e^A_\a\oplus e^A_\p m^B_\p e^A_\a m^B_\a \\
       &\oplus e^A_\p m^A_\p e^A_\a m^A_\a \oplus e^A_\p m^A_\p e^A_\a m^A_\a m^B_\a \\
       &\oplus e^A_\p m^A_\p m^B_\p e^A_\a m^A_\a \oplus e^A_\p m^A_\p m^B_\p e^A_\a m^A_\a m^B_\a   
\cr 
      \cL_{8}=&1\oplus e^A_\a m^B_\a \oplus m^B_\p m^B_\a \oplus m^B_\p e^A_\a \\
      &\oplus e^A_\p m^B_\a \oplus \underline{e^A_\p e^A_\a }\oplus e^A_\p m^B_\p \\
      &\oplus e^A_\p m^B_\p e^A_\a m^B_\a \oplus e^B_\p m^A_\p e^B_\a m^A_\a \\&\oplus e^B_\p m^A_\p e^A_\a e^B_\a m^A_\a m^B_\a \oplus e^B_\p m^A_\p m^B_\p e^B_\a m^A_\a m^B_\a \\
      &\oplus e^B_\p m^A_\p m^B_\p e^A_\a e^B_\a m^A_\a \oplus e^A_\p e^B_\p m^A_\p e^B_\a m^A_\a m^B_\a \\
      &\oplus e^A_\p e^B_\p m^A_\p e^A_\a e^B_\a m^A_\a \oplus e^A_\p e^B_\p m^A_\p m^B_\p e^B_\a m^A_\a \\
      &\oplus  e^A_\p e^B_\p m^A_\p m^B_\p e^A_\a e^B_\a m^A_\a m^B_\a   
\ea
\ee

Their charged local operators are the same, given by 
\be
\cO_{e^A_\p  e^A_\a } \,.
\ee
This indicates the breaking of $\Z_2^A$ strong to a weak $\Z_2^A$ symmetry. The weak symmetry is generated by 
\be
\Z_2^{A, \weak}:\qquad m_\p ^A m_\a ^A \,.
\ee
We still also have strong $\Z_2^B$ symmetry generated by 
\be
\Z_2^{B, \strong}:\qquad m_\p ^B \ \text{and} \ m_\a ^B \,.
\ee
Let's contrast the phases associated to these two algebras:
\begin{itemize}
\item $\cL_7$: $m_\p ^Am_\a ^A$, $m_\p ^B$ and $m_\a ^B$ are part of this algebra which generate the symmetries $\Z_2^{A, \weak}\times \Z_2^{B, \strong}$. These are trivial string-like order parameters. 
\item $\cL_8$: In contrast, this Lagrangian algebra contains 
\be
e_\p ^A m_\p ^B \oplus  e_\p ^A m_\a ^B \oplus  e_\a ^A m_\p ^B \oplus  e_\a ^A m_\a ^B  \,,
\ee
\end{itemize}
which give rise to string order parameters that are charged under the weak $\Z_2^A$, and have a string of the strong symmetry $\Z_2^B$ attached. In this sense this is a {\bf strong-weak-SPT} phase. A $\Z_2^\strong\times\Z_2^\weak$ SPT was provided in \cite{lee2025, Ma:2023rji}. The phase corresponding to $\cL_8$ with $\Z_2^A\times\Z_2^B$ strong symmetry is very similar to this SPT, except one of the $\mathbb{Z}_2$ symmetries is broken spontaneously rather than explicitly.

\vspace{2mm}
\noindent {\bf Density matrix.}
{Let $\mathbb{Z}_2^A$ be generated by $\prod_iX_{2i+1}$ and the $\mathbb{Z}_2^B$ symmetry be generated by $\prod_iX_{2i}$. Then a density matrix realizing $\mathcal{L}_8$ are respectively:
\begin{equation}
\begin{split}
       \rho &=\frac{1}{2^N}\prod_i\left(\bI+Z_{2i-1}X_{2i}Z_{2i+1}\right)\cdot \Bigl(\bI+\prod_iX_{2i+1}\Bigr)\,. 
\end{split}
\end{equation}
One can get this from the anyons in $\mathcal{L}_8$: the only anyon living entirely in $\mathcal{H}_L$ is $e_L^Am_L^B$ so if we trace out the spins forming $\mathcal{H}_R$ we only get the local projectors corresponding to $e^Am^B$ and the global projectors corresponding to the strong symmetries $(\bI+\prod_iX_{2i})$ and $(\bI+\prod_iX_{2i+1})$. The former is already included in the local projectors corresponding to $e^Am^B$ given by $(\bI+Z_{2i-1}X_{2i}Z_{2i+1})$.
This density matrix can be obtained from the pure cluster state $\rho=\frac{1}{2^N}\prod_i(\bI+Z_{i-1}X_iZ_{i+1})$ by applying the following channel on every pair of odd sites:
\begin{equation}
    E_{2i-1,2i+1}[\rho]=\frac{1}{2}\rho+\frac{1}{2}Z_{2i-1}Z_{2i+1}\rho Z_{2i-1}Z_{2i+1}\,.
\end{equation}}

\vspace{2mm}
\noindent
{\bf Diagnostic for the SPT.}
Let us recall how the patch operators corresponding to the anyons 
$e_L^{A}m_L^{B}, e_L^{A}m_R^{B}$ are represented on the doubled Hilbert space
\begin{equation}
\begin{split}
\cP_{e_L^{A}m_L^{B}}(2i+1\,, 2j+1)&= \left[Z_{2i+1}\left(\prod_{\ell=i+1}^{j}X_{2\ell}\right)Z_{2j+1}\right]_{L}\hspace{-5pt}\otimes 1_{R}\\
\cP_{e_L^{A}m_R^{B}}(2i+1\,, 2j+1)&=
\left[Z_{2i+1}Z_{2j+1
}\right]_{L}
\otimes 
\left[\prod_{\ell=i+1}^{j}X_{2\ell}\right]_{R} \,.
\end{split}
\end{equation}
It follows that the diagnostics for the mixed phase that corresponds to the Lagrangian algebra $\cL_8$ are
\begin{equation}
\begin{split}
\Tr\left[\rho Z_{2i+1}\left(\prod_{\ell=i+1}^{j}X_{2\ell}\right)Z_{2j+1}\right]&=1\,, \\
\Tr\left[\rho 
Z_{2i+1}Z_{2j+1
}
\rho 
\left(\prod_{\ell=i+1}^{j}X_{2\ell}\right)
\right]&=1 \,.
\end{split}
\end{equation}
These expectation values are 1 on the fixed point density matrices but can more generally serve as diagnostics for the mixed state gapped phase.

\subsection{Setup for Strong $S_3$ and Strong $\Rep (S_3)$} 
\label{sec:SymTFT_S3}
We now turn to the non-abelian group
\be
    S_3=\langle a,b \;\;|\;\;a^3=1\,,\quad b^2=1\,,\quad bab=a^2\, \rangle\,,
\ee
whose conjugacy classes are
\be
    [1]=\{1\}\,,\qquad [a]=\{a,\,a^2\}\,,\qquad [b]=\{b,\,ab,\,a^2b\}.
\ee
We denote its representations by
\be
    \Rep(S_3)=\{1,\,P,\,E\}
\ee
where $P$ is a non-trivial 1d irrep with character $-1$ on elements in $[b]$, while $E$ is the 2d dimensional irrep, with matrix representation:
\be\ba
    \cD_E(a)=
    \begin{pmatrix}
    \omega & 0 \\
    0 & \ol{\omega}
    \end{pmatrix}\,,\qquad 
     \cD_E(b)=
    \begin{pmatrix}
    0 & 1 \\
    1 & 0
    \end{pmatrix}\,,
\ea\ee
where $\omega=e^{2\pi i/3}$ and $\ol{\omega}=\omega^2$. \\

The tensor products (fusions) of irreps are:
\be
    P\otimes P=1\,,\qquad E\otimes E=1\oplus P\oplus E\,,
\ee
the latter being a non-invertible fusion rule. Again, we will write $S_3$ as short-hand for $\mathrm{Vec}_{S_3}$.

\vspace{2mm}
\noindent{\bf SymTFT Setup.} For both symmetries, we take the doubled SymTFT $\cZ (S_3\boxtimes \ol{S_3})\cong \cZ (S_3)\boxtimes \ol{\cZ (S_3)}$, where $\overline{S_3}$ has symmetry operators that are complex conjugated compared to those for $S_3$ (the fusion categories themselves are equivalent). We denote the anyons by $\ell_k$ for $k\in\{L,R\}$ and
\be
\ell\in\{1,\,P,\,E,\,\,a,\,a^{\omega},\,a^{\ol{\omega}},\,b,\,b^-\,\}
\ee
where we label the anyons in the conjugacy classes $[a]$ and $[b]$ simply by the representatives $a$ and $b$. 
$\omega$ and $-$ indicate the irreducible representations of the centralizers of $a$ and $b$, which are $\Z_3$ and $\Z_2$ respectively. Note that the anyons $a^\omega$ and $a^{\ol{\omega
}}$ are time-reversal partners, while  all others are time-reversal invariant.
See \cite{Bhardwaj:2023idu,Bhardwaj:2023fca} for instances of our conventions.

We start with the symmetry boundary given by the factorized Lagrangian algebras
\be\ba \label{eq:LS3strong}
&\cL_{S_3 \times \ol{S_3}} 
= \cL_{\Vec_{S_3}}^{\text{strong}} \\
=&\; (1\oplus P\oplus 2E)_\p  \otimes (1\oplus P \oplus 2 E)_\a  \\
=&\; 1\oplus P_\p  \oplus P_\a  \oplus P_\p  P_\a  \oplus 2 E_\p  P_\a  \oplus 2 P_\p   E_\a   \\
&\phantom{=} \oplus  2 E_\p  \oplus 2 E_\a  \oplus 4 E_\p  E_\a  
\ea\ee
\be\ba \label{eq:LRepS3strong}
&\cL_{\Rep (S_3)  \times \ol{\Rep (S_3)} } =\cL_{\Rep({S_3})}^{\text{strong}} \\
=&\;(1\oplus a\oplus b)_\p  \otimes (1\oplus a\oplus b)_\a  \\
=&\; 1\oplus a_\p  \oplus a_\a  \oplus a_\p  a_\a   \oplus b_\p  \oplus b_\a  \oplus b_\p  b_\a  \\
&\phantom{=}\oplus a_\p  b_\a  \oplus b_\p  a_\a  \,.
\ea\ee
There are four Lagrangian algebras that satisfy the hermiticity and positivity constraints and do not factorize:\footnote{Note that this does not  assume that the anyons are bosons in the $\mathcal{Z}(\mathcal{S}_3)$ and $\mathcal{Z}(\overline{\mathcal{S}_3})$ individually.}
\be\ba \label{eq:S3_Lags}
\cL_{1}=&\; 1\oplus P_\p  P_\a  \oplus E_\p  E_\a  \oplus   a_\p  a_\a  \oplus a_\p ^\omega a_\a ^{\overline{\omega}} \oplus  a_\p ^{\overline{\omega}} a_\a ^{\omega}  \\
&\phantom{=}\oplus b_\p  b_\a  \oplus b_\p ^- b_\a ^- \,,\cr
\cL_{2} =&\; 1\oplus P_\p   \oplus P_\a  \oplus P_\p  P_\a  \oplus 2 a_\p  a_\a  \oplus 2 a_\p ^\omega a_\a ^{\overline{\omega}} \oplus 2 a_\p ^{\overline{\omega}} a_\a ^{\omega}  \\
&\phantom{=}\oplus 2 E_\p  E_\a \,, \cr 
\cL_{3}=&\; 1\oplus P_\p  P_\a  \oplus E_\p  \oplus E_\a  \oplus 2 E_\p  E_\a  \oplus E_\p  P_\a  \oplus P_\p  E_\a  \\
&\phantom{=}\oplus  b_\p  b_\a  \oplus b_\p ^- b_\a ^- \,, \cr 
\cL_{4} =&\; 1\oplus a_\p  \oplus a_\a  \oplus 2 a_\p  a_\a  \oplus P_\p  P_\a  \oplus a_\p  P_\a  \oplus P_\p  a_\a  \\
&\phantom{=}\oplus b_\p  b_\a \oplus b_\p ^- b_\a ^-\,.
\ea\ee
For the lattice model realization, note that these algebras can be characterized uniquely in terms of subgroups $\wt{F}$ of $\wt{G}=S_3\times S_3$ with trivial 2-cocycles
\be \label{eq:Ftildes_S3}
\ba
\wt{F}_1 &= S_3^{\diag} \,,&\quad  
\wt{F}_2 &= \Z_3^{\diag} \,,\\
\wt{F}_3 &= \Z_2^{\diag} \,, &\quad 
\wt{F}_4&= (\Z_3 \times \Z_3 )\rtimes \Z_2  \,.
\ea
\ee

\vspace{2mm}
\noindent{\bf Lattice model setup.}
We will use the conventions for $S_3$ and $\Rep(S_3)$ of \cite[App. B]{Bhardwaj:2024kvy} (see also \cite{Albert:2021vts,Tantivasadakarn:2021usi, Delcamp:2023kew,Fechisin:2023dkj,Bhardwaj:2024wlr}).
Following the general approach of Sec. \ref{SubSec:G Chain}, we take the Hilbert space to be:
\be
\cH = \bigotimes_j \mathbb{C}_j^{6} \,,
\ee
on which we define operators $L^g_j$ and $R^g_j$ implementing left and right multiplication by $g\in S_3$.

The on-site diagonal representation operators take the form \cite{Bhardwaj:2024kvy}
\begin{align} \label{eq:ZS3}
    Z^{1}&=\mathbb I\,, \nn\\
    Z^{P}&=\ketbra{1}+\ketbra{a}+\ketbra{a^2}\\
    &-\ketbra{b} 
     -\ketbra{ab}-\ketbra{a^2b}\,, \nn\\
    Z^{E}_{11}&=\ketbra{1}+\omega\ketbra{a}+\omega^2\ketbra{a^2}\,, \nn\\ 
    Z^{E}_{12}&=\ketbra{b}+\omega\ketbra{ab}+\omega^2\ketbra{a^2b}\,, \nn\\ 
    Z^{E}_{21}&=\ketbra{b}+\omega^2\ketbra{ab}+\omega\ketbra{a^2b}\,, \nn\\ 
    Z^{E}_{22}&=\ketbra{1}+\omega^2\ketbra{a}+\omega\ketbra{a^2}\,, \hspace{-5cm}
\end{align}
from which one sees that $(Z^P)^\dagger=Z^P$.

We will use this lattice setup to realize density matrices for mixed phases with strong $S_3$ symmetry in Sec.~\ref{sec:strongS3} and non-invertible $\Rep(S_3)$ symmetry in Sec.~\ref{sec:strongRepS3}.

\subsection{Mixed State Phases for $S_3$} \label{sec:strongS3}
Let us start with strong $S_3$ symmetry. From the SymTFT, we can determine the possible strong and weak spontaneous symmetry breaking phases.  

\subsubsection{SymTFT derivation of SWSSB $S_3$ phases}
\vspace{2mm}
\noindent {\bf $\bm{S_3}$ SWSSB.}
Consider the sandwich $( \cL_{\Vec_{S_3}}^{\text{strong}} , \cL_1)$. The order parameters come from anyons ending on both SymTFT boundaries: 
\vspace{-2mm}
\be\begin{split} \label{eq:SymTFT_S3SW}
\begin{tikzpicture}
\node at (2,-1.5) {$\Longrightarrow \ S_3$ SWSSB};
\draw [thick] (0,-1) -- (0,1) ;
\draw [thick] (4,-1) -- (4,1) ;
\draw [thick] (0, 0.25) -- (4, 0.25) ;
\draw [thick] (4, -0.5) -- (0, -0.5) ;
\node[above] at (2,0.25) {$\bm{P_\p  P_\a}$} ;
\node[above] at (2,-0.5) {$\bm{4 E_\p  E_\a }$} ;
\node[above] at (0,1) {$\cL_{\Vec_{S_3}}^{\text{strong}}$}; 
\node[above] at (4,1) {$\cL_1$}; 
\draw [black,fill=black] (0,0.25) ellipse (0.05 and 0.05);
\draw [black,fill=black] (4,0.25) ellipse (0.05 and 0.05);
\draw [black,fill=black] (4,-0.5) ellipse (0.05 and 0.05);
\draw [black,fill=black] (0,-0.5) ellipse (0.05 and 0.05);
\end{tikzpicture}
\end{split}\ee
There are 6 order parameters ($1$, $P_\p  P_\a $ and the four components of $E_\p  E_\a $ -- we should really draw four lines with $E_\p  E_\a $ label ending on the symmetry boundary). The order parameters are for the strong symmetry that arises from projecting $a_k$ and $b_k$ to the symmetry boundary. The weak diagonal $S_3$ symmetry, whose order parameters would be $P_k$ and $E_k$, is instead unbroken, i.e. we have an $S_3$ SWSSB. 

Apart from the SWSSB for the full $S_3$ symmetry we have three other gapped phases. 

\vspace{2mm}
\noindent {\bf $\bm{\Z_2}$ SSB, $\bm{\Z_3}$ SWSSB.}
The sandwich $( \cL_{\Vec_{S_3}}^{\text{strong}} , \cL_2)$ gives 
\vspace{-2mm}
\be\begin{split} \label{eq:Z3SWSymTFT}
\begin{tikzpicture}
\node at (2,-1.5) {$\Longrightarrow \ \Z_2$ SSB, $\Z_3$ SWSSB};
\begin{scope}[shift={(0,0)}]
\draw [thick] (0,-1) -- (0,1) ;
\draw [thick] (4,-1) -- (4,1) ;
\draw [thick] (0, 0.25) -- (4, 0.25) ;
\draw [thick] (4, -0.5) -- (0, -0.5) ;
\node[above] at (2,0.25) {$P_\p , P_\a , \bm{P_\p  P_\a }$} ;
\node[above] at (2,-0.5) {$\bm{8 E_\p  E_\a }$} ;
\node[above] at (0,1) {$\cL_{\Vec_{S_3}}^{\text{strong}}$}; 
\node[above] at (4,1) {$\cL_2$}; 
\draw [black,fill=black] (0,0.25) ellipse (0.05 and 0.05);
\draw [black,fill=black] (4,0.25) ellipse (0.05 and 0.05);
\draw [black,fill=black] (4,-0.5) ellipse (0.05 and 0.05);
\draw [black,fill=black] (0,-0.5) ellipse (0.05 and 0.05);
\end{scope}
\end{tikzpicture}
\end{split}\ee
The lines $P_\p P_\a, P_\p , P_\a  $ ending give order parameters for both strong and weak $\Z_2$ symmetry arising from $b_\p b_\a $ which braids non-trivially with $P_\p $ and $P_\a $. However, only the identity and the $T$-symmetric operator  $P_LP_R$ can give rise to a density matrix. 

On the other hand $E_\p  E_\a $ has two ends on the symmetry boundary and four on the physical boundary (but neither $E_\p $ or $E_\a $ are order parameters). This breaks the $\Z_3$ strong symmetry but leaves the $\Z_3$ weak symmetry intact.

\vspace{2mm}
\noindent {\bf $\bm{\Z_3}$ SSB, $\bm{\Z_2}$ SWSSB.} Next consider $( \cL_{\Vec_{S_3}}^{\text{strong}} , \cL_3)$
\vspace{-2mm}
\be\begin{split} \label{eq:Z2SWSymTFT}
\begin{tikzpicture}
\node at (2,-1.5) {$\Longrightarrow \ \Z_3$ SSB, $\Z_2$ SWSSB};
\begin{scope}[shift={(0,0)}]
\draw [thick] (0,-1) -- (0,1) ;
\draw [thick] (4,-1) -- (4,1) ;
\draw [thick] (0, 0.25) -- (4, 0.25) ;
\draw [thick] (4, -0.5) -- (0, -0.5) ;
\node[above] at (2,0.25) {$2E_\p , 2E_\a , \bm{8E_\p  E_\a} $} ;
\node[above] at (2,-0.5) {$\bm{P_\p  P_\a }$} ;
\node[above] at (0,1) {$\cL_{\Vec_{S_3}}^{\text{strong}}$}; 
\node[above] at (4,1) {$\cL_3$}; 
\draw [black,fill=black] (0,0.25) ellipse (0.05 and 0.05);
\draw [black,fill=black] (4,0.25) ellipse (0.05 and 0.05);
\draw [black,fill=black] (4,-0.5) ellipse (0.05 and 0.05);
\draw [black,fill=black] (0,-0.5) ellipse (0.05 and 0.05);
\end{scope}
\end{tikzpicture}
\end{split}\ee
The order parameters from $E_\p , E_\a , E_\p E_\a $ imply that the $\Z_3$ is fully broken. $P_LP_R$ indicates breaking of the $\Z_2$ strong symmetry, but the $\Z_2$ weak symmetry is preserved. Again only the $T$-symmetric charged operators coming from $P_LP_R$  and $8 E_L E_R$ can give rise to Choi states that are associated to density matrices.

\vspace{2mm}
\noindent {\bf $\bm{\Z_2}$ SWSSB.}
Finally consider $( \cL_{\Vec_{S_3}}^{\text{strong}} , \cL_4)$
\vspace{-2mm}
\be\begin{split} \label{eq:Z3Z3Z2SymTFT}
\begin{tikzpicture}
\node at (2,-1) {$\Longrightarrow \ \Z_2$ SWSSB};
\draw [thick] (0,-0.5) -- (0,1) ;
\draw [thick] (4,-0.5) -- (4,1) ;
\draw [thick] (0, 0.25) -- (4, 0.25) ;
\node[above] at (2,0.25) {$\bm{P_\p  P_\a }$} ;
\node[above] at (0,1) {$\cL_{\Vec_{S_3}}^{\text{strong}}$}; 
\node[above] at (4,1) {$\cL_4$}; 
\draw [black,fill=black] (0,0.25) ellipse (0.05 and 0.05);
\draw [black,fill=black] (4,0.25) ellipse (0.05 and 0.05);
\end{tikzpicture}
\end{split}\ee
with order parameter $\cO_{P_\p P_\a }$ which indicates that the strong $\Z_2^\strong\subset S_3^\strong$ is SSB'ed but the weak $\Z_2^\weak$ is acting trivially. This corresponds to a $\Z_2$ SWSSB. Furthermore, the strong $\Z_3$ symmetry is preserved.

\subsubsection{Density matrices for Mixed $S_3$ Phases}
We take the $S_3$ symmetry to be generated by
\be
    U^g=\prod_jR^g_j\,.
\ee
And write the projectors as \cite{Bhardwaj:2024kvy}
\begin{align} \label{eq:VecS3_projectors}
    P^{(S_3)}_{j+\frac{1}{2}}&=\mathbb I_{j,j+1}\,, \nn\\
    P^{(\Z_3)}_{j+\frac{1}{2}}&=\frac{1}{2}\lbb\bbI_j\bbI_{j+1}+Z^P_jZ^P_{j+1}\rbb\,, \nn\\
    P^{(\Z_2)}_{j+\frac{1}{2}}&=\frac{1}{3}\Bigl[\mathbb I_{j,j+1}+\sum_{IJ} \left(Z^{E}_{j}\cdot (Z^E_{j+1})^{\dagger}\right)_{IJ}\Bigr]\,, \nn\\
    P^{(1)}_{j+\frac{1}{2}}&= \sum_{\Gamma}\frac{{\dim}(\Gamma)}{|G|}{\rm Tr}\left(Z^{\Gamma}_{j}\cdot (Z^{\Gamma}_{j+1})^{\dagger}\right)\,.
\end{align}

The doubled group is in this case
\be
    \wt{G}=(S_3)_L\times (S_3)_R\,,
\ee
with corresponding Hilbert space
\be
\wt{\cH} = \bigotimes_j (\mathbb{C}^{6}_L\times \mathbb{C}^{6}_R)_j \,.
\ee
Its subgroups given in equation \eqref{eq:Ftildes_S3} each correspond to a Lagrangian algebra whose condensation on the SymTFT physical boundary gives rise to a SWSSB phase as discussed above, for which we will now provide explicit density matrices, with strong $S_3$ symmetry, i.e. satisfying
\be\ba
 U^a\rho^\pm&=\rho^\pm=\rho^\pm U^a\,,\\
    U^b\rho^\pm&=\pm\rho^\pm=\rho^\pm U^b\,,
\ea\ee
where $\rho^+$ and $\rho^-$ carry respectively the trivial and sign 1d irreps of $S_3$.

\vspace{2mm}
\noindent {\bf $S_3$ SWSSB.}
The Lagrangian algebra is characterized by the subgroup $\wt{F}_1=S_3^{\text{diag}}$.
We write the projector (normalized to have unit trace) onto $S_3$-invariant states as
\be\ba
    P_\sym^\pm&=\frac{1}{6^N}(\bbI+U^a+U^{a^2}\pm U^b\pm U^{ab}\pm U^{a^2b})\,,\\
\ea\ee
where $P_\sym^-$ carries the non-trivial 1d irrep $P$.
The Hamiltonian in the doubled Hilbert space is
\be \label{eq:H_S3diag}
 \wt{H}_{(S_3^\diag)}=-\frac{1}{6}\sum_{j}\sum_{\wt{f}\in S_3^\diag} \wt{P}^{(S_3^\diag)}_{j-\frac{1}{2}}\,\wt{L}^{\wt{f}}_{j}\,\wt{P}^{(S_3^\diag)}_{j+\frac{1}{2}}\,.
\ee
The projector contains a sum over linear combinations of irreps in which $S_3^\diag$ is trivial, corresponding to the anyons ending on both boundaries of the SymTFT
(\ref{eq:SymTFT_S3SW}): in particular it is diagonal in $LR$. Following the general procedure of sec. \ref{SubSec:G Chain}, we can then write $\wt{\rho}_{S_3^{\diag}}$ as products of the terms in \eqref{eq:H_S3diag} multiplied by $P_\sym^{\pm,L}P_\sym^{\pm,R}$. Notice that all non-trivial terms in \eqref{eq:H_S3diag} are diagonal in $LR$, therefore, upon tracing out $\cH_R$, they evaluate to zero and we are left with
\be
    \rho_{S_3^{\diag}}^\pm=P_\sym^\pm \,.
\ee
$\rho_{S_3^{\diag}}^\pm$ is a physical density matrix with $S_3$ strong symmetry. Denote it now as $\rho$, it has the following properties:
\be
\ba
   \Tr(\rho \,Z^\Gamma_{IJ,i}(Z^\Gamma_{IJ,j})^\dagger) &=0\cr 
 \Tr(\rho \, Z^\Gamma_{IJ,i}(Z^\Gamma_{IJ,j})^\dagger\,\rho\, Z^\Gamma_{IJ,j}(Z^\Gamma_{IJ,i})^\dagger)&\neq 0\,,
\ea
\ee
for all $(Z^\Gamma_{IJ})\in\{Z^P,\,Z^E_{11},Z^E_{12},Z^E_{21},Z^E_{22}\}$. 
The first line implies that the weak $S_3$ symmetry is preserved, while the second line reveals spontaneous symmetry breaking of the strong $S_3$ symmetry, consistent with the SymTFT prediction \eqref{eq:SymTFT_S3SW}.

\vspace{2mm}
\noindent {\bf $\Z_2$ SSB, $\Z_3$ SWSSB.}
For the subgroup
\be
    \wt{F}=\wt{F}_2=\Z_3^{\text{diag}}=\{1,\,a_La_R,\,a_L^2a_R^2\}
\ee
we write the doubled-space Hamiltonian as
\be \label{eq:H_Z3diag}
 \wt{H}_{(\Z_3^\diag)}=-\frac{1}{3}\sum_{j}\sum_{\wt{f}\in \Z_3^\diag} \wt{P}^{(\Z_3^\diag)}_{j-\frac{1}{2}}\,\wt{L}^{\wt{f}}_{j}\,\wt{P}^{(\Z_3^\diag)}_{j+\frac{1}{2}}\,,
\ee
where $\wt{P}^{(\Z_3^\diag)}$ can be expressed as linear combinations of irreps in which $\Z_3^\diag$ is trivial, corresponding to the anyons ending on both SymTFT boundaries in eq. (\ref{eq:Z3SWSymTFT}). We can then write $\wt{\rho}_{\Z_3^{\diag}}$ as products of the terms in \eqref{eq:H_Z3diag} multiplied by $P_\sym^{\pm,L}P_\sym^{\pm,R}$ and trace out $\cH_R$. Note that, the Lagrangian algebra $\cL_2$ in \eqref{eq:S3_Lags} on the physical boundary of the SymTFT, contains $1\oplus P_L$ (not diagonal in $LR$) which survives the trace over $R$ and gives rise to non-trivial projectors $P^{(\Z_3)}_{j+\frac{1}{2}}$:
\be
    \rho_{\Z_3^{\diag}}^\pm=\frac{P_\sym^\pm}{2}\prod_j\lbb\bI+Z^P_jZ^P_{j+1}\rbb\,.
\ee
$\rho_{\Z_3^{\diag}}^\pm$ has strong $S_3$ symmetry and
the following properties (using $\rho$ as shorthand for $\rho_{\Z_3^{\diag}}^\pm$): 
\be
\ba
  \Tr(\rho \,Z^P_i Z^P_j) &\neq 0\cr 
  \Tr(\rho \,Z^E_{IJ,i}(Z^E_{IJ,j})^\dagger)&=0 \cr  
 \Tr(\rho \, Z^\Gamma_{IJ,i}(Z^\Gamma_{IJ,j})^\dagger\,\rho\, Z^\Gamma_{IJ,j}(Z^\Gamma_{IJ,i})^\dagger) &\neq 0\,,
\ea
\ee
for $(Z^\Gamma_{IJ})
\in\{Z^P,\,Z^E_{11},Z^E_{12},Z^E_{21},Z^E_{22}\}$. 
The first line implies that the weak $\Z_2$ symmetry is spontaneously broken, while the weak $\Z_3$ is preserved. The second line reveals spontaneous symmetry breaking of the strong $S_3$ symmetry. This is therefore a $\Z_2$ SSB, $\Z_3$ SWSSB phase, consistent with the SymTFT \eqref{eq:Z3SWSymTFT}.

\vspace{2mm}
\noindent {\bf $\Z_3$ SSB, $\Z_2$ SWSSB.}
Completely analogously to the previous case, for the subgroup
\be
    \wt{F}=\wt{F}_3=\Z_2^{\text{diag}}=\{1,\,b_Lb_R\}
\ee
we write the doubled-space Hamiltonian as
\be \label{eq:H_Z2diag}
 \wt{H}_{(\Z_2^\diag)}=-\frac{1}{3}\sum_{j}\sum_{\wt{f}\in \Z_2^\diag} \wt{P}^{(\Z_2^\diag)}_{j-\frac{1}{2}}\,\wt{L}^{\wt{f}}_{j}\,\wt{P}^{(\Z_2^\diag)}_{j+\frac{1}{2}}\,,
\ee
where $\wt{P}^{(\Z_2^\diag)}$ can be expressed as linear combinations of irreps in which $\Z_2^\diag$ is trivial, corresponding to the anyons ending on both SymTFT boundaries in eq. (\ref{eq:Z2SWSymTFT}).
We can then write $\wt{\rho}_{\Z_2^{\diag}}$ as products of the terms in \eqref{eq:H_Z2diag} multiplied by $P_\sym^{\pm,L}P_\sym^{\pm,R}$ and trace out $\cH_R$. The Lagrangian algebra $\cL_3$ in \eqref{eq:S3_Lags} on the physical boundary of the SymTFT, contains $1\oplus E_L$ (not diagonal in $LR$) which survives the trace over $R$ and gives rise to non-trivial projectors $P^{(\Z_2)}_{j+\frac{1}{2}}$:
\be
    \rho_{\Z_2^{\diag}}^\pm=\frac{P_\sym^\pm}{3}\prod_j\Bigl[\mathbb I_{j,j+1}+\sum_{IJ} \left(Z^{E}_{j}\cdot (Z^E_{j+1})^{\dagger}\right)_{IJ}\Bigr]\,.
\ee
$\rho_{\Z_2^{\diag}}^\pm$ has strong $S_3$ symmetry and
the following properties (denoting it as $\rho$):
\be
\ba
  \Tr(\rho \,Z^P_i Z^P_j) &= 0\cr 
  \Tr(\rho \,Z^E_{IJ,i}(Z^E_{IJ,j})^\dagger)&\neq 0\cr 
 \Tr(\rho \, Z^\Gamma_{IJ,i}(Z^\Gamma_{IJ,j})^\dagger\,\rho\, Z^\Gamma_{IJ,j}(Z^\Gamma_{IJ,i})^\dagger) &\neq 0\,,
\ea
\ee
for $(Z^\Gamma_{IJ})\in\{Z^P,\,Z^E_{11},Z^E_{12},Z^E_{21},Z^E_{22}\}$.
The first line implies that this time the weak $\Z_2$ symmetry is preserved, while the weak $\Z_3$ is spontaneously broken, as is the strong $S_3$ (second line). This is therefore a $\Z_3$ SSB, $\Z_2$ SWSSB phase, consistent with the SymTFT \eqref{eq:Z2SWSymTFT}.

\vspace{2mm}
\noindent {\bf $\Z_2$ SWSSB.}
Finally, for the subgroup
\be
    \wt{F}=\wt{F}_4= (\Z_3 \times \Z_3 )\rtimes \Z_2
\ee
we write the doubled-space Hamiltonian as
\be \label{eq:H_Z3Z3Z2}
 \wt{H}_{\wt{F}_4}=-\frac{1}{3}\sum_{j}\sum_{\wt{f}\in \wt{F}_4} \wt{P}^{(\wt{F}_4)}_{j-\frac{1}{2}}\,\wt{L}^{\wt{f}}_{j}\,\wt{P}^{(\wt{F}_4)}_{j+\frac{1}{2}}\,,
\ee
where $\wt{P}^{(\wt{F}_4)}$ can be expressed as linear combinations of irreps in which $\wt{F}_4$ is trivial, corresponding to the anyons ending on both SymTFT boundaries in eq. (\ref{eq:Z3Z3Z2SymTFT}): in this case, the only non-trivial anyon is the diagonal $P_LP_R$ under which the full $\Z_3^L\times\Z_3^R$ is uncharged. This means that the the ground states of \eqref{eq:H_Z3Z3Z2} are automatically $\Z_3^L\times\Z_3^R$ invariant and, for $\wt{\rho}_{\wt{F}_4}$, we need to multiply the product of its terms by $\frac{1}{4}P_{\sym,\Z_2}^{\pm L} P_{\sym,\Z_2}^{\pm R}$, where 
\be\ba
    P_{\sym,\Z_2}^\pm&=\frac{1}{2^N}(\bbI\pm U^b)\,,\\
\ea\ee
to take combinations invariant under strong $\Z_2$. Note that $\wt{F}_4$ now includes disordering terms $L^{\wt{f}}$ for non-diagonal group elements: specifically $L^{a_L}$ and $L^{a^2_L}$ survive the trace over $\cH_R$ and we obtain:
\be
    \rho_{\wt{F}_4}^\pm=\frac{P_{\sym,\Z_2}^\pm}{2}\prod_j\lbb\bI+L_j^a+L_j^{a^2}\rbb\,.
\ee
$\rho_{\wt{F}_4}^\pm$ has strong $S_3$ symmetry and the following properties (denoting it as $\rho$)
\begin{align}
   \Tr(\rho \,Z^\Gamma_{IJ,i}(Z^\Gamma_{IJ,j})^\dagger)=0\,, \\  
   \nn 
 \Tr(\rho \, Z^E_{IJ,i}(Z^E_{IJ,j})^\dagger\,\rho\, Z^E_{IJ,j}(Z^E_{IJ,i})^\dagger)= 0\,,\nn \\
  \Tr(\rho \, Z^P_iZ^P_j\,\rho\, Z^P_jZ^P_i)\neq 0\,.
\end{align}
for $(Z^\Gamma_{IJ})\in\{Z^P,\,Z^E_{11},Z^E_{12},Z^E_{21},Z^E_{22}\}$. These reveal that this is a $\Z_2$ SWSSB, in which the strong $\Z_2$ symmetry is spontaneously broken to weak while the $\Z_3$ strong symmetry is fully preserved.

\subsection{Mixed State Phases for $\Rep (S_3)$} \label{sec:strongRepS3}
We now change the symmetry boundary of the SymTFT $\cZ(S_3^L\times S_3^R)$ of Sec.~\ref{sec:SymTFT_S3} to represent the non-invertible strong $\Rep (S_3)$ symmetry. To study mixed-state phases with strong $\Rep (S_3)$ symmetry, we must pick the symmetry Lagrangian algebra to be \ref{eq:LRepS3strong}. The mixed-state phases we will derive will have strong $\Rep(S_3)$ symmetry, which has three elements
\be
    \Rep(S_3)=\{1,\,P,\,E\}\,.
\ee
The character table is
\be \label{eq:S3_chars}
\begin{array}{|r|ccc|}
    \hline
     & [1]  & [a] & [b] \\
    \hline
    1:& 1 & 1 & 1\\
    P:& 1 & 1 & -1\\
    E:& 2 & -1 & 0\\
     \hline
\end{array}
\ee
and the irreps fuse according to the following tensor products:
\be
    P\otimes P=1\,,\qquad E\otimes E=1\oplus P\oplus E\,,
\ee
the latter being a non-invertible fusion rule.

\subsubsection{SymTFT Derivation of $\Rep(S_3)$ Phases}
We shall now determine the strong-to-weak symmetry breaking patterns for $\Rep(S_3)$ from the SymTFT.

\vspace{2mm}
\noindent {\bf $\bm{\Rep(S_3)}$ SWSSB.}
Let us start with the sandwich $(\cL_{\Rep({S_3})}^{\text{strong}} , \cL_1) $. The charged local operators come from anyons ending on both SymTFT boundaries: 
\vspace{-2mm}
\be \label{SandoF1}
\begin{tikzpicture}
\node at (2,-1.5) {$\Longrightarrow \ \Rep(S_3)$ SWSSB};
\draw [thick] (0,-1) -- (0,1) ;
\draw [thick] (4,-1) -- (4,1) ;
\draw [thick] (0, 0.25) -- (4, 0.25) ;
\draw [thick] (4, -0.5) -- (0, -0.5) ;
\node[above] at (2,0.25) {$\bm{a_\p  a_\a }$} ;
\node[above] at (2,-0.5) {$\bm{b_\p  b_\a }$} ;
\node[above] at (0,1) {$\cL_{\Rep({S_3})}^{\text{strong}}$}; 
\node[above] at (4,1) {$\cL_1$}; 
\draw [black,fill=black] (0,0.25) ellipse (0.05 and 0.05);
\draw [black,fill=black] (4,0.25) ellipse (0.05 and 0.05);
\draw [black,fill=black] (4,-0.5) ellipse (0.05 and 0.05);
\draw [black,fill=black] (0,-0.5) ellipse (0.05 and 0.05);
\end{tikzpicture}
\ee
After compactifying the SymTFT interval, we will have charged local operators $\cO_{a_\p a_\a }$ and $\cO_{b_\p b_\a }$ which carry non-trivial charge under both $\Rep(S_3)_\p $ and $\Rep(S_3)_\a $ generated by $P_k$ and $E_k$, for $k\in\{L,R\}$ however we will see that the weak $\Rep(S_3)$ symmetry is preserved since $a_La_R$ and $b_Lb_R$ are diagonal. 

Recall from eq. \eqref{eq:Ftildes_S3} that the subgroup of $S_3\times S_3$ associated to $\cL_1$ is $\wt{F}=S_3^\diag$: this means that on the physical boundary the fluxes in $\cL_1$ are in conjugacy classes of $S_3^\diag$. The charged local operators, $\cO_{a_La_R}$ and $\cO_{b_Lb_R}$ correspond to
\be\ba
    [a_La_R]&=\{a_La_R,\,a_L^2a^2_R\}\,,\quad \text{and}\\
    [b_Lb_R]&=\{b_Lb_R,\,a_Lb_La_Rb_R,\,a_L^2b_La^2_Rb_R\}
\ea\ee
respectively. The fusion algebra of the charged local operators can then be computed from $S_3^\diag$ group multiplication to be 
\be\ba
    \cO_{a_La_R}\times\cO_{b_Lb_R}&=2\,\cO_{b_Lb_R}\,,\\
    \cO_{a_La_R}\times\cO_{a_La_R}&=2+\cO_{a_La_R}\,,\\
    \cO_{b_Lb_R}\times\cO_{b_Lb_R}&=3+3\,\cO_{a_La_R}\,,
\ea\ee
which is diagonalized by the linear combinations:
\be\ba
    v_1&=(1+\cO_{a_La_R}+\cO_{b_Lb_R})\,,\\
    v_2&=(1+\cO_{a_La_R}-\cO_{b_Lb_R})\,,\\
    v_3&=(2-\cO_{a_La_R})\,.
\ea\ee
The strong $\Rep(S_3)$ symmetry action on the local operators can be computed from the $S$-matrix in the SymTFT bulk, or equivalently from the character table \eqref{eq:S3_chars} and is:
\be\ba
 P_{L,R}:(\cO_{a_La_R}\,, \cO_{b_Lb_R})&\longrightarrow  (\cO_{a_La_R}\,, -\cO_{b_Lb_R})\,,\\
 E_{L,R}:(\cO_{a_La_R}\,, \cO_{b_Lb_R})&\longrightarrow  (-\cO_{a_La_R}\,, 0)\,.
\ea\ee

On the idempotents it is therefore:
\be\ba
 P_{L,R}:(v_1,\,v_2,\,v_3)&\longrightarrow  (v_2,\,v_1,\,v_3)\,,\\
 E_{L,R}:(v_1,\,v_2,\,v_3)&\longrightarrow  (v_3,\,v_3,\,v_1+v_2+v_3)\,.
\ea\ee
The strong $\Rep(S_3)$ symmetry is thus SSB'ed.

For the diagonal symmetry generators we instead have
\be\ba
 P_{L}P_{R}:(\cO_{a_La_R}\,, \cO_{b_Lb_R})&\longrightarrow  (\cO_{a_La_R}\,, \cO_{b_Lb_R})\,,\\
 E_{L}E_{R}:(\cO_{a_La_R}\,, \cO_{b_Lb_R})&\longrightarrow  (\cO_{a_La_R}\,, 0)\,,
\ea\ee
Each charged operator will thus give rise to a $\Rep(S_3)$ SWSSB density matrix. Note that the eigenvalue of $\cO_{b_Lb_R}$ under $E_LE_R$ will be zero: this is possible for non-invertible symmetries.

\vspace{2mm}
\noindent {\bf $\bm{\Rep(S_3)/\Z_2}$ SWSSB.}
Let us now consider the sandwich $(\cL_{\Rep({S_3})}^{\text{strong}} , \cL_2)$. We will see that in this phase the non-invertible $E$ symmetry is SWSSB'ed while the $\Z_2$ symmetry $P$ is fully preserved.
\be \label{SandoF2}
\begin{tikzpicture}
\node at (2,-1) {$\Longrightarrow \ (\Rep(S_3)/\Z_2)$ SWSSB};
\draw [thick] (0,-0.5) -- (0,1) ;
\draw [thick] (4,-0.5) -- (4,1) ;
\draw [thick] (0, 0.25) -- (4, 0.25) ;
\node[above] at (2,0.25) {$\bm{2a_\p  a_\a }$} ;
\node[above] at (0,1) {$ \cL_{\Rep({S_3})}^{\text{strong}}$}; 
\node[above] at (4,1) {$\cL_2$}; 
\draw [black,fill=black] (0,0.25) ellipse (0.05 and 0.05);
\draw [black,fill=black] (4,0.25) ellipse (0.05 and 0.05);
\end{tikzpicture}
\ee

Here, the factor of two means that the anyon $a_La_R$ has two ends on the physical boundary and thus splits into two local operators $\cO_{a_La_R}$ and $\cO_{a^2_La^2_R}$. Indeed, these are different conjugacy classes of $\Z_3^\diag$, which, from eq. \eqref{eq:Ftildes_S3}, is the subgroup corresponding to $\cL_2$ and labeling the fluxes on the physical boundary.

The fusion algebra of the charged local operators is simply given by $\Z_3^\diag$ group multiplication:
\be\ba
    \cO_{a_La_R}\times\cO_{a_La_R}&=\cO_{a^2_La^2_R}\,,\\
    \cO_{a_La_R}\times\cO_{a^2_La^2_R}&=1\,.
\ea\ee
and is diagonalized by:
\be\ba
    v_1&=(1+\cO_{a_La_R}+\cO_{a^2_La^2_R})\,,\\
    v_2&=(1+\omega\, \cO_{a_La_R}+\ol{\omega}\,\cO_{a^2_La^2_R})\,,\\
    v_3&=(1+\ol{\omega}\,\cO_{a_La_R}+\omega\,\cO_{a^2_La^2_R})\,,
\ea\ee
where $\omega+\ol{\omega}=-1$.
The strong $\Rep(S_3)$ symmetry action on the local operators can again be computed from \eqref{eq:S3_chars}:
\be\ba
 P_{L,R}:(\cO_{a_La_R}\,, \cO_{a^2_La^2_R})&\longrightarrow  (\cO_{a_La_R}\,, \cO_{a^2_La^2_R})\,,\\
 E_{L,R}:(\cO_{a_La_R}\,, \cO_{a^2_La^2_R})&\longrightarrow  (-\cO_{a_La_R}\,, -\cO_{a^2_La^2_R})\,.
\ea\ee
On the vacua we therefore have
\be\ba
 P_{L,R}:(v_1,\,v_2,\,v_3)&\longrightarrow  (v_1,\,v_2,\,v_3)\,,\\
 E_{L,R}:(v_1,\,v_2,\,v_3)&\longrightarrow  (v_2+v_3,\,v_1+v_3,\,v_1+v_2)\,,
\ea\ee
From which we see that the strong $\Z_2$ symmetry generated by $P_L$ and $P_R$ is preserved, while the strong non-invertible $E_L,\,E_R$ is SSB'ed. The action of the weak $E_LE_R$ non-invertible  symmetry is trivial:
\be\ba
 E_{L}E_{R}:(\cO_{a_La_R}\,, \cO_{a^2_La^2_R})&\longrightarrow  (\cO_{a_La_R}\,, \cO_{a^2_La^2_R})\,.
\ea\ee
Each local operator will thus give rise to a ${\Rep(S_3)/\Z_2}$ SWSSB density matrix, in which the non-invertible strong $E$ symmetry is broken to weak, while the strong $\Z_2$ symmetry generated by $P$ is fully preserved.

\vspace{2mm}
\noindent {\bf $\bm{\Z_2}$ SWSSB.}
Consider the sandwich $(\cL_{\Rep({S_3})}^{\text{strong}} , \cL_3)$
\vspace{-2mm}
\be \label{Sandoz2}
\begin{tikzpicture}
\node at (2,-1) {$\Longrightarrow \ \Z_2$ SWSSB};
\draw [thick] (0,-0.5) -- (0,1) ;
\draw [thick] (4,-0.5) -- (4,1) ;
\draw [thick] (0, 0.25) -- (4, 0.25) ;
\node[above] at (2,0.25) {$\bm{b_\p  b_\a} $} ;
\node[above] at (0,1) {$ \cL_{\Rep({S_3})}^{\text{strong}}$}; 
\node[above] at (4,1) {$\cL_3$}; 
\draw [black,fill=black] (0,0.25) ellipse (0.05 and 0.05);
\draw [black,fill=black] (4,0.25) ellipse (0.05 and 0.05);
\end{tikzpicture}
\ee
Compactifying the SymTFT interval gives a local operator $\cO_{b_Lb_R}$ charged under the strong $\Z_2$ generated by $P_L,P_R$ but uncharged under the weak $P_LP_R$: this will therefore give a phase in which the non-invertible $E$ symmetry is preserved  while the $\Z_2$ symmetry is SWSSB'ed. 

The local operator $\cO_{b_Lb_R}$ corresponds to the generator of the subgroup $\Z_2^\diag$ characterizing $\cL_3$ in Eq. \eqref{eq:Ftildes_S3}, therefore:
\be
    \cO_{b_Lb_R}\times \cO_{b_Lb_R}=1
\ee
this is diagonalized by
\be\ba
    v_1=(1+\cO_{b_Lb_R})\,,\quad v_2=(1-\cO_{b_Lb_R})\,.
\ea\ee
The left and right symmetry generators act as
\be\ba
 P_{L,R}:\cO_{b_Lb_R},&\longrightarrow  -\cO_{b_Lb_R}\,,\\
 E_{L,R}:\cO_{b_Lb_R}&\longrightarrow  0\,.
\ea\ee
$P$ is SWSSB'ed.

\vspace{2mm}
\noindent {\bf $\bm{(\Rep(S_3)/\Z_2)}$ SSB, $\bm{\Z_2}$ SWSSB.} Finally, consider:
\be 
\begin{tikzpicture} \label{SandoF4}
\node at (2,-1.5) {$\Longrightarrow \ (\Rep(S_3)/\Z_2)$ SSB,  $\Z_2$ SWSSB};
\draw [thick] (0,-1) -- (0,1) ;
\draw [thick] (4,-1) -- (4,1) ;
\draw [thick] (0, 0.25) -- (4, 0.25) ;
\draw [thick] (4, -0.5) -- (0, -0.5) ;
\node[above] at (2,0.25) {$\bm{2a_\p  a_\a} ,\, a_\a,\,a_R$} ;
\node[above] at (2,-0.5) {$\bm{b_\p  b_\a} $} ;
\node[above] at (0,1) {$\cL_{\Rep({S_3})}^{\text{strong}}$}; 
\node[above] at (4,1) {$\cL_4$}; 
\draw [black,fill=black] (0,0.25) ellipse (0.05 and 0.05);
\draw [black,fill=black] (4,0.25) ellipse (0.05 and 0.05);
\draw [black,fill=black] (4,-0.5) ellipse (0.05 and 0.05);
\draw [black,fill=black] (0,-0.5) ellipse (0.05 and 0.05);
\end{tikzpicture}
\ee
This has genuine order parameters from $a_\p  a_\a,\,a_\p ,\, a_\a  $ anyons in the bulk which implies that both strong and weak $\Rep(S_3)/\Z_2$ symmetry is broken. On the other hand we only have the diagonal $b_\p  b_\a  $ anyon line, so the remaining $\Z_2$ is weakly broken.  

The group characterizing $\cL_4$ is $\wt{F}_4=(\Z_3\times\Z_3)\rtimes\Z_2$, generated by the group elements $a_L,a_R,b_Lb_R$.
On the physical boundary, the anyons in in \eqref{SandoF4} give rise to charged local operators $\cO_{g}$ corresponding a representative of each conjugacy class of $\wt{F}_4$:
\be\ba
[a_L]&=\{a_L,\,a^2_L\}\,,& [a_R]&=\{a_R,\,a^2_R\}\,,\\
[a_La_R]&=\{a_La_R,\,a_La^2_R\}\,,& [a_La^2_R]&=\{a_La^2_R,\,a^2La_R\}\,,\\
[b_Lb_R]&=b_Lb_R\,.
\ea\ee
Note that $\cO_{a_La_R}$ and $\cO_{a_La^2_R}$ come from the splitting of the anyon $a_La_R$ which has two ends on the physical boundary as indicated by the factor of 2 in the Lagrangian algebra.

The fusion algebra of the local operators follows from group multiplication and is diagonalized by:\footnote{Note that, like in the previous cases, there is an idempotent for each irrep of $\wt{F}$: the characters of each irrep provide the coefficients in the linear combinations of local operators.} 
%% v2 maybe recheck these
\begin{align}
    v_1&=(1+\cO_{a_L}+\cO_{a_R}+\cO_{a_La_R}+\cO_{a_La^2_R}+\cO_{b_Lb_R})\,,\nn\\
    v_2&=(1+\cO_{a_L}+\cO_{a_R}+\cO_{a_La_R}+\cO_{a_La^2_R}-\cO_{b_Lb_R})\,,\nn\\
    v_3&=(2+2\cO_{a_L}-\cO_{a_R}-\cO_{a_La_R}-\cO_{a_La^2_R})\,, \nn\\
     v_4&=(2-\cO_{a_L}+2\cO_{a_R}-\cO_{a_La_R}-\cO_{a_La^2_R})\,,\nn\\
     v_5&=(2-\cO_{a_L}-\cO_{a_R}+2\cO_{a_La_R}-\cO_{a_La^2_R})\,,\nn\\
     v_6&=(2-\cO_{a_L}-\cO_{a_R}-\cO_{a_La_R}+2\cO_{a_La^2_R})\,.
\end{align}
%The non-trivial symmetry action is: 
The symmetry operators act on the charged operators as dictated by the characters \eqref{eq:S3_chars}.
This is therefore a phase in which both strong and weak $E$ are SSB'ed, as can be inferred from the off-diagonal $a_L$ and $a_R$ anyons in \eqref{SandoF4} while $P$ is SWSSB'ed (indeed, there are no off-diagonal $b_L,b_R$ anyons).

Note that however only the $T$-symmetric charged operators arising from $2a_La_R$ and $b_Lb_R$ give rise to Choi states that correspond to density matrices, to which we now turn.

\subsubsection{Density Matrices for $\Rep (S_3)$ Mixed States}
\label{sec:RepS3Lat}
We can apply the general construction in Sec.~\ref{SubSec:Rep(G) Chain} to obtain density matrices for mixed-state
phases with strong $\Rep(S_3)$ symmetry, exhibiting various patterns of SSB such as SWSSB. 

We will use the lattice setup summarized at the end of Sec.~\ref{sec:SymTFT_S3}, with a local Hilbert space of dimension six, spanned by $\ket{g}$ for $g\in S_3$.
The diagonal operators $Z^P$ and $Z^E_{IJ}$ are written in equation \eqref{eq:ZS3}.

In this model, the $\Rep(S_3)$ symmetry generators are:
\be\ba
    \cS_P=\prod_jZ^P_j\,,\quad 
    \cS_E=Z^{E,\text{prod}}_{1,1}+Z^{E,\text{prod}}_{2,2}\,,
\ea\ee
where $Z^{E,\text{prod}}=\prod_j(Z^E)_j$ 
is the product of matrices of $Z^E_{IJ}$ blocks at each site.
The symmetry operator $\cS_\Gamma$ for $\Gamma\in\Rep(S_3)$ multiplies a basis state by the character of its `holonomy', which is the product along the lattice of all the group elements in the state. 
\be
    \cS_\Gamma\ket{g_1,g_2,...,g_N}=\chi_\Gamma(g_1g_2\cdots g_N)\ket{g_1,g_2,...,g_N}\,.
\ee
The $\Rep(S_3)$ character table is shown in equation \eqref{eq:S3_chars}.
The density matrices will furthermore involve on-site projectors onto subgroups of $S_3$, which can be written as \cite{Bhardwaj:2024kvy}
\begin{equation}
\begin{split}
    P^{(S_3)}_{j}&=\mathbb I_{j}\,, \\
    P^{(\Z_3)}_{j}&=\frac{1}{2} \sum_{\Gamma=1,P}Z^{\Gamma}_{j}\,, \\
    P^{(\Z_2)}_{j}&=\frac{1}{3}\left[\mathbb I_{j}+\sum_{IJ} \left(Z^{E}_{j}\right)_{IJ}\right]\,, \\
    P^{(1)}_{j}&= \sum_{\Gamma}\frac{{\dim}(\Gamma)}{|G|}{\rm Tr}\left(Z^{\Gamma}_{j}\right)\,.
\end{split}
\end{equation}
Following our general approach, we will write commuting projector Hamiltonians in the doubled Hilbert space $\wt{\cH}=\cH_L\otimes\cH_R$, multiply their terms to get a fixed point state on $\widetilde{\cH}$, multiply with projectors onto $\Rep(S_3)$ charged sectors (coming from diagonal anyons ending on both SymTFT boundaries) and then trace out the $\cH_R$ degrees of freedom to obtain strong $\Rep(S_3)$ symmetric density matrices on $\cH$. Note that if we did not multiply with projectors onto $\Rep(S_3)$ sectors then any Hamiltonian with ground state degeneracy would not give a strongly symmetric density matrix after tracing out the $\cH_R$ degrees of freedom.

Recall equation \eqref{eq:Ftildes_S3} for the subgroups $\wt{F}$ of $S_3^L\times S_3^R$ characterizing the Lagrangian algebras on the physical boundary of the SymTFT for each phase. In this case, all 2-cocycles are trivial, and the commuting projector Hamiltonian for the gapped phase labeled $\wt F$ has the form \eqref{DoubleHam}
\be \label{eq:RepS3H}
\wt{H}_{(\widetilde{F})}^{\Rep (S_3)} = - {1\over |F|} \sum_{\substack{j \\ f\in\wt{F}}}(R^{f^{-1}})_{j}  (L^{f})_{ j+1}  - \sum_{j} P_j^{(\widetilde{F})} \,.
\ee

\vspace{2mm}
\noindent {\bf $\bm{\Rep (S_3)}$ SWSSB.} 
The physical boundary has $\cL_1$ condensed giving the sandwich (\ref{SandoF1}).
To obtain the physical density matrices, one first takes the product of terms in \eqref{eq:RepS3H} for $\wt{F}=\widetilde{F}_1= S_3^{\diag}$, which is the subgroup of $S_3^L\times S_3^R$ corresponding to $\cL_1$.
We will then multiply by a projector corresponding to each diagonal local operator coming from \eqref{SandoF1}:
\be
    1,\,\quad \cO_{a_La_R},\,\quad \cO_{b_Lb_R}\,.
\ee
After tracing out $\cH_R$, this produces three density matrices strongly symmetric under $\Rep(S_3)$:
\be\ba
   \rho_1&\propto P^{(1)}_\sym\propto \sum_{\{g_n: \ \prod_{n} g_n =1\} }|g_1 \cdots g_N \rangle \langle g_1 \cdots g_N |  \,,\\
   \rho_a&\propto P^{([a])}_\sym\propto \sum_{\{g_n: \ \prod_{n} g_n \in [a]\} }|g_1 \cdots g_N \rangle \langle g_1 \cdots g_N |  \,,\\
   \rho_a&\propto P^{([b])}_\sym\propto \sum_{\{g_n: \ \prod_{n} g_n \in [b]\} }|g_1 \cdots g_N \rangle \langle g_1 \cdots g_N |  \,.
\ea\ee
where we recall that $[a]=\{a,a^2\}$ and $[b]=\{b,ab,a^2b\}$. Here
\be\ba
    P^{(1)}_\sym&=\frac{1}{6}(\bI+\cS_P+2\cS_E)\,,\\
    P^{([a])}_\sym&=\frac{1}{3}(\bI+\cS_P-\cS_E)\,,\\
    P^{([b])}_\sym&=\frac{1}{2}(\bI-\cS_P)\,.
\ea\ee
are projectors onto states with definite $\Rep(S_3)$ eigenvalues. Specifically, we have that:
\be\ba \label{eq:RepS3_projectors}
     \cS_P\,\rho_1&=\rho_1 \,,& 
     \cS_P\,\rho_a&=\rho_a \,,& 
     \cS_P\,\rho_b&=-\rho_b\,,\\
    \cS_E\,\rho_1&=2\,\rho_1 \,,& 
    \cS_E\,\rho_a&=-\rho_a \,,& 
     \cS_E\,\rho_b&=0\,,\\
\ea\ee
with same charge when acting from the right.

Each of the above density matrices, denoted as $\rho$, has the following properties:
\begin{align}
  \Tr(\rho \,L^g_i(L^g)_j^\dagger)=0\,, \\  
   \nn
 \Tr(\rho \, L^{g}_i(L_j^g)^\dagger\,\rho\, L_j^g(L_i^g)^\dagger)\neq 0\,.
\end{align}
We recall that $L^g_i$ multiplies the group element on site $i$ by $g\in S_3$.
The first line, which follows from the facts that $\rho$ is diagonal in this basis and that basis states with different group elements are orthogonal, 
implies that the weak $\Rep(S_3)$ symmetry is preserved. The second line holds since $L_i^g(L_j^g)^\dagger=L_i^gL_j^{g^-1}$ does not change the conjugacy class of the `holonomy' of the state and each $\rho$ contains the sum over states of this type: it reveals spontaneous symmetry breaking of the strong $\Rep(S_3)$ symmetry, consistent with the SymTFT prediction \eqref{SandoF1}.

\vspace{2mm}
\noindent {\bf $\bm{\Rep(S_3)/\Z_2}$ SWSSB.}
Taking the physical boundary to have $\cL_2$ condensed gives the sandwich (\ref{SandoF2}), for which $\wt{F}=\widetilde{F}_2= \Z_3^{\diag}$.
We again take the product of terms in the Hamiltonian \eqref{eq:RepS3H} for this $\wt{F}$ and then multiply by a projector corresponding to each diagonal local operator coming from \eqref{SandoF2}. Recall from the SymTFT discussion that the factor of two for the anyon $a_La_R$ means that it splits on $\Bphys$ into the two non-trivial conjugacy classes of $\Z_3^\diag$, and we have the local operators
\be
    1,\,\quad \cO_{a_La_R},\,\quad \cO_{a^2_La^2_R}\,.
\ee
After tracing out $\cH_R$, we obtain the following three density matrices strongly symmetric under $\Rep(S_3)$:
\be\ba
   \rho_1&\propto P^{(1)}_\sym\prod_jP^{(\Z_3)}\propto \hspace{-5mm}\sum_{\{I_n: \sum_{n} I_n = 0 \,\text{mod}\, 3\}}\hspace{-7mm} 
|a^{I_1} \cdots a^{I_N} \rangle \langle a^{I_1} \cdots a^{I_N} | \,,\\
   \rho_a&\propto P^{(a)}_\sym\prod_jP^{(\Z_3)}\propto\hspace{-5mm} \sum_{\{I_n: \sum_{n} I_n = 1 \,\text{mod}\, 3\}} \hspace{-7mm}
|a^{I_1} \cdots a^{I_N} \rangle \langle a^{I_1} \cdots a^{I_N} | \,,\\
   \rho_{a^2}&\propto P^{(a^2)}_\sym\prod_jP^{(\Z_3)}\propto \hspace{-5mm}\sum_{\{I_n: \sum_{n} I_n = 2 \,\text{mod}\, 3\}} 
\hspace{-7mm}|a^{I_1} \cdots a^{I_N} \rangle \langle a^{I_1} \cdots a^{I_N} | \,.
\ea\ee
where $\prod_jP^{(\Z_3)}$ came from the projector in \eqref{eq:RepS3H}, and we have furthermore projected onto each of the $\Z_3$ conjugacy classes as follows,
\be\ba
    P^{(a)}_\sym&=\frac{1}{3}(\bI+\ol{\omega}\,Z^{E,\text{prod}}_{1,1}+\omega\,Z^{E,\text{prod}}_{2,2})\,,\\
    P^{(a^2)}_\sym&=\frac{1}{3}(\bI+\omega\,Z^{E,\text{prod}}_{1,1}+\ol{\omega}\,Z^{E,\text{prod}}_{2,2})\,,\\
\ea\ee
since, from \eqref{eq:ZS3} we see that $Z^E_{1,1}$ and \eqref{eq:ZS3} are $\Z_3$ irreps on $\Z_3$ and enable us to distinguish $a$ from $a^2$.

We have that $\cS_P\rho=\rho_g$ for $g\in\Z_3$ and:
\be\ba
    \cS_E\,\rho_1&=2\,\rho_1 \,,& Z^{E,\text{prod}}_{1,1}\,\rho_1&=\rho_1\,,\\
    \cS_E\,\rho_a&=-\rho_a \,,& Z^{E,\text{prod}}_{1,1}\,\rho_a&=\omega\,\rho_a \,,\\
     \cS_E\,\rho_{a^2}&=-\rho_{a^2}\,, &
     Z^{E,\text{prod}}_{1,1}\,\rho_{a^2}&=\ol{\omega}\,\rho_{a^2}\,,\\
\ea\ee
with conjugate charge when acting from the right. Here, $Z^{E,\text{prod}}_{1,1}$ distinguishes $\rho_a$ from $\rho_{a^2}$.

Each of the above density matrices, denoted as $\rho$, has the following properties:
\begin{align}
  \Tr(\rho \,L^g_i(L^g)_j^\dagger)=0\,, \\  
   \nn
   \Tr(\rho \, L^{b}_i(L_j^b)^\dagger\,\rho\, L_j^b(L_i^b)^\dagger)= 0\,,\\
 \Tr(\rho \, L^{a}_i(L_j^a)^\dagger\,\rho\, L_j^a(L_i^a)^\dagger)\neq 0\,.
\end{align}
The first line, which again follows from the facts that $\rho$ is diagonal in this basis and that basis states with different group elements are orthogonal, 
implies that the weak $\Rep(S_3)$ symmetry is preserved. The second line holds since any state with with $b$ is orthogonal to those in $\rho$, so the strong $\Z_2$ symmetry generated by $P$ is preserved. The last line is true because we are summing over all states with group elements in $\Z_3$ and $ L^{a}_i(L_j^a)^\dagger$ does not change the conjugacy class of the `holonomy': it reveals spontaneous symmetry breaking of the strong non-invertible $E$ symmetry, consistent with the SymTFT prediction \eqref{SandoF2}.

\vspace{2mm}
\noindent {\bf $\bm{\Z_}2$ SWSSB.} 
We now consider the physical boundary to have $\cL_2$ condensed, with the sandwich (\ref{SandoF2}), for which $\wt{F}=\widetilde{F}_3= \Z_2^{\diag}$ and the local operators
\be
    1,\,\quad \cO_{b_Lb_R}\,.
\ee
Following the same steps as before, we obtain two density matrices strongly symmetric under $\Rep(S_3)$:
\be\ba
   \rho_1&\propto P^{(1)}_\sym\prod_jP^{(\Z_2)}\propto \hspace{-5mm}\sum_{\{I_n: \sum_{n} I_n = 0 \,\text{mod}\, 2\}}\hspace{-7mm} 
|b^{I_1} \cdots b^{I_N} \rangle \langle b^{I_1} \cdots b^{I_N} | \,,\\
   \rho_b&\propto P^{([b])}_\sym\prod_jP^{(\Z_2)}\propto\hspace{-5mm} \sum_{\{I_n: \sum_{n} I_n = 1 \,\text{mod}\, 2\}} \hspace{-7mm}
|b^{I_1} \cdots b^{I_N} \rangle \langle b^{I_1} \cdots b^{I_N} | \,.
\ea\ee
where $\prod_jP^{(\Z_2)}$ came from the projector in \eqref{eq:RepS3H}, and we have furthermore projected onto each of the $\Z_2$ conjugacy classes with the projectors in \eqref{eq:RepS3_projectors}.

The strong symmetry acts as:
\be\ba
    \cS_P\,\rho_1&=\rho_1\,,& \cS_P\,\rho_b&=-\rho_b\,,& \\
    \cS_E\,\rho_1&=2\,\rho_1\,,& \cS_E\,\rho_b&=0\,,
\ea\ee
with the same charges when acting from the right.
The correlators are:
\begin{align}
  \Tr(\rho \,L^g_i(L^g)_j^\dagger)=0\,, \\  
   \nn
 \Tr(\rho \, L^{a}_i(L_j^a)^\dagger\,\rho\, L_j^a(L_i^a)^\dagger)= 0\,,\\
    \Tr(\rho \, L^{b}_i(L_j^b)^\dagger\,\rho\, L_j^b(L_i^b)^\dagger)\neq 0\,,
\end{align}
with analogous reasoning to the previous phase. In this case, the non-invertible $E$ strong symmetry is preserved while the $\Z_2$ generator $P$ is SWSSB'ed, so we have a $\Z_2$ SWSSB consistent with \eqref{Sandoz2}.

\vspace{2mm}
\noindent {\bf $\bm{(\Rep(S_3)/\Z_2)}$ SSB,  $\Z_2$ SWSSB.}
We finally take the physical boundary to have $\cL_4$ condensed, with the sandwich (\eqref{SandoF4}), for which $\wt{F}=\widetilde{F}_4= (\Z_3\times\Z_3)\rtimes\Z_2$ and the swap* invariant local operators
\be
    1\,,\quad \cO_{a_La_R}\,,\quad \cO_{a_La^2_R}\,,\quad \cO_{b_Lb_R}\,.
\ee
Following the same steps as the previous cases, produces four density matrices strongly symmetric under $\Rep(S_3)$:
\be\ba
   \rho_g&\propto P_\sym^{(g)} \prod_j\left(\bbI_j\bbI_{j+1}+R^{a}_jL^{a^2}_{j+1}+R^{a^2}_jL^{a}_{j+1} \right) \,.
\ea\ee
for $g\in\{1,a,a^2,b\}$. 

Here, $\prod_j\left(\bbI_j\bbI_{j+1}+R^{a}_jL^{a^2}_{j+1}+R^{a^2}_jL^{a}_{j+1} \right)$ comes from the disordering term in \eqref{eq:RepS3H}, and we have furthermore projected onto each of the symmetry charge sectors, with the projectors we have already encountered.
The symmetry acts as:
\be\ba
    \cS_P\,\rho_b&=-\rho_b\,, & \cS_P\,\rho_g&=\rho_g \,\;\; g\in\{1,a,a^2\}\,,\\
    \cS_E\,\rho_1&=2\,\rho_1 \,,& Z^{E,\text{prod}}_{1,1}\,\rho_1&=\rho_1\,,\\
    \cS_E\,\rho_a&=-\rho_a \,,& Z^{E,\text{prod}}_{1,1}\,\rho_a&=\omega\,\rho_a \,,\\
     \cS_E\,\rho_{a^2}&=-\rho_{a^2}\,, &
     Z^{E,\text{prod}}_{1,1}\,\rho_{a^2}&=\ol{\omega}\,\rho_{a^2}\,,\\
\ea\ee
The correlators are:
\begin{align}
  \Tr(\rho \,L^b_i(L^b)_j^\dagger)=0\,,\\
  \Tr(\rho \,L^a_i(L^a)_j^\dagger)\neq 0\,, \\  
   \nn
    \Tr(\rho \, L^{g}_i(L_j^g)^\dagger\,\rho\, L_j^g(L_i^g)^\dagger)\neq 0\,,
\end{align}
The last line shows that the strong $\Rep(S_3)$ symmetry is spontaneously broken. The second line, which is non-zero thanks to the $1+L^a+L^{a^2}$ terms in $\rho$, indicates spontaneous symmetry breaking also of weak $E$ symmetry, while the first reveals that the $\Z_2$ weak symmetry is preserved. This is therefore a ${(\Rep(S_3)/\Z_2)}$ SSB,  $\Z_2$ SWSSB, consistent with the SymTFT \eqref{SandoF4}.

\subsection{Ising Strong Symmetry Mixed Phases}
We now describe mixed phases with strong Ising category symmetry.
The Ising fusion category has three simple objects which we denote as $1,\,P,\, D$.
These satisfy the fusion rules
\begin{align}
    P\otimes P&=1\,,\nn\\
    P\otimes D&= D\otimes P= D\,,\\
     D\otimes D&=1\oplus P\,. \nn
\end{align}
$P$ generates an invertible $\Z_2$ symmetry, while the Kramers-Wannier duality defect $D$ is non-invertible.
In the Drinfeld center, which we write as the double $\cZ (\Ising) = \Ising \boxtimes \overline{\Ising}$, the objects are in the Cartesian product of sets $\{1,\psi\,,\sigma\}\times \{\ol 1,\ol \psi\,,\ol \sigma\}$. There is a unique Lagrangian algebra :
\begin{align}
    \cL_\Ising=1\oplus \psi \ol{\psi}\oplus  \sigma\ol{ \sigma}\,,
\end{align}
on which the category of defects is Ising.
The generalized charges, i.e., properties of the ends of the bulk SymTFT lines on this boundary were detailed in \cite{Bhardwaj:2023ayw}.
We recall that the $\psi\ol\psi$ line ended on an untwisted sector operator $\cO_{-}$ uncharged under $P$ but carrying a $-1$ charge under $D$.
Meanwhile the $\sigma \ol \sigma$ line corresponds to a doublet of operators transforming irreducibly under the Ising symmetry.
This doublet contains an untwisted operator $\cO_{\sigma}$ that is charged under $P$ and a $P$-twisted sector operator $\cO_{\mu}^{P}$ that is uncharged under $P$.
This doublet of operators become the familiar order ($\sigma$) and disorder ($\mu$) operators in the Ising CFT.
The operators $\cO_{\sigma}$ and $\cO_{\mu}^{P}$ are swapped under the action of $D$.

For brevity we do not describe all the other generalized charges but instead refer the reader to section~4.4.5 of \cite{Bhardwaj:2023ayw}.
For the present purpose, we are interested in studying mixed phases with strong Ising symmetry and therefore consider the doubled SymTFT 
\begin{equation}
    \cZ(\Ising \boxtimes \Ising)=\cZ(\Ising)_L\boxtimes \cZ(\Ising)_R\,.
\end{equation}
We denote the anyons $\cZ(\Ising)_L$ and $\cZ(\Ising)_R$ with subscripts $L$ and $R$ respectively.
There are three Lagrangian algebras in the SymTFT
\be\label{LIsingIsing}
\ba
    \cL_{1}^{\Ising^2}=&1\oplus \psi_\p  \ol{\psi}_\p \oplus \ol{\psi}_\a  \psi_\a 
    \oplus \psi_\p  \ol{\psi}_\p  \ol{\psi}_\a  \psi_\a  \oplus \psi_\p  \ol{\psi}_\p  \ol{ \sigma}_\a   \sigma_\a \oplus \\
&\phantom{=}\oplus  \sigma_\p  \ol{ \sigma}_\p  \ol{\psi}_\a  \psi_\a 
\oplus \ol{ \sigma}_\a   \sigma_\a \oplus  \sigma_\p  \ol{ \sigma}_\p  \oplus  \sigma_\p  \ol{ \sigma}_\p  \ol{ \sigma}_\a   \sigma_\a \,,\\
    \cL_{2}^{\Ising^2}=&1\oplus \psi_\p  \ol{\psi}_\a \oplus \ol{\psi}_\p  \psi_\a \oplus \psi_\p  \ol{\psi}_\p  \ol{\psi}_\a  \psi_\a \oplus \psi_\p  \ol{ \sigma}_\p  \ol{\psi}_\a   \sigma_\a  \oplus \\
&\phantom{=}\oplus  \sigma_\p  \ol{\psi}_\p  \ol{ \sigma}_\a  \psi_\a \oplus  \sigma_\p  \ol{ \sigma}_\a \oplus \ol{ \sigma}_\p   \sigma_\a 
    \oplus  \sigma_\p  \ol{ \sigma}_\p  \ol{ \sigma}_\a   \sigma_\a \,,\\
    \cL_{3}^{\Ising^2}=&1\oplus \psi_\p  \ol{\psi}_\p \oplus \ol{\psi}_\a  \psi_\a \oplus \psi_\p  \ol{\psi}_\a \oplus \ol{\psi}_\p  \psi_\a \oplus \psi_\p  \psi_\a \oplus \\
&\phantom{=}\oplus \ol{\psi}_\p  \ol{\psi}_\a \oplus \psi_\p  \ol{\psi}_\p  \ol{\psi}_\a  \psi_\a \oplus 2 \sigma_\p  \ol{ \sigma}_\p  \ol{ \sigma}_\a   \sigma_\a \,.
\ea
\ee
All three respect (\ref{hermcond}) and (\ref{poscond}) and therefore may furnish mixed phases with strong Ising symmetry.
The symmetry Lagrangian for the strong Ising symmetry is 
\be
\cL_\Ising^\text{strong} = \cL_1^{\Ising^2}= \cL_{\Ising_L}\otimes \cL_{\Ising_R} \,.
\ee
While $\cL_2$ and $\cL_3$ are not decomposable into products. 
Let us now in turn describe the gapped phases obtained by picking the three Lagrangian algebras to define the physical boundaries of the SymTFT.

\medskip \noindent {\bf Strong to nothing SSB Phase.} For this we pick the physical boundary to be defined via $\cL_1$, and obtain the following SymTFT sandwich:
\be
\begin{tikzpicture}
\begin{scope}[shift={(0,0)}]
\draw [thick] (-2,-1) -- (-2,1) ;
\draw [thick] (4,-1) -- (4,1) ;
\draw [thick] (-2, 0.5) -- (4, 0.5) ;
\draw [thick] (-2, 0) -- (4, 0) ;
\draw [thick] (-2, -0.5) -- (4, -0.5) ;
\node[above] at (1,0.5) {$
1, \  \psi_\p  \ol{\psi}_\p ,\  \ol{\psi}_\a  \psi_\a 
    ,\  \bm{\psi_\p  \ol{\psi}_\p  \ol{\psi}_\a  \psi_\a} 
    $} ;
\node[above] at (1,0) {$ \sigma_\p  \ol{ \sigma}_\p  \ol{\psi}_\a  \psi_\a ,\   \psi_\p  \ol{\psi}_\p  \ol{ \sigma}_\a   \sigma_\a$} ;
\node[above] at (1,-0.5) {$\ol{ \sigma}_\a   \sigma_\a ,\  
\sigma_\p  \ol{ \sigma}_\p  ,\ 
\bm{\sigma_\p  \ol{ \sigma}_\p  \ol{ \sigma}_\a   \sigma_\a}$} ;
\node[above] at (-2,1) {$\cL_\Ising^\text{strong}$}; 
\node[above] at (4,1) {$\cL_1^{\Ising^2}$}; 
\draw [black,fill=black] (4,0.5) ellipse (0.05 and 0.05);
\draw [black,fill=black] (-2,0.5) ellipse (0.05 and 0.05);
\draw [black,fill=black] (4,0) ellipse (0.05 and 0.05);
\draw [black,fill=black] (-2,0) ellipse (0.05 and 0.05);
\draw [black,fill=black] (4,-0.5) ellipse (0.05 and 0.05);
\draw [black,fill=black] (-2,-0.5) ellipse (0.05 and 0.05);
\end{scope}
\end{tikzpicture}
\ee

Clearly, each line in $\cL_1$ can end on both boundaries, therefore one obtains nine charged local operators upon compactifying the SymTFT. 
We denote the local operators that come from compactifying the identity line, $\psi_L\ol{\psi}_L$ or $\sigma_L\ol{\sigma}_L$ as $\cO_{P_L}$ and $\cO_{D_L}$ and similarly for $R$.
It can be checked that these satisfy the algebra \cite{Bhardwaj:2023idu}
\begin{equation}
\begin{split}
    \cO_{P_L} \times \cO_{P_L} =&\,1 \,, \\   
    \cO_{D_L} \times \cO_{P_L} = \cO_{P_L} \times \cO_{D_L}=&\,\cO_{D_L} \,, \\ 
    \cO_{D_L} \times \cO_{D_L}=&\; 1 + \cO_{P_L}\,,
\end{split}
\end{equation}
and similarly for the $R$ operators.
The pure state phase breaks the $\Ising_L\boxtimes\Ising_R$ symmetry completely since it contains charged local operators that transform under very symmetry generator.
Specifically $\cO_{P_{I}}$ transforms under $D_I$ and $\cO_{D_{I}}$ transforms under $P_{I}$ for $I\in \{L,R\}$.
Correspondingly the mixed state phase breaks the strong $\Ising$ symmetry to nothing.

To construct density matrices within this phase, we consider a state that transforms in a definite $\Ising_L\boxtimes\Ising_R$ irrep.
These states are in one-to-one correspondence with local operators $\cO_{a_{L}b_R}$ with $a\,,b\in \{1,P,D\}$\footnote{Note, that the states $|a_Lb_R\drangle$ are not vacua as these local operators are not idempotents, however $|a_Lb_R\drangle$ transform in definite irreps of the $\Ising_L\boxtimes \Ising_R$ symmetry.} and we denote the corresponding states as
\begin{equation}
    |a_Lb_R\drangle\,.
\end{equation}
The corresponding fixed-point density matrix in this phase is obtained by constructing the fixed-point ground state in irrep $a_Lb_R$, i.e., $|a_Lb_R\drangle\dlangle a_Lb_R|$ and tracing over the right degrees of freedom.

\medskip \noindent {\bf Ising Strong-to-Weak SSB.} We now move onto the more interesting Lagrangian algebras which do not be factorize into left and right Lagrangians.
Consider the SymTFT sandwich with $\cL_\phys=\cL_2^{\Ising^2}$:
\be
\begin{tikzpicture}
\begin{scope}[shift={(0,0)}]
\draw [thick] (-2,-1) -- (-2,1) ;
\draw [thick] (3,-1) -- (3,1) ;
\draw [thick] (-2, 0.25) -- (3, 0.25) ;
\draw [thick] (-2, -0.5) -- (3, -0.5) ;
\node[above] at (0.5,0.25) {$1, \ \bm{\psi_L\ol\psi_L\psi_R\ol\psi_R}$} ;
\node[above] at (0.5,-0.5) {$ \bm{\sigma_L\ol\sigma_L\sigma_R\ol\sigma_R} $} ;
\node[above] at (-2,1) {$\cL_\Ising^\text{strong}$}; 
\node[above] at (3,1) {$\cL_2^{\Ising^2}$}; 
\draw [black,fill=black] (3,0.25) ellipse (0.05 and 0.05);
\draw [black,fill=black] (-2,0.25) ellipse (0.05 and 0.05);
\draw [black,fill=black] (3,-0.5) ellipse (0.05 and 0.05);
\draw [black,fill=black] (-2,-0.5) ellipse (0.05 and 0.05);
\end{scope}
\end{tikzpicture}
\ee
Let us describe the pure state phase. There are three charged local operators that come from compactifying SymTFT lines $1$, $\psi_L\ol\psi_L\psi_R\ol\psi_R$ and $\sigma_L\ol\sigma_L\sigma_R\ol\sigma_R$.
We denote the corresponding local operators as 1, $\cO_{P}$ and $\cO_{D}$ respectively.
By compatibility with bulk fusion, one can deduce that these operators satisfy the algebra relations 
\begin{equation}
\begin{split}
    \cO_{P} \times \cO_{P} =&\,1 \,, \\   
    \cO_{D} \times \cO_{P} = \cO_{P} \times \cO_{D}=&\,\cO_{D} \,, \\ 
    \cO_{D} \times \cO_{D}=&\; 1 + \cO_{P}\,.
\end{split}
\end{equation}
Diagonalizing this algebra, one obtains idempotents 
\begin{equation}
\begin{split}
    v_0=& \frac{1}{2}(1-\cO_P)\,, \\
    v_1=& \frac{1}{4}(1+\cO_P+\sqrt{2}\cO_D)\,, \\
    v_2=& \frac{1}{4}(1+\cO_P-\sqrt{2}\cO_D)\,,
\end{split}
\end{equation}
which correspond to the vacua in this pure state gapped phase.
The symmetry charges carried by $\cO_P$ and $\cO_D$ can be easily deduced since these are nothing but the (diagonal) product of the above described Ising charges.
Therefore the left and right Ising categories act identically on these operators.
$P_L$ and $P_{R}$ act as
\begin{equation}
    P_{L,R}:(\cO_{P}\,, \cO_{D})\longrightarrow  (\cO_{P}\,, -\cO_{D})\,,
\end{equation}
Meanwhile the non-invertible elements $S_L$ and $S_R$ act as
\begin{equation}
D_{L,R}:(1\,, \cO_P\,, \cO_D)\longrightarrow (\sqrt 2\,, -\sqrt 2 \cO_P\,, 0)\,.    
\end{equation}
Using these relations, it can be seen that the vacua transform as follows under the diagonal Ising action.
\begin{equation}
\begin{split}
    P_LP_R&:v_i \longmapsto v_{i} \\
    D_LD_R&:(v_0\,, v_1\,, v_2)  \longmapsto (2v_0\,, v_1+v_2\,, v_1+v_2) \,.
\end{split}
\end{equation}
Therefore the diagonal $\Rep(D_8)$ symmetry within $\Ising^2$ generated by $D_LD_R$ and $P_L$ and $P_R$ is preserved in the vacua $v_0$ and $v_1+v_2$.
One can also consider mixed-state diagnostics obtained from patch operators corresponding to the other anyons in the Lagrangian algebra $\cL_2$.
For instance, the patch operators labeled by $\psi_L\ol{\psi}_R$ corresponds to a $P_LP_R$ disorder line with local operators at its two ends.
The local operators transform with a minus sign under the $P_L$ and $P_R$ symmetry action and with a $+ i$ (respectively $-i$) under the $D_L$ (respectively $D_R$) at one end of the patch operator and its inverse at the other end.

\medskip \noindent {\bf Duality SSB and $\Z_2$ Strong-to-Weak SSB.} Finally, let us consider $\cL_\phys=\cL_3$:
\be
\begin{tikzpicture}
\begin{scope}[shift={(0,0)}]
\draw [thick] (-2,-1) -- (-2,1) ;
\draw [thick] (4,-1) -- (4,1) ;
\draw [thick] (-2, 0.25) -- (4, 0.25) ;
\draw [thick] (-2, -0.5) -- (4, -0.5) ;
\node[above] at (1,0.25) {$1,\; \psi_\p  \ol{\psi}_\p \,,\; {\psi}_\a  \ol\psi_\a \,,\; \bm{\psi_\p  \ol{\psi}_\p  \ol{\psi}_\a  \psi_\a }$} ;
\node[above] at (1,-0.5) {$\bm{ 2\,\sigma_\p  \ol{ \sigma}_\p  \ol{ \sigma}_\a   \sigma_\a} $} ;
\node[above] at (-2,1) {$\cL_\Ising^\text{strong}$}; 
\node[above] at (4,1) {$\cL_3^{\Ising^2}$}; 
\draw [black,fill=black] (4,0.25) ellipse (0.05 and 0.05);
\draw [black,fill=black] (-2,0.25) ellipse (0.05 and 0.05);
\draw [black,fill=black] (4,-0.5) ellipse (0.05 and 0.05);
\draw [black,fill=black] (-2,-0.5) ellipse (0.05 and 0.05);
\end{scope}
\end{tikzpicture}
\ee
This pure state gapped phase has six charged local operators.
Four of these come from compactifying the invertible SymTFT lines $1,\psi_L\ol{\psi}_L, \psi_R\ol{\psi}_R$ and $\psi_L\ol{\psi}_L\psi_R\ol{\psi}_R$.
Let us denote the corresponding operators as $1,\cO_L,\cO_R$ and $\cO_L\cO_R$ which have a $\Z_2\times \Z_2$ product structure.
These operator $\cO_L$ is charged under $D_L$ and $D_LD_R$.
Similarly $\cO_R$ is charged under $D_R$ and $D_LD_R$.
Therefore the duality symmetry is completely broken.
Next, let us inspect the pair of charged local operator obtained from compactifying the line $\sigma_L\ol\sigma_L\sigma_R\ol\sigma_R$ which has two ends on the physical boundary and a single end on te symmetry boundary.
Let us denote the corresponding local operators as $\cO_1$ and $\cO_2$.
These operators are charged under $P_L$ and $P_R$ but uncharged under the diagonal $P_LP_R$ symmetry.
This implies that the $\Z_2$ strong sub-symmetry is spontaneously broken to a weak symmetry.

{It would be interesting to construct anyon chain realizations of these phases, which we will leave for future work. }

\section{SymTFT for Mixed Strong and Weak Symmetries} 
\label{sec:WeakStrong}

\noindent
We now turn to the case where the density matrix has weak symmetries distinct from its strong symmetries. In the invertible case, the strong symmetry is given by $\mathcal{S}_L\boxtimes\mathcal{S}_R$ where $\mathcal{S}_L=\mathrm{Vec}_K^{\omega}$ and $\mathcal{S}_R=\mathrm{Vec}_K^{\overline{\omega}}$. In the doubled Hilbert space any additional weak invertible $G$ symmetry always forms $\mathcal{W}=\mathrm{Vec}_G$ with trivial anomaly $\omega$.

Strong and weak symmetries may also form a nontrivial extension $1\to K\to H \to G\to 1$ where $K$ is the normal subgroup \cite{Ma:2023rji}. When the extension is trivial, the symmetry of the Choi state is just $(K\times K)\rtimes T\times G$. The semi-direct product comes from the fact that $T$ permutes the strong symmetries, but leaves the weak symmetries invariant. In the following, we will drop $T$ when describing the symmetries and take it as implicit. In the non-invertible case, there may be no factorization of the total symmetry into a direct product of the strong symmetries and the weak symmetries. We will study in examples density matrices with a strong $\mathbb{Z}_2$ symmetry and weak non-invertible duality symmetry. {The symmetry pattern is similar to the extension described above, because the weak duality symmetry fuses to objects generated by the strong symmetries.}

Throughout this section we will use $\cS$ to denote a strong symmetry (that comes along with a weak symmetry) and $\cW$ denotes the weak symmetry with no corresponding strong operators.

Below, we will propose a method for constructing consistent patterns of strong and weak non-invertible symmetries in the doubled Hilbert space. The basic idea is that we will start with a larger strong symmetry, and consider the ways we can explicitly break parts of the strong symmetry to be weak by considering various condensations to smaller topological orders (TOs), where part of the symmetry will be only weak. 
The advantage of this approach is that it allows us to use the hermiticity and positivity conditions (\ref{hermcond}) and (\ref{poscond}) for the case where all of the symmetries are strong and the symmetry in the doubled Hilbert space is of the form $\mathcal{S}_L\boxtimes\mathcal{S}_R$ with $\mathcal{S}_R=\overline{\mathcal{S}}_L$. Extending to the larger $\mathcal{S}_L\boxtimes\mathcal{S}_R$ symmetry also allows us to naturally construct Choi states, which live in a doubled Hilbert space.

This approach also makes a connection to the quantum channels used to obtain density matrices with various strong and weak symmetries, starting from pure states whose symmetries are all strong. From the perspective of SymTFT, this explicit symmetry breaking is obtained by condensing a condensable (but non-Lagrangian) algebra in the bulk SymTFT  to get a smaller topological order. This topological order is smaller because explicitly breaking some strong symmetries to be only weak symmetries reduces the number of symmetry operators, and since the symmetry is smaller its center is also smaller. 

We will see that this construction naturally shows that not all patterns of strong and weak symmetries are consistent, because not all collections of bosons can be condensed. For example, we can have a strong $\mathbb{Z}_2$ symmetry with a weak duality symmetry, but we cannot have a weak $\mathbb{Z}_2$ symmetry with a strong duality symmetry. Intuitively, this comes from the observation that the strong duality symmetry would generate (by fusion) a strong $\mathbb{Z}_2$ symmetry, so the $\mathbb{Z}_2$ symmetry cannot just be weak.

\subsection{SymTFT for Weak Symmetries}

We would now like to set up a framework to study starting points where the system has weak symmetry (including weak non-invertible symmetries) without any corresponding strong symmetry. To achieve this, we use the SymTFT, starting with a larger strong symmetry. We then condense a non-maximal algebra {comprised of diagonal charges}, which defines an interface to a reduced topological order. The reduced SymTFT may not factorize into $\mathcal{Z}(\mathcal{S}_L)\boxtimes\mathcal{Z}(\mathcal{S}_R)$, and therefore can describe more general patterns of strong and weak symmetries. 

\subsubsection{Warmup: from strong to weak groups.}
Let us first discuss how to incorporate weak symmetries when all of the symmetries are invertible. Instead of considering $\cS= \Vec_K \boxtimes \Vec_K$ and $\cW= \Vec_G$  at the outset, and constructing the Drinfeld center for this symmetry, we will begin with a larger symmetry and its center. This will allow for a more direct generalization to the non-invertible case. In the following, we assume that the symmetry groups have no anomalies, and will just write $K_L,K_R, $ etc rather than $\mathrm{Vec}_K$. 

Consider the larger symmetry group $K_L\times K_R\times G_L \times G_R$.\footnote{Note that one subtlety is that this is a many-to-one mapping, i.e. we can consider anomalous $G_L,G_R$. The diagonal (weak) symmetry is always anomaly-free.} The SymTFT is an untwisted $K\times K\times G\times G$ gauge theory:
\begin{equation}\ba
    &\mathcal{Z}(K_L\times K_R\times G_L\times G_R)\cr 
    & \ =\mathcal{Z}(K_L)\boxtimes \mathcal{Z}(K_R)\boxtimes \mathcal{Z}(G_L)\boxtimes\mathcal{Z}(G_R) \,.
\ea \end{equation}
 To explicitly break the strong symmetry $G\times G$ down to a diagonal weak symmetry, we condense in this gauge theory diagonal $G$ charges, given by the condensable algebra 
\begin{equation}
    \mathcal{L}_{G\times G\to G}=\oplus_{\Gamma\in \Rep(G)}\Gamma_L\ol{\Gamma}_R \,,
\end{equation}
where $\Gamma_L,\,\Gamma_R$ are anyons corresponding to pure gauge charges (irreps) of $\cZ (G_L)$ and $\cZ (G_R)$ respectively.  This condensable algebra also defines a gapped boundary condition between $\cZ (G_L\times G_R)$ and $\cZ (G)$, the SymTFT of the diagonal $G$ symmetry. 

Condensing this algebra in the bulk SymTFT removes the strong $G_L,G_R$ symmetries because $\Gamma_L\ol{\Gamma}_R$
confine the $G_L,G_R$ gauge fluxes. However, the diagonal fluxes survive, because they braid trivially with the condensed anyons. As a result, this condensation directly takes the center of the enlarged symmetry group to the center of the desired $K_L\times K_R\times G$ symmetry. {The Choi state constructed from this approach would have a $(|K|\times |G|)^2$ dimensional local Hilbert space, and the condensation corresponds to explicitly breaking the strong $G$ symmetry to weak. The resultind density matrix lives in a spin chain with a $|K|\times |G|$ dimensional local Hilbert space.}

We can also include in this framework symmetries that form a nontrivial extension. The simplest invertible example is where a weak $\mathbb{Z}_2$ symmetry squares to the diagonal symmetry coming from a strong $\mathbb{Z}_2$. To get such a symmetry pattern, we start with a strong $\mathbb{Z}_4$ symmetry with trivial cocycle, so the SymTFT is a $\mathbb{Z}_4\times\mathbb{Z}_4$ gauge theory with anyons generated by $e_L, e_R, m_L, m_R$ with $e_L^4=e_R^4=m_L^4=m_R^4=1$. If we condense the diagonal $\mathbb{Z}_2$ charges $\mathcal{L}=1\oplus e_L^2e_R^2$ then we get a $\mathbb{Z}_2\times\mathbb{Z}_4$ gauge theory. The condensation breaks the strong $\mathbb{Z}_4$ symmetries associated with $m_L,m_R$ but there are strong $\mathbb{Z}_2$ symmetries $m_L^2, m_R^2$ and a weak $\mathbb{Z}_4$ symmetry generated by $m_Lm_R$. As desired, the weak $m_L m_R$ squares to the diagonal symmetry generated by the strong $\mathbb{Z}_2$.

{This idea extends to non-invertible symmetries: we propose that all  partial bulk condensations of diagonal charges in SymTFTs of the form $\mathcal{Z}(\mathcal{S}_L)\boxtimes\mathcal{Z}(\mathcal{S}_R)$ give rise to consistent patterns of strong and weak non-invertible symmetries.}

\subsubsection{From Strong to Weak Categories}
  
We again start with $\cZ (\cS_L \boxtimes \cS_R)$, with the strong symmetry boundary condition $\Bsym = \cL_{\cS}^\strong$ and 
\begin{equation}
    \mathcal{L}_{\mathcal{S}}^{\mathrm{strong}} = \mathcal{L}_{\cS,L}\otimes\mathcal{L}_{cS,R}=\bigoplus_{a_Lb_R}n_{a_Lb_R,\mathrm{sym}}\, a_Lb_R\,.
\end{equation}
We then condense diagonal charges, i.e. a condensable algebra of the form
\begin{equation}\label{chargecond}
    \mathcal{A}=\oplus_{a_Lb_R}n_{a_Lb_R}a_Lb_R\,,
\end{equation}
where $T(a_L)=b_R$ and $n_{a_Lb_R}\neq 0$ only if $n_{a_Lb_R,\mathrm{sym}}\neq 0$. This means that any $a_Lb_R$ with nonzero coefficient in $\mathcal{A}$ is a \emph{diagonal charge} of $\mathcal{Z}(\mathcal{S}_L\boxtimes\mathcal{S}_R)$. The condensation of $\mathcal{A}$ leads to a reduced TO that does not in general factorize into $\mathcal{Z}(\mathcal{S_L'})\boxtimes\mathcal{Z}(\mathcal{S}_R')$; it removes some of the strong symmetries but maintains the corresponding weak ones. 

More generally, one might consider the condensation of other condensable (but not Lagrangian) algebras with the following constraints:
\begin{enumerate}
\item {\bf $T$ Invariance:} $n_{a_L, b_R}= n_{\overline{b}_L,\overline{a}_R}$  
\item {\bf Positivity:}
$n_{a_L,b_R}\leq n_{a_L,\overline{a}_R}n_{\overline{b}_L,b_R}$
\item {\bf Non-Factorization}: The algebra does not factorize into $\cA_L\otimes \cA_R$. 
\end{enumerate}
These conditions are all satisfied automatically for condensable algebras of the form (\ref{chargecond}). However, the physical meaning of these more general condensations are unclear. For example, condensing $1\oplus e_Le_R$ in $\mathbb{Z}_2\times\mathbb{Z}_2$ gauge theory corresponds to breaking the strong $\mathbb{Z}_2$ symmetry to a weak $\mathbb{Z}_2$ symmetry explicitly. On the other hand, condensing $1\oplus m_Lm_R$ appears to give a single $\mathbb{Z}_2$ symmetry that is strong, because $m_L\sim m_R$ where the worldline of $m_L$ in the vicinity of $\mathfrak{B}_{\mathrm{sym}}$ gives the strong $\mathbb{Z}_2$ symmetry. We leave an exploration of possible applications of these other condensations to future work, and will focus here on the condensation of diagonal charges.

\subsubsection{Weak Symmetry Alone}\label{weakalone}

Let us consider the simplest case of having just a weak symmetry $\cW$. The first important conclusion that we will reach is that we cannot make a non-invertible symmetry category completely weak. 
Later on, we will discuss more intricate symmetries, with weak non-invertible symmetries and strong invertible symmetries, that form a single category (and does not factorize in separate weak and strong parts). 

Let us borrow intuition from the invertible case and try to realize the $\cW$ symmetry by starting with a larger strong $\cW_L \boxtimes \cW_R$ symmetry. We can then try to condense to the diagonal $\cW$ symmetry:
\be\label{condredTO}
\cZ( \cW_L \boxtimes \cW_R) \to \cZ (\cW) \,,
\ee
Let $\cL_\cW^\strong = \cL_{\mathcal{W},L} \otimes \cL_{\cW,R}$ as before, where $\cL_{\cW,L} = \bigoplus_{w} n_w w_L$ are the Lagrangian algebras for the $\cW$ symmetry. 
Then in order to get the weak $\cW$ symmetry we would condense the ``diagonal" algebra  
\be
\cA_\cW^\weak = \bigoplus_{w} n_{w} w_L \ol{w_R} \,.
\ee
This is notably not maximal. 
However, crucially, this is not always a condensable algebra. For $\cW= \mathrm{Vec}_G$ a finite group symmetry, abelian or non-abelian, this is indeed an algebra. In the classification of condensable algebras by subgroups $H$ and normal subgroups $N \triangleleft H$ this would correspond to $H= G^{\text{diag}}$ and $N=1$.\footnote{Note that on the other hand in order to condense all the fluxes (conjugacy classes) $w= [g]$, we would have to have $H= G \times G$ and $N= G^{\text{diag}}$. However this group $N$ is only a normal subgroup for $G$ abelian. }

The condensation can be depicted as follows in terms of a club sandwich\footnote{The systematic study of these interfaces is useful in the SymTFT description of gapless phases, where these were dubbed club sandwiches \cite{Bhardwaj:2023bbf,Bhardwaj:2024qrf}.} type SymTFT configuration 
\be\label{ClubSando}
\begin{tikzpicture}
\begin{scope}[shift={(0,0)}]
\draw [cyan,  fill=cyan] 
(0,0) -- (0,3) --(3,3) -- (3,0) -- (0,0) ; 
\draw [Green, fill= Green](3,0) -- (3,3) --(6,3) -- (6,0) -- (3,0) ; 
\draw [very thick] (0,0) -- (0,3) ;
\draw [very thick] (3,3) -- (3,0) ;
\draw [very thick] (6,3) -- (6,0) ;
\node at (1.5,1.5) {$\cZ(\cW_L\boxtimes \overline{\cW_R})$} ;
\node at (4.5,1.5) {$\cZ(\cW')$} ;
\node[above] at (0,3) {$\Bsym = \cL_{\cW}^{\text{strong}}$}; 
\node[above] at (3,3) {$\cA_{\cW}^\weak$}; 
\node[above] at (6,3) {$\Bphys ={\cL}_{\phys} $}; 
\end{scope}
\end{tikzpicture}
\ee
In order to analyze the possible symmetries, we can restrict to 
{the study of interface only and the map between the two topological orders that it defines. In the SymTFT literature this is also known as 
the ``club quiche".}

This implies that not all symmetries can be made weak, because a condensation relating the left and right hand sides of (\ref{condredTO}) does not in general exist. From the point of view of the density matrix we can try to derive conditions on the fusion rules that are consistent with a weak symmetry. Denote by $W_a$ the weak symmetry generators, with fusion $W_a \otimes W_b = \oplus_c N_{ab}^c W_c$. 
They satisfy (\ref{weaksymnon}). 
Then we can require the action of the symmetry to be consistent with the fusion: 
\be\label{Rows}
\ba
|c_a|^2 |c_b|^2 \rho &= W_a (W_{b} \rho W_b^\dagger) W_a^\dagger \cr 
&= \left(\bigoplus_c n_{ab}^c  W_c \right)\rho \left(\bigoplus_d n_{ab}^d W_d^\dagger \right) \,.
\ea
\ee
{Here we consider the singlet density matrix, for which $|c_a|^2= d_a^2$ the square of the quantum dimension.}
If all of the $W_c$ are weak symmetries, then the only way this is consistent is if the fusion has specific properties:
the RHS has to equal $\bigoplus_f m_f W_f \rho W_f^\dagger = \sum_f |c_f|^2 \rho$. This can only  be achieved if the fusion is invertible, in which case there is no sum over $f$. On the other hand, if we allow for the symmetry to be partially weak and partially strong, then the equation can be satisfied if all operators that appear as the outcome of a non-invertible fusions are strong symmetries. 

\vspace{2mm}
\noindent{\bf Examples.}
For example, $\Rep (S_3)$ has the fusion rule $P^2=1$ $P E = E P = E$ and $E^2= 1\oplus P \oplus E$. This cannot be made weak. However we cannot seem to make $P$ strong and $E$ weak either: {if $E$ were weak then 
$E \rho E = |c|^2 \rho$ and $E^2 \rho E^2 = |c|^4 \rho$. On the other hand this also has to equal  $(1+ P + E) \rho (1+ P +E)$. However if we assume that $E$ is weak, the terms with single $E$s acting on $\rho$ cannot be further simplified. }
This symmetry can only be realized as a strong symmetry. 

Instead, consider the Ising fusion category generated by $U,D$, with fusion  $U^2=1$,  $UD = DU =D$ and $D^2 = 1+U$. Notably $D$ does not appear in the fusion $D^2$. Then 
making the $\Z_2$ generated by $U$ strong and the symmetry generated by the Kramers-Wannier duality $D$ weak can be consistent. We will explore this example in detail in Sec.~\ref{weakkwz2}.

Finally, consider $\Rep (H_8)$ or $\Rep (D_8)$ symmetry. In this case the non-invertible 2d irrep has fusion $E^2 = 1\oplus 1_a \oplus 1_b \oplus 1_c$ and $E 1_a= E$  and $1_a^2 = 1$ etc, where $1_a$ are 1d irreps and thus invertible. Again (\ref{Rows}) cannot be satisfied if all the symmetries are weak. However we can make $1_a, 1_b, 1_c$ strong and $E$ weak. In this case we get (\ref{Rows}) to be satisfied as $|c_E|^2=4)$ and the RHS becomes $ (1\oplus 1_a \oplus 1_b \oplus 1_c) \rho (1\oplus 1_a \oplus 1_b \oplus 1_c) = 16 \rho$. We will discuss this in Sec.~\ref{sec:GeneralWeakDu}.

\subsection{Examples: Invertible Weak Symmetries}
In the following we will first discuss invertible weak symmetries such as $\Z_2$ and $S_3$ to illustrate our approach. We will then study examples involving non-invertible weak symmetries.

\subsubsection{Weak $\Z_2$ Symmetry}
Let $\cW= \mathrm{Vec}_{\Z_2}$ with no anomaly and start with $\cZ(\cW_L\boxtimes \cW_R)$ which is $\mathbb{Z}_2\times\mathbb{Z}_2$ gauge theory. We can condense the algebra 
\be\label{AweakZ2}
\cA_{\Z_2}^\weak = 1 \oplus e_L e_R \,.
\ee
We thus have the following configuration (where we only show the club quiche with no physical boundary condition in order to discuss the symmetry properties):  
\be\label{ClubSandoZ2}
\begin{tikzpicture}
\begin{scope}[shift={(0,0)}]
\draw [cyan,  fill=cyan] 
(0,0) -- (0,3) --(3,3) -- (3,0) -- (0,0) ; 
\draw [Green, fill=Green](3,0) -- (3,3) --(6,3) -- (6,0) -- (0,0) ; 
% \draw [very thick] (0,0) -- (0,3) ;
\draw [very thick] (3,3) -- (3,0) ;
%\draw [very thick] (6,3) -- (6,0) ;
\draw [thick] (0,2) -- (6,2) ;
\node[above] at  (1.5, 2) {$1+ e_L e_R$} ; 
\node[above] at (4.5,2) {$ 1$} ; 
\draw [thick] (0,1.5) -- (6,1.5) ;
\node[above] at  (1.5, 1.5) {$e_L,\,  e_R$} ; 
\node[above] at (4.5,1.5) {$e $} ; 
\draw [thick] (0,1) -- (6,1) ;
\node[above] at  (1.5, 1) {$m_L m_R$} ; 
\node[above] at (4.5,1) {$m$} ; 
\draw [thick] (0,0.5) -- (2.9,0.5) -- (2.9, 0);
\node[above] at  (1.5, 0.5) {$m_L , \, m_R$} ; 
% \node[above] at (0,3) {$\Bsym = (1+ e_L)(1+e_R)$}; 
\node[above] at (3,3) {$\cA_{\cW}^\weak$}; 
%\node[above] at (6,3) {$\Bphys ={\cL}_{\phys} $}; 
\end{scope}
\end{tikzpicture}
\ee
The reduced TO is $\cZ(\cW') = \cZ(\mathrm{Vec}_{\Z_2})$ and is realized in terms of the diagonal symmetry, generated by $m_L m_R \sim m$ acting both from the left and the right. We can now insert the symmetry boundary $\Bsym = (1+ e_L)(1+e_R)$ on the LHS, which corresponds to the following symmetry boundary of the reduced TO: 
\be
\cL'= 1+ e \,.
\ee
For this simple example, we can immediately write down states that demonstrate weak $\mathbb{Z}_2$ symmetry SSB and preservation, for a weak $\mathbb{Z}_2$ symmetry given by $U^\dagger\rho U=\rho$ with $U=\prod_iX_i$ on a chain of qubits. An SSB state is given by
\begin{equation}
    \rho=\frac{1}{2}(|\uparrow\rangle\langle \uparrow|+|\downarrow\rangle\langle\downarrow|) \,,
\end{equation}
where $|\uparrow/\downarrow\rangle$ are product states where every state is in the $\pm$ eigenstate of $\Z_i$. This state demonstrates weak symmetry SSB because $\mathrm{Tr}(\rho Z_iZ_j)=1$. A state that preserves the weak $\mathbb{Z}_2$ is given by
\begin{equation}
    \rho=\frac{1}{2^N}\mathbf{1} \,.
\end{equation}

\subsubsection{Strong $\mathbb{Z}_2$ and Weak $\Z_2$}

We now study the case where there is both a strong $\mathbb{Z}_2$ symmetry and a weak $\mathbb{Z}_2$ symmetry. We can take the symmetries to have the following unitary representation on a 1+1d spin chain where each unit cell contains two qubits. $U_S$ acts on the $S$ qubits and $U_W$ acts on the $W$ qubits:
\begin{equation}
    U_{S}=\prod_{i}X_{i,S}\qquad U_{W}=\prod_iX_{i,W}\,,
\end{equation}
The two $\mathbb{Z}_2$ symmetries are simply global spin flips on the $S$ and $W$ qubits respectively. We can break the strong $W$ symmetry down to a weak symmetry using the channel acting on the $W$ spins:
\begin{equation}
E[\rho]=\bigotimes_iE_{i}[\rho]\qquad E_i[\rho]=\frac{1}{2}\rho +\frac{1}{2}Z_{i,W}\rho Z_{i,W}\,.
\end{equation}
The Krauss operators are not individually $W$ symmetric, so this channel breaks the $W$ symmetry explicitly: $U_WE[\rho]\neq e^{i\theta}E[\rho]$. However, a weak $W$ symmetry is preserved, so
\begin{equation}
    U_SE[\rho]=E[\rho] U_S=e^{i\theta}E[\rho]\qquad U_W^\dagger E[\rho] U_W=E[\rho]\,.
\end{equation}
In the doubled Hilbert space, we begin with symmetry $(\mathbb{Z}_2)^4$, and the SymTFT is $(\mathbb{Z}_2)^4$ gauge theory. The gauge charges are generated by $e_{W,L},e_{W,R},e_{S,L},e_{S,R}$, each of which has $\mathbb{Z}_2$ fusion rules. We explicitly break the strong $W$ symmetry to a weak one by condensing $1\oplus e_{W,L}e_{W,R}$ as in \eqref{AweakZ2} in the bulk SymTFT, resulting in $(\mathbb{Z}_2)^3$ gauge theory generated by the anyons
\begin{equation}
    \{e_{S,L},e_{S,R},m_{S,L},m_{S,R},e_{W,L}\sim e_{W,R}, m_{W,L}m_{W,R}\}\,.
\end{equation}
The diagonal $W$ flux $m_{W,L}m_{W,R}$ survives the condensation but $m_{W,L}$ and $m_{W,R}$ individually are confined. 

To see why the above strong symmetry breaking channel corresponds to this condensation, note that in the doubled Hilbert space,
\begin{equation}
    |E[\rho]\RR=\prod_i\left(\frac{1+Z_{i,W,L}Z_{i,W,R}}{2}\right)|\rho\RR\,.
\end{equation}
This corresponds precisely to the effective boundary action from condensing $1+e_{W,L}e_{W,R}$ in the bulk 2+1d SymTFT, i.e. giving the open anyon string operator for $e_{W,L}e_{W,R}$ a nonzero expectation value.

As an example, we can consider the pure state $\rho=|+_S\rangle\langle +_S|\otimes |+_{W}\rangle\langle+_W|$. Then, clearly, 
\begin{equation}
    E[\rho]\propto|+_S\rangle\langle +_S|\otimes\mathbf{1}_W \,.
\end{equation}
This state has a weak $\mathbb{Z}_2$ symmetry acting on the $W$ qubits and a strong $\mathbb{Z}_2$ symmetry acting on the $S$ qubits.

\subsubsection{Weak $S_3$ Symmetry}

Similarly, we can construct a weak invertible group-symmetry for any group $G$. It is useful to illustrate this for $G=S_3$, the simplest non-abelian group. We start with the symmetry boundary that realizes $S_3\times S_3$
\be
\cL_{S_3 \times S_3} = (1\oplus P\oplus E)_L\otimes (1\oplus P \oplus E)_R \,.
\ee
We do a partial anyon condensation with the algebra
\be
\cA_{\cW=S_3}^\weak = 1 \oplus P_L P_R \oplus E_L E_R \,,
\ee
which gives rise to an $S_3$ reduced TO $\cZ(\cW')= \cZ(S_3)$, with the map of anyons given by the club quiche 
\be\label{ClubSandoS3}
\begin{tikzpicture}
\begin{scope}[shift={(0,0)}]
\draw [cyan,  fill=cyan] 
(0,0) -- (0,3) --(3,3) -- (3,0) -- (0,0) ; 
\draw [Green, fill=Green](3,0) -- (3,3) --(6,3) -- (6,0) -- (0,0) ; 
% \draw [very thick] (0,0) -- (0,3) ;
\draw [very thick] (3,3) -- (3,0) ;
%\draw [very thick] (6,3) -- (6,0) ;
\draw [thick] (0,2.5) -- (6,2.5) ;
\node[above] at  (1.5, 2.5) {$1+ P_L P_R + E_L E_R$} ; 
\node[above] at (4.5,2.5) {$ 1$} ; 
\draw [thick] (0,2) -- (6,2) ;
\node[above] at  (1.5, 2) {$P_L,\,  P_R$} ; 
\node[above] at (4.5,2) {$P $} ; 
\draw [thick] (0,1.5) -- (6,1.5) ;
\node[above] at  (1.5, 1.5) {$E_L\,,\ E_R$} ; 
\node[above] at (4.5,1.5) {$E$} ; 
\draw [thick] (0,1) -- (6,1) ;
\node[above] at  (1.5, 1) {$ a_L a_R$} ; 
\node[above] at (4.5,1) {$a$} ; 
\draw [thick] (0,0.5) -- (6,0.5) ;
\node[above] at  (1.5, 0.5) {$ b_L b_R$} ; 
\node[above] at (4.5,0.5) {$b$} ; 
% \node[above] at (0,3) {$\Bsym $}; 
\node[above] at (3,3) {$\cA_{\cW}^\weak$}; 
%\node[above] at (6,3) {$\Bphys ={\cL}_{\phys} $}; 
\end{scope}
\end{tikzpicture}
\ee
The weak symmetry, is generated by
\be
a_La_R \sim a \,,\quad b_L b_R \sim b \,,
\ee
and acts on the order parameters $P$ and $E$.

\subsection{SymTFT for Weak Duality and \\
Strong Invertible Symmetry}\label{sec:WeakDStrongZ}

We will now discuss a class of examples where the weak symmetry is non-invertible. Consider Tambara-Yamagami categories associated to an abelian group $\mathbb{A}$ \cite{TambaYama}.  The structure of these fusion categories is that there is one non-invertible duality $D$ generator 
\be
D\otimes D  = \oplus_{a\in \mathbb{A}}a  \,.
\ee
The remaining generators $a$ fuse according to the abelian group $\mbA$ and $D a =a D = D$ for all $a$.
The following analysis will produce phases with 
\be
\ba
D &\quad \text{weak} \cr 
a&\quad \text{strong}\,.
\ea
\ee
For a lightning summary of $\TY$ categories and their centers see Appendix \ref{app:TambaYama}. 
We start with the simplest example of this type in this section. 
Let the abelian group be $\mathbb{Z}_2$. In this case, there are two $\TY(\mathbb{Z}_2)$ fusion categories. We will choose the one given by the $\Ising$ fusion category. In this case, 
\be
\cZ(\Ising_L \boxtimes  \Ising_R ) = \cZ(\Ising_L)\boxtimes  \cZ(\Ising_R) \,,
\ee
where $\cZ(\Ising_L)$ is doubled Ising topological order. On the symmetry boundary we choose the Lagrangian that gives rise to $\Ising\boxtimes \Ising$ 
\be
  \mathcal{L}_{\sym}^{\strong}=(1+\psi_L\bar{\psi}_L+\sigma_L\bar{\sigma}_L)(1+\psi_R\bar{\psi}_R+\sigma_R\bar{\sigma}_R) \,.
\ee
We denote the anyons of $\cZ(\Ising)$ by $1, \psi, \sigma$ and $\ol{\psi}, \ol{\sigma}$ and products thereof. Here, $1, \psi, \sigma$ fuse according to the fusion rules of the Ising fusion category with the non-invertible fusion $\sigma^2= 1 \oplus \psi$, $\psi^2=1$, and $\psi\sigma=\sigma\psi =\sigma$. 

We can determine all the mixed condensable algebras of charges, which are listed in Table \ref{tab:Ising_CondAlgs} and find the following symmetries after condensation:
\begin{itemize}
    \item $\mathcal{A}=1$: strong $\TY({\mathbb{Z}_2})$ symmetry
    \item $\mathcal{A}^{\sw}=1\oplus\psi_L\bar{\psi}_L\psi_R\bar{\psi}_R$: strong $\mathbb{Z}_2$ and weak duality
    \item $\mathcal{A}=1\oplus\psi_L\bar{\psi}_L\oplus\psi_R\bar{\psi}_R\oplus\psi_L\bar{\psi}_L\psi_R\bar{\psi}_R$: strong $\mathbb{Z}_2$
    \item $\mathcal{A}=1\oplus\psi_L\bar{\psi}_L\oplus\psi_R\bar{\psi}_R\oplus\psi_L\bar{\psi}_L\psi_R\bar{\psi}_R\oplus\sigma_L\bar{\sigma}_L\sigma_R\bar{\sigma}_R$: weak $\mathbb{Z}_2$
    \item $\mathcal{A}=(1\oplus\psi_L\bar{\psi}_L\oplus\sigma_L\bar{\sigma}_L)(1\oplus\psi_R\bar{\psi}_R\oplus\sigma_R\bar{\sigma}_R)$: no symmetry
\end{itemize}
The most interesting condensation is given by the $\mathcal{A}^{\mathrm{sw}}$, which is an algebra composed of diagonal charges. We will now study the symmetry associated with the reduced TO obtained from this condensation in detail.

\subsubsection{Weak Kramers-Wannier and Strong $\Z_2$}\label{weakkwz2}

We condense the dimension 2 algebra 
\be
\cA^{\sw} = 1 \oplus \psi_L \ol{\psi}_L  \psi_R \ol{\psi}_R \,.
\ee
Note that $\psi\ol{\psi}$ is simply  the dual $\Z_2$ anyon that we obtain after gauging the em-duality in the toric code to get the $\cZ(\Ising)$ topological order. The key observation is that each $\sigma_{L/R}$ (and barred) anyon will braid non-trivially with the anyon in the algebra and therefore will be confined. 
We find that the reduced TO from the condensation is $\cZ(\Rep(H_8))$, where $\Rep(H_8)$ is $\TY(\mathbb{Z}_2\times\mathbb{Z}_2)$ with a particular choice of $F$ symbol. In particular, we obtain the following map of anyons between $\cZ(\Ising)\boxtimes\cZ(\Ising)$ and $\cZ(\Rep(H_8))$ (see Appendix~\ref{app:RepH8} for a review of the notation we use).
The abelian anyons $\psi_{L/R}$ and barred, all pass through the interface and are identified with:
\be\label{ClubSandoIsing}
\begin{tikzpicture}
\begin{scope}[shift={(0,0)}]
\draw [cyan,  fill=cyan] 
(0,-1.5) -- (0,4.2) --(3,4.2) -- (3,-1.5) -- (0,-1.5) ; 
\draw [Green, fill=Green]
(3,-1.5) -- (3,4.2) --(6,4.2) -- (6,-1.5) -- (0,-1.5) ; 
% \draw [very thick] (0,-0.5) -- (0,3) ;
\draw [very thick] (3,4.2) -- (3,-1.5) ;
%\draw [very thick] (6,3) -- (6,0) ;
\draw [thick] (0,2.5) -- (6,2.5) ;
\node[below] at (1.5,3.8) {$\cZ (\Ising_L \boxtimes  \Ising_R)$};
\node[below] at (4.5,3.8) {$\cZ (\TY (\Z_2^L\times  \Z_2^R))$};
\node[above] at (3,4.2) {$\cA^\sw$}; 
\node[above] at  (1.5, 2.5) {$1+ \psi_L \ol{\psi}_L {\psi}_R \ol{\psi_R}$} ; 
\node[above] at (4.5,2.5) {$ 1$} ; 
\draw [thick] (0,2) -- (6,2) ;
\node[above] at  (1.5, 1.95) {$\psi_L$} ; 
\node[above] at (4.5,2) {$ X_{a_L, i}$} ; 
\draw [thick] (0,1.5) -- (6,1.5) ;
\node[above] at  (1.5, 1.45) {$ \psi_R$} ; 
\node[above] at (4.5,1.5) {$X_{a_R,i}$} ; 
\draw [thick] (0,1) -- (6,1) ;
\node[above] at  (1.5, 0.95) {$ \ol{\psi_L}$} ; 
\node[above] at (4.5,1) {$X_{a_L, -i}$} ; 
\draw [thick] (0,0.5) -- (6,0.5) ;
\node[above] at  (1.5, 0.45) {$\ol{\psi_R}$} ; 
\node[above] at (4.5,0.5) {$X_{a_R,-i}$} ; 
\draw [thick] (0,0) -- (6,0) ;
\node[above] at  (1.5, -0.05) {$ \psi_L\ol{\psi_L}$} ; 
\node[above] at (4.5,0) {$X_{1, -1}$} ; 
\draw [thick] (0,-0.5) -- (6,-0.5) ;
\node[above] at  (1.5, -0.55) {$  \psi_L {\psi_R}$} ; 
\node[above] at (4.5,-0.5) {$X_{a_La_R, -1}$} ; 
\draw [thick] (0,-1) -- (6,-1) ;
\node[above] at  (1.5, -1.05) {$  \psi_L\ol{\psi_R}$} ; 
\node[above] at (4.5,-1) {$X_{a_La_R, 1}$} ; 
\end{scope}
\end{tikzpicture}
\ee
These are all the dim 1 anyons in the reduced SymTFT.
{Note that anyons related by fusion with the condensed anyon $\psi_L \ol{\psi}_L {\psi}_R \ol{\psi_R}$ are identified in the reduced TO, so for example $\psi_R\ol{\psi_R}$ is also mapped to $X_{1,-1}$.}

Meanwhile, the map for the non-invertible anyons is:
\be\label{ClubSandoIsing}
\begin{tikzpicture}
\begin{scope}[shift={(0,0)}]
\draw [cyan,  fill=cyan] 
(0,-2.5) -- (0,4) --(3,4) -- (3,-2.5) -- (0,-2.5) ; 
\draw [Green, fill=Green]
(3,-2.5) -- (3,4) --(6,4) -- (6,-2.5) -- (0,-2.5) ; 
% \draw [very thick] (0,-0.5) -- (0,3) ;
\draw [very thick] (3,4) -- (3,-2.5) ;
%\draw [very thick] (6,3) -- (6,0) ;
\draw [thick] (0,2.5) -- (6,2.5) ;
\node[below] at (1.5,3.8) {$\cZ (\Ising_L \boxtimes  \Ising_R)$};
\node[below] at (4.5,3.8) {$\cZ (\TY (\Z_2^L\times  \Z_2^R))$};
\node[above] at (3,4) {$\cA^\sw$}; 
\node[above] at  (1.5, 2.5) {$1+ \psi_L \ol{\psi}_L {\psi}_R \ol{\psi_R}$} ; 
\node[above] at (4.5,2.5) {$ 1$} ; 
\draw [thick] (0,2) -- (6,2) ;
\node[above] at  (1.5, 2) {$\sigma_L \sigma_R$} ; 
\node[above] at (4.5,2) {$ Z_{\rho_1, \zeta_8} $} ; 
\draw [thick] (0,1.5) -- (6,1.5) ;
\node[above] at  (1.5, 1.5) {$ \ol{\sigma_L} \sigma_R$} ; 
\node[above] at (4.5,1.5) {$Z_{\rho_3, 1}$} ; 
\draw [thick] (0,1) -- (6,1) ;
\node[above] at  (1.5, 1) {$ {\sigma_L} \ol{\sigma_R}$} ; 
\node[above] at (4.5,1) {$Z_{\rho_4,1}$} ; 
\draw [thick] (0,0.5) -- (6,0.5) ;
\node[above] at  (1.5, 0.5) {$ \ol{\sigma_L} \ol{\sigma_R}$} ; 
\node[above] at (4.5,0.5) {$Z_{\rho_2,\zeta_8^7}$} ; 
\draw [thick] (0,0) -- (6,0) ;
\node[above] at  (1.5, 0) {$ {\sigma_L} \ol{\sigma_L}$} ; 
\node[above] at (4.5,0) {$Y_{1, a_L}$} ; 
\draw [thick] (0,-0.5) -- (6,-0.5) ;
\node[above] at  (1.5, -0.5) {$ {\sigma_R} \ol{\sigma_R}$} ; 
\node[above] at (4.5,-0.5) {$Y_{1,a_R}$} ; 
\draw [thick] (0,-1) -- (6,-1) ;
\node[above] at  (1.5, -1) {$ {\sigma_L} \ol{\sigma_L}{\sigma_R} \ol{\sigma_R}$} ; 
\node[above] at (4.5,-1) {$Y_{1,a_La_R} + Y_{a_L,a_R}$} ; 
\end{scope}
\begin{scope}[shift={(0,-1.5)}]
\draw [thick] (0,0) -- (2.9,0) -- (2.9, -0.8);
\node[above] at  (1.5, 0) {$ \sigma_L, \, \ol{\sigma_L},  \, \sigma_R,\,  \ol{\sigma_R}$} ; 
\end{scope}
\end{tikzpicture}
\ee
On the right hand side we adopted the standard notation for the anyons in the center $\cZ (\TY(\mathbb{Z}_2\times\mathbb{Z}_2))$ {related to the physical picture of gauging a $\mathbb{Z}_2$ anyon permuting duality symmetry in $\mathbb{Z}_2\times\mathbb{Z}_2$ gauge theory. In particular, $Z_{\rho, s}$ are dimension 2 anyons of spin $s$ related to duality defects, and $Y_{m,n}$ are dimension 2 anyons that arise from duality invariant linear combinations of anyons in $\mathbb{Z}_2\times\mathbb{Z}_2$ gauge theory. We review this notation in Appendix \ref{app:TambaYama}}. Note that $Y_{1, a_L} \otimes Y_{1, a_R} = Y_{a_L, a_R} \oplus Y_{1, a_L a_R}$, so that $\sigma_L \ol{\sigma_L} \sigma_R \ol{\sigma_R}$ maps to $Y_{a_L, a_R} \oplus Y_{1, a_L a_R}$ in $\TY (\Z_2\times \Z_2)$. Taking products then determines the map on the remaining anyons. 
This anyon fusion and braiding match that of $\cZ (\TY (\Z_2 \times \Z_2))$ where $\TY (\Z_2 \times \Z_2)$ has an $F$ symbol given by the diagonal bi-character and trivial Frobenius-Schur indicator, i.e. $\Rep(H_8)$ (see Appendix~\ref{app:RepH8} for more details). The symmetry boundary is now given by
\be\label{algebrapsicond}
\cL_{\Rep(H_8)}\equiv
1\oplus X_{1, -1} \oplus Y_{1, a_L} \oplus Y_{1, a_R} \oplus Y_{1, a_La_R}   \,.
\ee

The strong $\Z_2$, realized by $\Z_2^L\times \Z_2^R$ with generators $P_L$ and $P_R$, come from projecting $X_{a_L, i}$ and $X_{a_R, -i}$ with Neumann boundary conditions onto the symmetry boundary, and the weak duality is generated by the projection of $Z_{\rho_4, 1}= \sigma_L\ol{\sigma_R}$ onto the symmetry boundary. In summary the symmetry generators for the $\TY (\Z_2\times \Z_2)$ arise from projecting the anyons 
\be\label{RepH8SymGens}
X_{a_L, i} \to P_L \,, \ X_{a_R, -i} \to P_R\,,\  Z_{\rho_4, 1}  \to D \,,
\ee
with the fusion 
\be
D^2 = 1 \oplus P_L \oplus P_R \oplus P_L P_R \,.
\ee

We obtain a symmetry that has weak duality symmetry $D$, and strong $\Z_2$. These combine into the fusion category $\Rep (H_8) = \TY (\Z_2\times \Z_2)$, where the duality defect acts weakly and the $\Z_2^L\times \Z_2^R$ realize the left and right actions of a strong $\Z_2$ symmetry. We should therefore call this an Ising symmetric setup, where the duality is weak and the $\Z_2$ is strong.

\subsubsection{Gapped Mixed Phases with Weak Duality and Strong $\Z_2$}
Now that we have the reduced TO and $\mathfrak{B}_{\mathrm{sym}}$, we can consider the possible choices of $\mathfrak{B}_{\mathrm{phys}}$ and the meaning of the corresponding phases. The condensable algebras for $\cZ (\Rep (H_8))$ were determined in \cite{GaiSchaferNamekiWarman}. There are 
 six Lagrangian condensable algebras, in the notation of $\cZ (\TY (\Z_2^L\times \Z_2^R))$: 
\be\ba \label{H8Lags}
\cL_1&=1 \oplus X_{1,-1} \oplus Y_{1,a_L} \oplus Y_{1,a_R} \oplus Y_{1,a_La_R}\cr 
\cL_2&=1 \oplus X_{1,-1} \oplus Y_{1,a_L} \oplus Y_{1,a_R} \oplus Y_{a_L,a_R}\cr
\cL_3&=1 \oplus X_{1,-1}\oplus X_{a_La_R,1}  \oplus X_{a_La_R,-1} \oplus 2Y_{a_L,a_R}\cr 
\cL_4&=1\oplus X_{1,-1} \oplus X_{a_La_R,1}  \oplus X_{a_La_R,-1} \oplus 2Y_{1,ab}\cr 
\cL_5&=1 \oplus X_{a_La_R,1} \oplus Y_{1,a_La_R} \oplus Z_{\rho_3,1} \oplus Z_{\rho_4,1}\cr 
\cL_6&=1 \oplus X_{a_La_R,1} \oplus Y_{a_L,a_R} \oplus Z_{\rho_3,1} \oplus Z_{\rho_4,1}\,.
\ea\ee
where $\mathcal{L}_1=\mathcal{L}_{\Rep(H_8)}$ specifies $\mathfrak{B}_{\mathrm{sym}}$.

The viable Lagrangian algebras that will give rise to gapped phases with weak duality and strong $\Z_2$ symmetry need to be $T$ symmetric where the $T$ symmetry is inherited from $\cZ(\Ising)\boxtimes\cZ(\Ising)$ (since the condensation of $\mathcal{A}$ is $T$ symmetric). They must also obey constraints implied by positivity. All  six Lagrangians in (\ref{H8Lags}) satisfy these conditions.

 Note that $\mathcal{L}_1-\mathcal{L}_6$ come from the three  $\cZ (\Ising \boxtimes \Ising)$ mixed state Lagrangian algebras in  (\ref{LIsingIsing}), using the map that the interface $\cA^\sw$ specifies.
We will now study in detail the different phases given by $\mathcal{L}_1-\mathcal{L}_6$. $\mathcal{L}_1-\mathcal{L}_5$ all demonstrate (partial) symmetry breaking, while $\mathcal{L}_6$ is an SPT. As mentioned in Sec.~\ref{multiplet}, spontaneous symmetry breaking leads to locally indistinguishable density matrices in addition to the density matrix that is a singlet under all of the symmetries. These other states are obtained by acting with charged operators given by anyon lines that tunnel between $\mathfrak{B}_{\mathrm{sym}}$ and $\mathfrak{B}_{\mathrm{phys}}$. These anyon lines are constrained to be $T$ symmetric by hermiticity of the density matrix, so we will bold-face the anyon lines that are $T$ symmetric.

\vspace{2mm}
\noindent{\bf Phase with $\cL_\phys= \cL_{1}= \cL_{\sym}$.}
The most obvious phase is the complete SSB phase, where the physical boundary is the same as the symmetry boundary. This choice of $\mathcal{L}_{\mathrm{phys}}$ always leads to complete SSB of the symmetry. In this case we have five operators tunneling between $\mathfrak{B}_{\mathrm{sym}}$ and $\mathfrak{B}_{\mathrm{phys}}$, but only three of them are $T$ symmetric:
\be
\begin{tikzpicture}
\begin{scope}[shift={(0,0)}]
\draw [thick] (0,-1) -- (0,1) ;
\draw [thick] (5,-1) -- (5,1) ;
\draw [thick] (0, 0) -- (5, 0) ;
\node[above] at (2.5,0.1) {$\bm{1, X_{1, -1}}, Y_{1, a_L}, Y_{1, a_R}, \bm{Y_{1, a_La_R}}$} ;
\node[above] at (0,1) {$\cL_{\TY (\Z_2^L \times \Z_2^R)}$}; 
\node[above] at (5,1) {$\cL_{1}$}; 
\draw [black,fill=black] (5,0) ellipse (0.05 and 0.05);
\draw [black,fill=black] (0,0) ellipse (0.05 and 0.05);
\end{scope}
\end{tikzpicture} 
\ee
The pure state phase can be described as follows. Let us
denote the charged operators coming from the anyons that end on both boundaries by $\cO_-$, $\cO_{1, a_L}$, $\cO_{1, a_R}$, $\cO_{1, a_La_R}$. The  linear combinations of local operators that diagonalize the fusion algebra is \footnote{This is entirely analogous to the analysis for $\Rep (D_8)$ in \cite{Warman:2024lir}.} 
\be
\ba
v_1 & = 1+ \cO_{-} + \cO_{1, a_L}  + \cO_{1, a_R} +  \cO_{1, a_La_R}  \cr 
v_2 & = 1+ \cO_{-} - \cO_{1, a_L}  + \cO_{1, a_R} -  \cO_{1, a_La_R}  \cr 
v_3 & = 1+ \cO_{-} + \cO_{1, a_L}  - \cO_{1, a_R} -  \cO_{1, a_La_R}  \cr 
v_4 & = 1+ \cO_{-} - \cO_{1, a_L}  - \cO_{1, a_R} +  \cO_{1, a_La_R}  \cr 
v_5 & = 2 -2  \cO_{-}  \,. 
\ea
\ee
The symmetry generators (\ref{RepH8SymGens}) act on these as follows: the $\Z_2^{L/R}$ are obvious e.g. $\Z_2^L$ swaps $v_1 \leftrightarrow v_2$ and $v_3 \leftrightarrow v_4$, and $v_5$ is invariant. The duality symmetry acts as  
\be
\ba
D: \quad v_i \to v_5 \,,\ i=1, \cdots, 4\,, \quad 
v_5 \to \sum_{i=1}^4 v_i \,.
\ea
\ee

The anyon $Y_{1, a_{L/R}}$  braids non-trivially with $X_{a_{L/R}, i}$ and $X_{1,-1}$ braids nontrivially with $Z_{\rho_4,1}$ (this can be seen from the mapping to $\cZ(\Ising)\boxtimes\cZ(\Ising)$), causing the strong $\mathbb{Z}_2$ symmetry and weak duality to be completely broken. 
Although  $Y_{1, a_L}, Y_{1, a_R}$ give rise to charged operators, they are not $T$ symmetric and thus they do not give rise to new density matrices. 
The two charged operators that give rise to genuinely new density matrices come from {$1$}, $X_{1, -1}$ and $Y_{1, a_L a_R}$.
This is the {\bf weak duality and $\Z_2$ strong complete SSB} phase.

\vspace{2mm}
\noindent{\bf Phase with $\cL_\phys= \cL_2$.}  There are four local operators arising from the anyons, two of which are $T$ symmetric: 1 and $X_{1, -1}$:
\be
\begin{tikzpicture}
\begin{scope}[shift={(0,0)}]
\draw [thick] (0,-1) -- (0,1) ;
\draw [thick] (3.5,-1) -- (3.5,1) ;
\draw [thick] (0, 0) -- (3.5, 0) ;
\node[above] at (1.75,0.1) {$\bm{1, X_{1, -1}}, Y_{1, a_L}, Y_{1, a_R}$} ;
\node[above] at (0,1) {$\cL_{\TY (\Z_2^L \times \Z_2^R)}$}; 
\node[above] at (3.5,1) {$\cL_{2}$}; 
\draw [black,fill=black] (3.5,0) ellipse (0.05 and 0.05);
\draw [black,fill=black] (0,0) ellipse (0.05 and 0.05);
\end{scope}
\end{tikzpicture} 
\ee
The weak duality is again spontaneously broken by $X_{1,-1}$. $Y_{1, a_{L,R}}$ again breaks the $\Z_2$ strong symmetry, but like for $\mathcal{L}_1$, it does not give rise to new density matrices. 

This phase has a spontaneously broken weak duality and strong $\Z_2$. However it is distinct from the $\cL_1$ full SSB phase:
to see the precise structure consider the pure state phase specified by this SymTFT. There are now three charged operators $\cO_{-}, \cO_{1, a_L},  \cO_{1, a_R}$ coming from the anyons that end on both boundaries rather than four, and we can determine the linear combinations of local operators that diagonalize the fusion algebra:
\be
\ba
v_1 & = 1+ \cO_{-} + \cO_{1, a_L}  + \cO_{1, a_R} \cr 
v_2 & = 1+ \cO_{-} - \cO_{1, a_L}  + \cO_{1, a_R}   \cr 
v_3 & = 1+ \cO_{-} + \cO_{1, a_L}  - \cO_{1, a_R}  \cr 
v_4 & = 2 -2  \cO_{-}  \,. 
\ea
\ee
The symmetry $P_L$ and $P_R$ act again in the obvious way, but $P_LP_R v_1 = -v_1 + v_2+ v_3$. The duality acts as 
\be
Dv_i = v_4 \,,\quad i=1, 2, 3 \,,\ 
D v_4 = 2 (v_2 + v_3) \,.
\ee
Note that the breaking pattern is different from the full SSB phase. Without the $T$ symmetric condition, there would be four states in the multiplet rather than five. Roughly speaking, in this case, there is a ``fractional" breaking of the non-invertible symmetry. 

In summary the phase $\cL_{2}$ is a 
{\bf weak duality SSB, strong $\Z_2$ SSB.}

\vspace{2mm}
\noindent{\bf Phase with $\cL_\phys=\cL_3$.} This phase has one non-trivial charged operator comes from the $X_{1, -1}$ anyon. 
\be
\begin{tikzpicture}
\begin{scope}[shift={(0,0)}]
\draw [thick] (0,-1) -- (0,1) ;
\draw [thick] (3,-1) -- (3,1) ;
\draw [thick] (0, 0) -- (3, 0) ;
\node[above] at (1.5,0.1) {$\bm{1, X_{1, -1}}$} ;
\node[above] at (0,1) {$\cL_{\TY (\Z_2^L \times \Z_2^R)}$}; 
\node[above] at (3,1) {$\cL_{3}$}; 
\draw [black,fill=black] (3,0) ellipse (0.05 and 0.05);
\draw [black,fill=black] (0,0) ellipse (0.05 and 0.05);
\end{scope}
\end{tikzpicture} 
\ee
The charged operator is $\cO_-$ and the linear combinations of local operators that diagonalize the fusion algebra is 
\be
v_1= 1+ \cO_-\,, \quad v_2 = 1- \cO_-
\ee
with $D v_1= 2 v_2$ and $D v_2= 2 v_1$ and the invertible symmetries acting trivially. 

This is simply the {\bf SSB phase for the weak duality symmetry}. 

\vspace{2mm}
\noindent{\bf Phase with $\cL_\phys=\cL_4$.} The charged operators that arise from the ends of anyons $ X_{1,-1}$ and $2 Y_{1, a_La_R}$  are all $T$ invariant: 
\be
\begin{tikzpicture}
\begin{scope}[shift={(0,0)}]
\draw [thick] (0,-1) -- (0,1) ;
\draw [thick] (4,-1) -- (4,1) ;
\draw [thick] (0, 0) -- (4, 0) ;
\node[above] at (2,0.1) {$\bm{1, X_{1, -1}, 2 Y_{1, a_L a_R}}$} ;
\node[above] at (0,1) {$\cL_{\TY (\Z_2^L \times \Z_2^R)}$}; 
\node[above] at (4,1) {$\cL_{4}$}; 
\draw [black,fill=black] (4,0) ellipse (0.05 and 0.05);
\draw [black,fill=black] (0,0) ellipse (0.05 and 0.05);
\end{scope}
\end{tikzpicture} 
\ee
Again, lets consider the pure state ground state structure again: the charged operators are $\cO_-$ from the anyon $X_{1, -1}$, and two operators $\cO_{1, a_L a_R}^{(1)}$ and $\cO_{1,a_L a_R}^{(2)}$ from $2 Y_{1, a_La_R}$. The  linear combinations of local operators that diagonalize the fusion algebra is
\be
\ba
v_1 & =  1+  \cO_{1, a_L a_R}^{(1)} + \cO_{1, a_L a_R}^{(2)} + \cO_- \cr 
v_2 & =  1+ i\, \cO_{1, a_L a_R}^{(1)} -i \, \cO_{1, a_L a_R}^{(2)} - \cO_- \cr 
v_3 & =  1- \cO_{1, a_L a_R}^{(1)} - \cO_{1, a_L a_R}^{(2)} + \cO_- \cr 
v_4 & =  1 -i  \,\cO_{1, a_L a_R}^{(1)} + i\,\cO_{1, a_L a_R}^{(2)} +-\cO_- \,.
\ea
\ee
The symmetry action is 
\be
\ba
P_L v_p &= P_R v_p = v_{p+2} \,,\quad p \in \Z_4 \cr 
D v_1 &= D v_3 = v_2 + v_4 \cr 
D v_2 &= D v_4 = v_1 + v_3 \,.
\ea
\ee
and represents correctly the fusion. 

This phase breaks the weak duality symmetry spontaneously.  Furthermore, it breaks the strong $\Z_2$ to a weak one as $Y_{1, a_La_R}$ is charged under both $\Z_2^L$ and $\Z_2^R$, but not under the diagonal, weak symmetry. In summary we have 
{\bf SSB for the weak duality, and SWSSB for the strong $\Z_2$. }

\vspace{2mm}
\noindent{\bf Phase with $\cL_\phys=\cL_5$.} The non-trivial charged operator comes from the anyon $Y_{1, a_La_R}$:
\be
\begin{tikzpicture}
\begin{scope}[shift={(0,0)}]
\draw [thick] (0,-1) -- (0,1) ;
\draw [thick] (3,-1) -- (3,1) ;
\draw [thick] (0, 0) -- (3, 0) ;
\node[above] at (1.5,0.1) {$\bm{1, Y_{1, a_La_R}}$} ;
\node[above] at (0,1) {$\cL_{\TY (\Z_2^L \times \Z_2^R)}$}; 
\node[above] at (3,1) {$\cL_{5}$}; 
\draw [black,fill=black] (3,0) ellipse (0.05 and 0.05);
\draw [black,fill=black] (0,0) ellipse (0.05 and 0.05);
\end{scope}
\end{tikzpicture} 
\ee
This operator $\cO_{1, a_La_R}$ is charged  $\Z_2^L$ and $\Z_2^R$, but is uncharged under the diagonal. Thus the strong $\Z_2$ is broken to a weak.
However, this means that the weak duality symmetry has to also be partially broken. 
The  linear combinations of local operators that diagonalize the fusion algebra is
\be
v_1 = 1+ \cO_{1, a_La_R} \,,\quad v_2 = 1- \cO_{1, a_La_R}
\ee
with the symmetry acting as $P_LP_R: \  v_1 \leftrightarrow v_2$ and 
\be
D v_i = v_1+ v_2 \,,\qquad i=1, 2 \,.
\ee
This satisfies again the fusion rules. 
But the duality is only partially  broken, as it maps $v_1$ both to itself and to $v_2$ etc. 
So we have a {\bf SWSSB for the strong $\Z_2$, and duality SSB.} 

\vspace{2mm}
\noindent{\bf Phase with $\cL_\phys=\cL_6$.} {\bf Trivial  phase}.  There are no local charged operators.  There are twisted sector operators, which will give rise to string-order parameters. In the present case the lines in $\cL_5$ becomes all symmetry lines:
\be
\begin{tikzpicture}
\begin{scope}[shift={(0,0)}]
\draw [thick] (0,-1) -- (0,1) ;
\draw [thick] (5,-1) -- (5,1) ;
\draw [very thick, orange] (0,-1) -- (0, 0) -- (5, 0) ;
\node[above] at (2.5,0.1) {$X_{1, a_La_R} \,, Y_{a_L, a_R} Z_{\rho_3,1}\,, Z_{\rho_4, 1} $} ;
\node[above] at (0,1) {$\cL_{\TY (\Z_2^L \times \Z_2^R)}$}; 
\node[above] at (5,1) {$\cL_{6}$}; 
\draw [black,fill=black] (5,0) ellipse (0.05 and 0.05);
%\draw [orange,fill=orange] (0,0) ellipse (0.05 and 0.05);
\end{scope}
\end{tikzpicture} 
\ee
This phase is an SPT of the symmetry, with {\bf no SSB}.

\subsubsection{Lattice model: weak duality and strong $\Z_2$}
\label{strz2weakd}

For the mixed states obtained from $\mathcal{L}_2$ and $\mathcal{L}_5$, we can work in a simpler setting consisting of a 1D chain of qubits with strong $\mathbb{Z}_2$ symmetry action $\prod_iX_i$. The fusion rules of the duality operators in this setting are slightly modified. In particular, the fusion rules of the strong duality operator includes translations \cite{seiberg2024}. In this setting, the Hilbert of the density matrix is $\mathcal{H}=\bigotimes_i\mathbb{C}_i^2$ and we consider density matrices satisfying

\begin{equation}
    U\rho = \pm \rho\qquad D\rho D^\dagger =2\rho\,.
\end{equation}
where $U=\prod_iX_i$ and $D$ can be written explicitly as\cite{seiberg2024}
\begin{equation}
D=e^{-2\pi i N/8}(d_1^xd_1^z)\cdots (d_1^xd_1^z)d_N^xP^+
\end{equation}
on a chain of $N$ qubits, where
\begin{equation}
    d_i^x=e^{i\pi X_i/4}\qquad d_i^z=e^{2\pi i Z_iZ_{i+1}/4}
\end{equation}
One can check that $DX_i=Z_iZ_{i+1}D$ and $DZ_iZ_{i+1}=X_{i+1}D$ for all $i=1,2,\dots N$. The fusion rules are
\begin{equation}
    D^2=P^+S\qquad U^2=1\qquad DU=UD=D
\end{equation}
where $S$ is a unit translation (shift) operator and $P^+=\frac{1+U}{2}$ is the projector onto the $\mathbb{Z}_2$ even sector, as in previous sections. In addition, $D^\dagger = DS^{-1}$. In the doubled Hilbert space, we have the symmetry operators $U_L\otimes\mathbf{1},\mathbf{1}\otimes U_R, $ and $D\otimes \overline{D}$. $\overline{D}$ has the same fusion properties as $D$: $\overline{D}X_i=Z_iZ_{i+1}\overline{D}$ and $\overline{D}Z_iZ_{i+1}=X_{i+1}\overline{D}$. One can also check by explicit calculation that $D\overline{D}=P^+S$. In the doubled Hilbert space, these operators form a $\Rep(H_8)$ fusion category (modulo the mixing with translation) \cite{seifnashri2024}.\footnote{Note that there is a very similar diagonal duality operator in the $\Rep(D_8)$ fusion category but in this lattice implementation that operator has an additional translation operator and takes the form $\tilde{S}D\otimes D$ where $S$ shifts qubits in the doubled Hilbert space in a ``snaking" way between $\mathcal{H}_L$ and $\mathcal{H}_R$. In other words, $\tilde{S}$ is a unit translation if we make the qubits in $\mathcal{H}_L$ the even sites and the qubits in $\mathcal{H}_R$ the odd sites of a single qubit chain. $\tilde{S}$ makes the duality operator of $\Rep(H_8)$ fail to be $T$ invariant  \cite{seifnashri2024}.}

$\mathcal{L}_2$ describes a mixed state phase in which the weak duality is SSB'd and the strong $\mathbb{Z}_2$ is SSB'd to nothing. A density matrix demonstrating this can be written as
\begin{align}
\begin{split}
    \rho&=\frac{1}{2^{N+1}}\left[\prod_i(1+X_i)+(1+\prod_iX_i)\prod_i(1+Z_iZ_{i+1})\right]\\
    &=\frac{1}{2}(|+\rangle\langle+|+|\mathrm{GHZ}+\rangle\langle \mathrm{GHZ}+|)
\end{split}
\end{align}
This mixed state has long range linear order parameter: $\mathrm{Tr}(\rho Z_iZ_j)\neq 0$. Therefore, the strong $\mathbb{Z}_2$ symmetry is broken, the weak $\mathbb{Z}_2$ symmetry is also broken, and the weak duality is also broken; all of the symmetries are SSB.\footnote{One can try to get a similar state with weak duality and weak $\mathbb{Z}_2$ by using $\rho\sim |+\rangle\langle +|+\frac{1}{2}(|\uparrow\rangle\langle\uparrow|+|\downarrow\rangle\langle\downarrow|)$ but this won't quite work due to the non-invertible fusion rules of $D$, which guarantee a strong $\mathbb{Z}_2$ symmetry.}

For $\mathcal{L}_5$, one can actually use the density matrix demonstrating SWSSB of $\mathbb{Z}_2$ symmetry. We can get this state by starting with the strong-to-trivial SSB pure state $\frac{1}{\sqrt{2}}(|+\rangle+|\mathrm{GHZ}+\rangle)$, which has a strong Ising symmetry, and explicitly breaking the strong duality to a weak duality symmetry.
We do this using a sequence of two commuting quantum channels:\footnote{Interestingly, in this tensor product Hilbert space, it seems not possible to construct a local channel (besides a unitary one) invariant under strong $\TY(\mathbb{Z}_2\times\mathbb{Z}_2)$. This is because the duality symmetry mixes with translation, so any Krauss operator $K_i$ must also come with its translated versions. As a result, any duality invariant Krauss operator appears to require support along the entire system. }
\begin{equation}
    E[\rho]=(E_{ZZ}\circ E_X)[\rho] \,,
\end{equation}
where 
\begin{equation}
\begin{aligned}
    E_{ZZ/X}[\rho]&=\bigotimes_iE_{i,ZZ/X}[\rho]\ \\
    E_{i,ZZ}[\rho]&=\frac{1}{2}\rho +\frac{1}{2}Z_{i}Z_{i+1}\rho Z_{i}Z_{i+1}\\
    E_{i,X}[\rho]&=\frac{1}{2}\rho +\frac{1}{2}X_{i}\rho X_{i}\,.
\end{aligned}
\end{equation}
We need both channels in order to preserve the weak duality symmetry; otherwise, we would explicitly break the strong duality symmetry completely. It is not hard to see that
\begin{align}
\begin{split}
    &E\left[\frac{1}{2}(|+\rangle+|\mathrm{GHZ}+\rangle)(\langle\mathrm{GHZ}+|+\langle +|)\right]\\
    &=\frac{1}{2^{N-1}}P^+=\rho_{wd}^+\,.
\end{split}
\end{align}
This density matrix clearly has a strong $\mathbb{Z}_2$ symmetry. It does not have a strong duality symmetry because $D\rho_{wd}^+\neq c\rho_{wd}^+$ for any scalar $c$. However, there is a weak duality symmetry:
\begin{equation}
D\rho_{wd}^+D^\dagger=2\rho_{wd}^+\,.
\end{equation}

There are two density matrices locally indistinguishable under operators that are symmetric under the strong $\mathbb{Z}_2$ symmetry and weak duality symmetry. Indeed, in addition to $\rho_{wd}^+$, we have $\frac{1}{2^N}\left(1-\prod_iX_i\right)=\rho_{wd}^{-}$. Notice that $\rho_{wd}^-$ and $\rho_{wd}^+$ not only have different eigenvalues under the strong $\mathbb{Z}_2$ symmetry, but they also have different eigenvalues (2 vs 0) under the weak duality symmetry:
\begin{align}
\begin{split}
&\prod_iX_i\rho_{wd}^+=\rho_{wd}^+\qquad \prod_iX_i\rho_{wd}^-=-\rho_{wd}^-\\
   & D\rho_{wd}^+D^\dagger = 2\rho_{wd}^+\qquad D\rho_{wd}^-D^\dagger = 0\cdot \rho_{wd}^-\,.
\end{split}
\end{align}
Therefore, even though there are no linear order parameters, the weak duality symmetry is \emph{spontaneously broken}. This phenomenon of a weak symmetry getting spontaneously broken even in the absence of a linear order parameter is unique to non-invertible symmetries, where the weak symmetry can generate strong symmetry operators.

Interestingly, in this Hilbert space, there is {\bf no valid Choi state that is an SPT of the $\Rep(H_8)$ symmetry, even though it is anomaly-free} (indeed, $\mathcal{L}_6$ gives an SPT). This is because in this Hilbert space, any SPT of $\Rep(H_8)$ is not a trivial product state: it is a cluster state of the $\mathbb{Z}_2\times\mathbb{Z}_2$ symmetry \cite{seifnashri2024}. The cluster state is not the Choi state of any hermitian, positive density matrix \cite{Ma:2024kma}. Indeed, the cluster state would correspond to the Lagrangian algebra $1+e_Lm_R+e_Rm_L+e_Le_Rm_Lm_R$ of $\mathbb{Z}_2\times\mathbb{Z}_2$ gauge theory, which fails to satisfy (\ref{poscond}). However, in the larger Hilbert space obtained from the anyon-chain construction, one should be able to construct a valid Choi state that is an SPT of $\Rep(H_8)$, since $\mathcal{L}_6$ satisfies both (\ref{hermcond}) and (\ref{poscond}).

It is worth mentioning that this kind of symmetry pattern occurs naturally when considering the disordered Ising model \cite{fisher1992,fisher1995}
\begin{equation}
    H(\{h_i,J_{i,i+1}\})=-\sum_ih_iX_i-\sum_{i}J_{i,i+1}Z_iZ_{i+1}\,,
\end{equation}
where $\{h_i\},\{J_{i,i+1}\}$ are drawn from Gaussian distributions centered at $h_0$ and $J_0$ respectively. At $h_0=J_0$, it was shown in \cite{fisher1992,fisher1995} that the model flows to an ``infinite randomness fixed point," where several properties can be computed exactly. 

Any particular $H(\{h_i,J_{i,i+1}\})$ does not have a duality symmetry. However, the \emph{ensemble} of Hamiltonians does, for $\{h_i\}$, $\{J_{i,i+1}\}$ drawn from the same distribution. We can build a diagonal density matrix using the ensemble of ground states; every ground state is orthogonal to the others in the thermodynamic limit. This density matrix then has a strong $\mathbb{Z}_2$ symmetry and a weak duality symmetry.

\subsubsection{General Weak Duality Symmetry}
\label{sec:GeneralWeakDu}

We now provide a generalization of the story above to other abelian groups. Consider the $\TY (\mbA)$ categories, which generalize the $\Ising=\TY(\Z_2)$ to an arbitrary abelian group $\mbA$. 
We summarize the salient features of these in Appendix \ref{app:TambaYama}. 

$\TY(\mbA)$ categories are obtained from $\mbA$ by considering its quantum double, (i.e. the toric code for $\mbA=\Z_2$), and gauging a $\Z_2$ outer automorphism (i.e. the em-duality). This gives the center of $\TY (\mbA)$. In this construction we have various choices of additional data, which for any finite abelian group $\mathbb{A}$ consist of $\chi: \mathbb{A} \times \mathbb{A} \to U(1)$ a symmetric bicharacter and the Frobenius–Schur (FS) indicator indicator $\tau = \pm 1/{\sqrt{|\mathbb{A}|}}$.
The resulting category is $\TY (\mathbb{A}, \chi, \tau)$.
Our proposal is to start with $\mathbb{A}$ of order $|\mathbb{A}|=n$ and with the doubled TO 
\be\label{Donuts}
\cZ(\TY (\mbA)_L \boxtimes {\TY(\mbA)_R}) \cong \cZ(\TY (\mbA)_L ) \boxtimes \cZ({\TY(\mbA)_R})\,,
\ee
where we choose throughout this section the bi-character to satisfy $\chi_R=\ol{\chi_L}$ and be symmetric, and the FS indicator is trivial.
We will use the notation of \cite{Gelaki:2009blp} (see also \cite{TambaYama,  Kaidi:2022cpf, Antinucci:2023ezl}) and label the anyons of the theory above as $X_{a, \epsilon}$, $Y_{a,b}$ and $Z_{\rho, \Delta}$ (see Appendix~\ref{app:TambaYama}). In this notation, we define the symmetry boundary using the Lagrangian algebra\cite{Bhardwaj:2023idu})
\be\ba
 \mathcal{L}_{\sym}^\strong= 
 &\left(1\oplus X_{1, -}^L \oplus \bigoplus_{1\not=a_L\in \mbA} Y_{1, a_L}\right)\cr 
 \otimes &
 \left(1\oplus X_{1, -}^R \oplus \bigoplus_{1\not= a_R\in \mbA} Y_{1, a_R}\right) \,.
\ea\ee
The dimension is $(2n)^2$ as required, {since $X_{1,-}^{L/R}$ is abelian and every $Y_{1,a_L}/Y_{1,a_R}$ has quantum dimension two}.  

In $\cZ(\TY (\mbA)_L \boxtimes {\TY(\mbA)_R})$ we condense the following dimension $2$ algebra 
\be\label{AntinucciAlgebra}
\cA^\sw= 1\oplus  X^L_{1 , -}   
{X^R_{1 ,-}} \,.
\ee
The non-trivial anyon in this algebra  is simply the product of the dual $\Z_2$ lines for the L and R $\TY$ categories. This is a straight-froward extension of the  Ising example above.

The anyons $Z_{\rho, \Delta}^{L/R}$, braid non-trivially with this algebra, and thus confine. However products 
\be
Z_{\rho, \Delta}^{L} {Z_{\rho, \Delta}^{R}}
\ee
will braid trivially and become anyons in the reduced TO, of dimension $n$. These generate non-invertible symmetries, that in the physical theory act simultaneously from left and right, and thus are a {\bf weak non-invertible} symmetry. On the other hand the abelian anyons form a strong, group-like symmetry $\mbA_L \times \mbA_R$. This condensation therefore gives
\be
\cZ (\TY (\mbA_L \times \mbA_R))\,,
\ee
where $\TY(\mathbb{A}_L\times\mathbb{A}_R)$ only has the diagonal duality operator $D_L\otimes  D_R$ in $\TY(\mathbb{A}_L)\boxtimes\TY(\mathbb{A}_R)$. 

{The strong symmetry boundary condition $\cL_{\sym}^\strong$, becomes, after compactifying the interval occupied by $\cZ (\TY (\mbA_L) \boxtimes\TY (\mbA_R))$, a gapped boundary condition for the reduced TO:}
\be\ba\label{Marzipan}
\cL_{\sym}^\strong  \stackrel{\cA^{\sw}}{\longrightarrow} 
1\oplus &X_{1, -1} \cr 
\oplus &\bigoplus_{1\not= a_L \in \mathbb{A}} Y_{1,a_L} \oplus \bigoplus_{1\not= a_R \in \mathbb{A}} Y_{1,a_R} \cr 
\oplus &\bigoplus_{1\not= a_L  \in \mathbb{A}} \bigoplus_{1\not= a_R  \in \mathbb{A}} Y_{1,a_L a_R} 
\cr
&= \cL_{\TY (\mbA_L\times \mbA_R)}\,.
\ea\ee
This has to be of dimension  $ 1+ 1+ 2 ( n-1) + 2 (n-1)+ 2(n-1)^2 = 2n^2$ to be  a gapped boundary condition of $\cZ(\TY (\mbA_L\times \mbA_R))$. Also it is fully $T$-symmetric, because $\mathcal{L}_{\mathrm{sym}}^{\mathrm{strong}}$ and $\mathcal{A}^{\mathrm{sw}}$ is $T$ symmetric. This is precisely the symmetry boundary condition for $\TY (\mbA_L\times \mbA_R)$ (again, see \cite{Bhardwaj:2023idu} for the  gapped boundary condition that gives the TY category symmetry boundary). 

This setup allows now a classification of all gapped $\TY (\mbA)$ phases where 
duality is weak and $\mbA$ is strong. The gapped phases are determined again by studying all of the choices of $\mathfrak{B}_{\mathrm{phys}}$ with all gapped boundary conditions of $\cZ(\TY (\mbA_L\times \mbA_R))$. We expect the structure to be similar to the Ising case, and leave a more detailed analysis to future work.

\subsection{Mixed Strong-Weak Non-invertible Symmetry with SPT
} 
\label{sec:NonInvSPT}

Finally, we will consider a non-invertible weak duality symmetry and strong $\mbA=\Z_2\times\Z_2$, where one of the 
$\Z_2$s forms an SPT with the weak duality, while the other $\Z_2$ is SWSSB’ed.\footnote{For $\Rep(G)$ symmetry categories, as considered in this example, there is always a unique `trivial SPT' whose Lagrangian algebra is the $\Vec_G$ symmetry boundary of the SymTFT, i.e. $\cL_{\Vec_G}=\oplus_\Gamma([1],\Gamma)$ which contains only pure irreps. If there are other Lagrangian algebras that give rise to $\Rep(G)$ SPT phases these will contain non-trivial dyons, that correspond to symmetry strings attached to local operators charged under $\Rep(G)$: these are thus `non-trivial' $\Rep(G)$-SPTs.}
We start with the setup in (\ref{Donuts}) with  $\mbA= \Z_2\times\Z_2$ and an off-diagonal bi-character $\chi_L=\overline{\chi_R}=\chi$, given by
\be\ba
\chi (e_1, e_1)&= \chi (e_2,e_2)= \chi (e_1 e_2,e_1 e_2)=1\,,\\
\chi (e_1, e_2)&=-1 \,,
\ea\ee
where $e_1, e_2$ generate $\Z_2\times\Z_2$, and $\tau=+1/2$.\\
This gives the fusion category \cite{TambaYama}
\be
\TY (\Z_2\times\Z_2, \chi, +1/2)=\Rep(D_8)\,,
\ee
of representations of the group $D_8=(\Z_2^a \times \Z_2^b)\rtimes \Z_2^c$ of symmetries of a square.
The irreps are
\be \label{eq:D8_irreps}
   \Rep (D_8)=\{ 1,\;1_c,\;1_a,\;1_{ca},\;E \}\,,
\ee
with $\dim (1_k)=1$, $\dim (E)=2$. These irreps indeed satisfy the $\TY(\Z_2\times\Z_2)$ fusion rules (with $k=c,a, ca$):\vspace{-1mm}
\be\label{fusions}
\ba
    1_a\otimes1_c&=1_c\otimes1_a=1_{ca}\,,\qquad  1_k \otimes 1_k =1\,,\qquad     \\
    E\otimes 1_k&=1_k\otimes E=E\,,  \\
    E\otimes E&=1\oplus 1_a\oplus 1_c \oplus 1_{ca}\,.
\ea
\ee
We will use the group-theoretical notation for this category and its center. A lattice model for all $\Rep(D_8)$ pure-state phases was developed in \cite{Warman:2024lir}.\footnote{A complementary approach based on the cluster state \cite{seifnashri2024, Li:2024gwx} can only capture a subset of pure-state phases.}

The character table for $D_8$ is:
\be\label{eq:D8_chars}
\text{\begin{tabular}{|r|ccccc|}
    \hline
     & $[1]$ & $[ab]$ & $[ca]$ & $[c]$ & $[a]$ \\
     \hline
     1:\;& $1$ & $1$ & $1$ & $1$ & $1$ \\
     $1_c$:\;& $1$ & $1$ & $-1$ & $1$ & $-1$ \\
     $1_a$:\; & $1$ & $1$ & $-1$ & $-1$ & $1$ \\
     $1_{ca}$:\; & $1$ & $1$ & $1$ & $-1$ & $-1$ \\
     $E$:\; & $2$ & $-2$ & $0$ & $0$ & $0$ \\
     \hline
    \end{tabular}}
\ee
where the conjugacy classes are
\be 
[ab]=\{ab\}\,,\,
[a]=\{a,b\}\,, \,
[c]=\{c,cab\}\,,\,
[ca]=\{ca,cb\}\,.
\ee
Note that $[ab]$ is charged only under the duality symmetry $E$ and is of dimension 1: it thus corresponds to $X_{1,-1}$ in the $\TY$ notation.

By following the general approach described above, starting now with $\cZ (\Rep(D_8)^L\times \Rep(D_8)^R)$ 
and condensing the following algebra
\be
\cA^{\weak} = 1\oplus X_{1,-}^LX^R_{1,-}=1\oplus [ab]_L [ab]_R  \,.
\ee
we get the reduced TO\footnote{This category is group theoretical \cite{davydov2017lagrangian,GaiSchaferNamekiWarman}, and we used GAP for certain calculations \cite{gruen2021computing,GAP4}.}
\be\label{RedTOZ25}
\cZ (\TY (\Z_2^4)) \cong
\cZ ((\Z_2^{a_L}\times \Z_2^{ab} \times \Z_2^{a_R})\rtimes (\Z_2^{c_L} \times \Z_2^{c_R})) \,,
\ee
This confines the 2d irreps $E_L$ and $E_R$, since they braid non-trivially with $[ab]_L [ab]_R$, whereas the product 
$E_LE_R$ becomes the dimension $\sqrt{n}=4$ $\TY$ symmetry generator.
The 1d irreps are instead all deconfined. Therefore, the reduced TO will have the symmetry generators:
\be
    1_{a_L},\; 1_{c_L},\; 1_{a_R},\; 1_{c_R},\; E_LE_R\,,
\ee
and products thereof, forming $\TY(\Z_2^4)$ and describing a strong $\Z_2\times\Z_2$ and weak duality symmetry since $E_LE_R$ is diagonal.

All the $D_8^L\times D_8^R$ conjugacy classes are also deconfined:
since $[ab]_L [ab]_R$ is condensed, in the reduced TO we will have $[ab]_L\sim[ab]_R$, that we denote as $[ab]$.
The elementary conjugacy classes $[a]_L,\,[c]_L,\,[ca]_L,\,[a]_R,\,[c]_R,\,[ca]_R$ also pass through while the conjugacy classes of dimension 4 split into two conjugacy classes of dimension 2.
The symmetry Lagrangian algebra for this weak-strong symmetry category is the sum of all the conjugacy classes (all of dim=2 expect $[1]$ and $[ab]$ of dim=1):
\be\ba
    \cL_\sym^{\TY(\Z_2^4)}&=[1]\oplus [ab]\oplus [a]_L\oplus [a]_R\oplus [c]_L\oplus [c]_R \\
    &\oplus [ca]_L \oplus [ca]_R \oplus [a_La_R] \oplus [c_Lc_R]\oplus [a_Lc_R]\\
    &\oplus [c_La_R] \oplus [c_La_La_R] \oplus [c_La_Lc_R] \oplus [c_Lc_Ra_R] \\
    &\oplus [a_Lc_Ra_R]\oplus [c_La_Lc_Ra_R]\,.
\ea\ee

 A non-trivial SPT phase for this mixed weak-strong symmetry corresponds to subgroup $F=(\Z_2^{c_Lc_R} \times\Z_2^{ab})$  with non-trivial 2-cocycle $\beta\in H^2(\Z_2\times\Z_2,U(1))$. Here, $\Z_2^{c_Lc_R}$ is generated by the diagonal group element $c_Lc_R$, which is charged under the strong $\Z_2$ generated by $1_{a_L},1_{a_R}$), while $[ab]$ came from $[ab]_L\sim[ab]_R$ in $\cZ(D_8^L\times D_8^R)$. For $\beta$, we choose the representative:
\be \label{eq:betaZ2Z2}
\ba
    \beta((p_1,q_1),(p_2,q_2))=(-1)^{p_1q_2}\,.
\ea
\ee
The corresponding Lagrangian algebra is:\footnote{Here $[c_Lc_R]_2$ is a boson that carries the 2-dimensional representation of its centralizer $D_8\times\Z_2$.}
\begin{align} \label{eq:non-triv-SPT}
 \cL_{\text{SPT}}^{\TY(\Z_2^4)}&=(1\oplus 1_{a_L} 1_{a_R})(1\oplus 1_{c_L}\oplus 1_{c_R}\oplus  1_{c_L} 1_{c_R})\nn\\
    &\oplus [ab](1_{a_L}\oplus 1_{c_La_L}\oplus 1_{c_La_R}\oplus 1_{c_La_Lc_R})\nn\\
    &\oplus [ab](1_{a_R}\oplus 1_{c_Ra_R}\oplus 1_{a_Lc_R}\oplus 1_{c_Lc_Ra_R})\nn\\
    &\oplus 4[c_Lc_R]_2\,,\hspace{-10mm}
\end{align}
We see that \eqref{eq:non-triv-SPT} is an SPT that has non-trivial string order parameters: for example $[ab]1_{a_L}$ is comprised of a group element with charge $-4$ under the non-invertible weak $E_LE_R$, attached to a symmetry string under which $c_L$ has charge $-1$.

To provide a density matrix for this phase, we will use the 3-qubit $\Rep(D_8)$-symmetric spin-chain model of \cite{Warman:2024lir}.
The Hilbert space is $\cH=\bigotimes_i\bC_2^3$, whose basis states are identified with the generators of $D_8=(\Z_2^a\times\Z_2^b)\rtimes\Z_2^c$ as:
\be\ba
|a\rangle &\mapsto |010\rangle\,, \;\; 
&
|b\rangle &\mapsto |001\rangle\,, \;\;
  |c\rangle &\mapsto |100\rangle\,.
\ea\ee
The doubled Hilbert space for the Choi state is $\wt{\cH}=\cH_L\otimes\cH_R$.
When mapping the Lagrangian algebra \eqref{eq:non-triv-SPT} back to $\cZ(D_8^L\times D_8^R)$, the anyon $[ab]\to[ab]_L+[ab]_R$. We thus see that the only only anyons in $\cL_{\text{SPT}}$ supported entirely in $\cH_L$ are $[ab]_L1_{a_L}$ and $1_{c_L}$. Upon tracing out $\cH_R$, and using the lattice realizations \cite{Warman:2024lir}
\be\ba
    [ab]\mapsto X^\II X^\III,\quad 1_{a}\mapsto Z^\I,\quad 1_{c}\mapsto Z^\II Z^\III\,,
\ea\ee
we obtain the density matrix
\be\ba
    \rho\propto\prod_i&\big[\bI+(Z^{\I}X^{\II}X^{\III})_{i}(X^{\II}X^{\III})_{i+1}\bigr]\cdot\\[-1mm]
    \cdot&\bigl[\bbI+ \prod_i(Z^{\II}_iZ^{\III}_i)\bigr]\,.
\ea\ee
Note that this density matrix coincides with the product of the terms of the commuting projector Hamiltonian $H_{(\Z_2^{c}\times\Z_2^{ab})^-}$ for the non-trivial $\Rep(D_8)$ SPT pure state of \cite{Warman:2024lir}, but $\rho$ describes a mixed state since it is not idempotent. 

$\rho$ is strongly symmetric under the $\Z_2\times\Z_2$ generators 
\be
\cS_{1_a}=\prod_iZ^\I_i,\quad \cS_{1_c}=\prod_iZ^\II_iZ^\III_i\,.
\ee
For $\cS_E$ can use a simplified form of its general expression provided in \cite{Warman:2024lir}, obtained by restricting to the subspace of states spanned by $\Z_2^c\times\Z_2^{ab}=\{\ket{000},\ket{100},\ket{011},\ket{111}\}$ (which are the only $D_8$ group elements appearing in the $\cL_{\text{SPT}}$ and hence in $\rho$), on which the symmetry generators are realized as:
\be
    \ba
         \cS_{1_c}|_{\Z_2^c\times\Z_2^{ab}}&=\prod_i \bbI_{i}\,,\quad 
         \cS_{1_a}|_{\Z_2^c\times\Z_2^{ab}}=\prod_i Z^{\I}_{i}\cr 
       \cS_E|_{\Z_2^c\times\Z_2^{ab}}&=\prod_i Z^{\II}_{i}+\prod_i (Z^{\I}_{i} Z^{\II}_{i})\,,
    \ea
\ee
in agreement with the fact that $\cS_E\ket{g_1..g_N}=\chi_\Gamma(g_1\cdots g_N)\ket{g_1..g_N}$ and the character table \ref{eq:D8_chars}, from which one can verify that $\rho$ is only weakly symmetric under the non-invertible duality symmetry $\cS_{E}$
\be \label{eq:rhoE}
    \cS_E\,\rho\not\propto\rho\,,\qquad  \cS_E\,\rho\,\cS_E^\dagger=4\,\rho\,.
\ee
$\rho$ obeys the following correlators:
\begin{align}
    \Tr(\rho\,(X^\II X^\III)_i\,\rho\, (X^\II X^\III)_i)&\neq 0\,,\\
    \Tr(\rho \, X^{\II}_iX_j^\II\,\rho\, X_j^\II X_i^\II)&\neq 0\,,\quad  \Tr(\rho \,X^\II_i X^\II_j)=0\,.\nn
\end{align}
Since $X^\II X^\III$ is charged under $\cS_E$, the first line indicates explicit breaking of strong $\cS_E$, in agreement with \eqref{eq:rhoE}. The second line shows that the strong $\cS_{1_c}$, under which $X^\II$ is charged, is spontaneously broken, however the weak symmetry is preserved in this phase: we thus have SWSSB of $\cS_{1_c}$. The strong $\cS_{1_a}$ is instead preserved, as is the weak non-invertible duality symmetry $\cS_E$.
The anyon $[ab]1_a$ furnishes a non-trivial string order parameter:
\be
    \mathcal{Q}_{([ab],1_a)}^{(i_0,i_0+N)}= (X^{\II}X^{\III})_{i_0}(   \prod_{i=i_0}^{i_{0}+N}Z^{\I}_i)(X^{\II}X^{\III})_{i_0+N}\,,
\ee
and the following correlator is non-zero:
\be
\ba
    \Tr(\rho\,\mathcal{Q}_{([ab],1_a)})\neq 0
\ea
\ee
This detects that there is a non-trivial strong-weak SPT between the strong $\cS_{1_a}$ and weak duality (under which $[ab]$ is charged), similar to the $\Z_2\times\Z_2$ case we discussed in section \eqref{sec:Z2Z2}, but now for strong $\Z_2$ and weak non-invertible duality symmetry.

\section{Discussion and Outlook}

In this work, we presented a framework for studying mixed states with general, possibly non-invertible, symmetries via the SymTFT of a doubled symmetry. This doubled symmetry represents strong symmetry operators acting on the left and on the right of a density matrix. We also propose a method for incorporating additional weak symmetries by starting in a larger SymTFT and condensing a condensable (but not Lagrangian) algebra of diagonal charges. We present a set of conditions on Lagrangian and more generally condensable algebras that ensure that the resulting doubled space analysis yields a consistent density matrix in the original physical Hilbert space. Such mixed state algebras are constrained from the hermiticity and positivity of the density matrix. 

The framework utilizing the SymTFT is general, in particular not limited to 1+1d systems, and is amenable to many generalizations, which are being explored successfully for pure states. We expect the main idea of the current work to be amenable to these extensions. 

Let us mention a number of concrete interesting future directions worth exploring. 
One important question is to more rigorously prove that the hermiticity and positivity conditions (\ref{hermcond}) and (\ref{poscond}) are \emph{sufficient} for the doubled state Lagrangian algebra to give a valid fixed-point Choi state. One possible way to prove this is by showing that if those two conditions are satisfied, $\mathrm{Tr}_R(|\rho_{\mathcal{L}}\RR\LL\rho_{\mathcal{L}}|)$ has the desired order and disorder parameters, as discussed in section~\ref{sec:Complete}. 

We considered in this work non-invertible symmetries that satisfy $D\rho D^\dagger\propto\rho$. However, one can also consider the more general quantum-channel-like condition $\sum_iD_i\rho D_i^\dagger\propto\rho$. While the former gives an operator in the doubled Hilbert space that is a tensor product $D\otimes \overline{D}$, the latter gives a more general $T$ invariant operator in the doubled Hilbert space $\sum_iD_i\otimes \overline{D}_i$. It may be interesting to study consistent patterns of such symmetries, and in which contexts they might arise.

The SymTFT approach for pure state gapped phases has been applied in higher dimensions \cite{Bhardwaj:2023ayw, Bhardwaj:2023fca, Bhardwaj:2024qiv, Xu:2024pwd,  Antinucci:2024ltv, Bhardwaj:2025piv,  Inamura:2025cum}, and we expect that some of the ideas here can be naturally generalized to higher dimensions. 
In higher dimensions, we can also have mixed state topological order \cite{hastings2011,bao2023,fan2024,Ellison:2024svg,chen2024separability,sohal2025,sala20251}, which can be studied from the perspective of strong and weak higher-form symmetries. Mixed state topological order is related to the stability of topological order to decoherence and finite temperature, and therefore has applications in quantum computation. However there are subtleties with studying mixed state phases with higher-form symmetries, arising from having to consider what it means to have ``emergent symmetries" (as higher-form symmetries are often emergent) and the mismatch between Choi state phases and mixed-state phases defined by two-way channel connectivity \cite{sang2025,zhang2025swssb}.

Non-invertible symmetries akin to the Kramers-Wannier duality or more generally Tambara-Yamagami categories as well as their SymTFTs have been developed in higher dimensions \cite{Choi:2021kmx, Kaidi:2021xfk, Bhardwaj:2022yxj, Kaidi:2022cpf,Apte:2022xtu, Antinucci:2022vyk, Antinucci:2023ezl,Cordova:2023bja, Decoppet:2023bay, Bhardwaj:2024xcx}, which could be starting points for a higher-dimensional generalization of our weak-strong phases in Sec.~\ref{sec:WeakStrong}.  {Another extension is to study symmetries of mixed states in actual (as opposed to topological) holography using the description in terms of wormholes in the gravitational bulk theory that model the purified state, i.e. the thermofield double \cite{Maldacena:2001kr}.}

Another directions  that the SymTFT has been instrumental in closed quantum system has been in the study of gapless phases in closed quantum systems \cite{Chatterjee:2022tyg, Bhardwaj:2023bbf, Wen:2023otf, Bhardwaj:2024qrf,  Wen:2024qsg,  Bhardwaj:2025jtf, Wen:2025thg} and fermionic systems \cite{Bhardwaj:2024ydc, Huang:2024ror}. Extensions of these ideas to study ``gapless" phases open quantum systems \cite{lee2023crit,lu2023crit,chen2024crit,murciano2023crit} clearly would be very interesting to develop in the future. 

Finally, we presented one approach for obtaining patterns of mixed weak/strong non-invertible symmetries by starting in a larger strong symmetry, constructing the SymTFT for the doubled strong symmetry, and condensing an algebra of diagonal charges. It would be interesting to either prove that all consistent patterns of weak/strong non-invertible symmetries can be obtained in this way, or to provide an example that cannot be obtained in this way. It would also be interesting to develop a way to determine which symmetry patterns are consistent directly from the original symmetry rather than from extending the symmetry to the form of $\mathcal{S}_L\boxtimes\mathcal{S}_R$ first. In other words, if we just start with a fusion category, how does one directly check that it can be the Choi state symmetry of some density matrix? Certainly this fusion category must admit a $T$ symmetry. We also showed that there are some constraints from fusion rules in Sec.~\ref{weakalone}. However, there are likely other constraints, for example coming from the $F$ symbol. 
We leave a more complete characterization of consistent weak/strong symmetry patterns in terms of their fusion rules and $F$ symbols for future work.

\begin{acknowledgments}
     We thank the KITP for hospitality during the program GenSym25, during which this collaboration was initiated. We thank Andrea Antinucci, Lakshya Bhardwaj, Yu-Hsueh Chen, Meng Cheng, Yuhan Gai, Tarun Grover, Kyle Kawagoe, Mark Mezei, Sanjay Moudgalya, Salvatore Pace, Shinsei Ryu, and Yichen Xu for discussions. We thank the authors of \cite{Luo:2025}  and \cite{Qi:2025} for coordinating submission.
     The work of  SSN and AW is supported by the UKRI Frontier Research Grant, underwriting the ERC Advanced Grant ``Generalized Symmetries in Quantum Field Theory and Quantum Gravity”. 
     The work of AT is funded by Villum Fonden Grant no. VIL60714. CZ is supported by the Harvard Society of Fellows.
     This research was supported in part by grant NSF PHY-2309135 to the Kavli Institute for Theoretical Physics (KITP). 
\end{acknowledgments}

\appendix

\section{Tambara-Yamagami Fusion Categories}
\label{app:TambaYama}

\subsection{The Center of TY Categories}

Let $\mbA$ be a finite abelian group. The Tambara-Yamagami (TY) fusion categoris are obtained from $\mbA$ by considering the quantum double, i.e. the toric code, and gauging a $\Z_2$ outer automorphism (e.g the em-duality). 
This gives the center of $\TY (\mbA)$. This has a canonical gapped boundary condition that realizes the TY fusion category with non-invertible duality defect $D$ satisfying 
\be
D\otimes D = \bigoplus_{a\in \mbA} a \,,
\ee
where $a$ represents the abelian anyons labeled by $a$.

In the construction we have various choices of so-called Frobenius-Schur indicators and bicharacters. For any finite abelian group $\mathbb{A}$ we have 
\be
\chi: \mathbb{A} \times \mathbb{A} \to \mathbb{C}^\times
\ee
a symmetric bicharacter and the 
Frobenius Schur (FS) indicator 
\be
\tau = \pm 1/{\sqrt{|\mathbb{A}|}}\,.
\ee
The resulting fusion category  is denoted by
\be
\TY (\mathbb{A}, \chi, \tau) \,.
\ee
The different choices of bicharacter and $\tau$ do not change the fusion, but the associators, see \cite{Gelaki:2009blp}. 
In terms of the data defining the category the non-trivial associators are given for $a, b, \in \mbA$
\be
\ba
\alpha_{a, D, b} &=\chi(a, b)  \cr 
\alpha_{D, a, D}&=\bigoplus_{b \in \mbA} \chi(a, b) \id_b \cr 
 \alpha_{D, D, D}&=\bigoplus_{a, b \in \mbA} \tau \chi(a, b)^{-1} \id_{D} \,.
\ea
\ee

The Drinfeld center of $\TY (\mbA)$ is constructed as follows:
Start with the toric code for $\mbA$ and gauge a $\Z_2$ outer automorphism. 
The center of $\TY(\mbA)$ has anyons that come from invariant abelian anyons in the $\mbA$ toric code, as well as those stacked with the dual line (obtained from the gauging of the 0-form symmetry of the toric code). 
These have quantum dimension 1, and are denoted by 
\be 
X_{a, \epsilon}\,,\ 
\epsilon^2= {1\over \chi (a,a)}\,.
\ee
There are also quantum dimension $2$ lines 
\be 
Y_{a,b}\,,\quad a\not=b \in \mbA \,.
\ee 
Finally there are the twisted lines, which  originate in the $\mbA$ toric code as non-genuine lines at the end of surfaces that generate the $\Z_2$ automorphism. After the $\Z_2$ gauging, they are genuine lines in the $\TY$ category. Denote these by 
\be
Z_{\rho, \Delta}\,,\ \text{with  }\rho: \mbA \to \mathbb{C}^\times:  
\rho(ab)=  {\rho(a)\rho(b)\over \chi(a,b)}\,,
\ee 
and 
\be
\Delta^{2}=\tau\sum_{a\in \mathbb{A}}\rho(a)^{-1}\,.
\ee
Their quantum dimension is $\sqrt{n}$. 
The quantum dimension of $\TY (\mbA)$ is $2|\mbA|$.

The spins are \cite{Gelaki:2009blp}
\be\ba
T_{X_{a, \epsilon}}&=\chi(a, a)^{-1} \,,\quad 
T_{Y_{a, b}} =\chi(a, b)^{-1} \,,\quad 
T_{Z_{\rho, \Delta}}=\Delta 
\ea\ee
and the S-matrices  are 
\be
\ba
S_{X_{a, \epsilon}, X_{b, \epsilon'}} & =\chi\left(a, b \right)^2 \cr 
S_{X_{a, \epsilon}, Y_{b, c}} & =2 \chi(a, bc), \\
S_{X_{a, \epsilon}, Z_{\rho, \Delta}} & =\epsilon \sqrt{n} \rho(a) \cr 
S_{Y_{a, b}, Y_{c, d}} & =2(\chi(a, d) \chi(b, c)+\chi(a, c) \chi(b, d)) \cr 
S_{Y_{a, b}, Z_{\rho, \Delta}} & =0 \cr 
S_{Z_{\rho, \Delta}, Z_{\rho^{\prime}, \Delta^{\prime}}} & =\frac{1}{\Delta \Delta^{\prime}} \sum_{a \in A} \chi(a, a)^2 \rho(a) \rho^{\prime}(a) .
\ea
\ee

\subsection{$\Rep (H_8)$}
\label{app:RepH8}

For the weak duality, strong $\Z_2$ mixed phase in Sec.~\ref{sec:WeakDStrongZ}, we will need the fusion category, that is the representations of the Hopf algebra $H_8$. The details of this are e.g. worked out (including extensions in \cite{GaiSchaferNamekiWarman}) which corresponds to the $\TY$ category
\begin{equation}
    \cS=\Rep(H_8)=\TY(\Z_2\times\Z_2,\chi_{ab},+\tfrac{1}{2})\,,
\end{equation}
with bicharacter specified by
\begin{align} \label{eqn:bichaab}
    \chi_{ab}(a,a)=\chi_{ab}(b,b)=-1,\quad \chi_{ab}(a,b)=+1\,.
\end{align}
The simple objects in its Drinfeld center are in the standard notation of $\TY$ categories summarized in Sec.~\ref{sec:GeneralWeakDu}, see also \cite{Gelaki:2009blp}. In the present case they are given in table \ref{tab:ZRepH8}.

This lists the anyons in $\cZ(\Rep(H_8))$ (first column), their quantum dimensions (second column) and diagonal $T$-matrix elements, which encode the anyon spins (third column). $\zeta_8$ is the eighth root of unity $\zeta_8=e^{\frac{2\pi i}{8}}$.
  
The $\rho_n$ are given by
    \begin{equation}
        \begin{array}{|c|c c c c|c|}\hline
         & 1 & a & b & ab & \Delta^{2} \\
         \hline
 \rho_1 & 1 & -i & -i & -1 & i \\
 \rho_2 & 1 & i & i & -1 & -i \\
 \rho_3 & 1 & -i & i & 1 & 1 \\
 \rho_4 & 1 & i & -i & 1 & 1 \\
 \hline
        \end{array}
    \end{equation}
% There are thus 10 bosons:
% \be \ba
%    &1,  \quad X_{1,-1}, \quad X_{ab, 1}, \quad X_{ab,-1}, \cr 
%    &Y_{1,a}, \quad
%     Y_{1,b},  \quad Y_{1,ab}, \quad Y_{a,b}, \quad Z_{\rho_{3}, 1}, \quad Z_{\rho_{4}, 1}\,.
% \ea\ee
% We discuss the algebras formed by these in the main text. 

\onecolumngrid

\begin{table}
\centering
$
\begin{array}{|c|c|c|}
    \hline
    \text{Label in $\cZ(\TY (\Z_2\times \Z_2))$} & \text{Dim} & T \\
    \hline\hline
1=X_{1,1} & 1 & 1\\
X_{a,i} & 1 & -1\\
X_{b,i} & 1 & -1\\
X_{ab,1} & 1 & 1\\
X_{1,-1} & 1 & 1\\
X_{a,-i} & 1 & -1\\
X_{b,-i} & 1 & -1\\
X_{ab,-1} & 1 & 1\\
\hline
Y_{1,a} & 2 & 1\\
Y_{1,b} & 2 & 1\\
Y_{1,ab} & 2 & 1\\
Y_{a,b} & 2 & 1\\
Y_{a,ab} & 2 & -1\\
Y_{b,ab} & 2 & -1\\
\hline
Z_{\rho_1,\zeta_8} & 2 & \zeta_8\\
Z_{\rho_1,\zeta_8^5} & 2 & \zeta_8^5\\
Z_{\rho_2,\zeta_8^7} & 2 & \zeta_8^7\\
Z_{\rho_2,\zeta_8^3} & 2 & \zeta_8^3\\
Z_{\rho_3,1} & 2 & 1\\
Z_{\rho_3,-1} & 2 & -1\\
Z_{\rho_4,1} & 2 & 1\\
Z_{\rho_4,-1} & 2 & -1\\ \hline
\end{array}
$
\caption{The anyons of $\cZ(\TY (\Z_2 \times \Z_2))$, their quantum dimension and spin.  \label{tab:ZRepH8}}
\end{table}

\section{{Mixed Condensable algebras for strong $\Z_2\times\Z_2$ and Ising}}
\subsection{Lagrangian Algebras for $\Z_2\times \Z_2$ Strong Symmetry}\label{app:LagsZ2Z2}
The Lagrangian algebras  for the analysis of the $\Z_2\times \Z_2$ strong symmetric phases in Sec.~\ref{sec:Z2Z2} are:
\begin{align}
    \cL_{1}&=1\oplus e^A_\a \oplus m^B_\p m^B_\a \oplus m^B_\p e^A_\a m^B_\a \oplus e^A_\p \oplus e^A_\p e^A_\a \oplus e^A_\p m^B_\p m^B_\a \oplus e^A_\p m^B_\p e^A_\a m^B_\a \oplus e^B_\p e^B_\a \oplus \nn\\
    &\oplus e^B_\p e^A_\a e^B_\a \oplus e^B_\p m^B_\p e^B_\a m^B_\a \oplus e^B_\p m^B_\p e^A_\a e^B_\a m^B_\a \oplus e^A_\p e^B_\p e^B_\a \oplus \nn\\
    &\oplus e^A_\p e^B_\p e^A_\a e^B_\a \oplus e^A_\p e^B_\p m^B_\p e^B_\a m^B_\a \oplus e^A_\p e^B_\p m^B_\p e^A_\a e^B_\a m^B_\a \\[1mm]
    \cL_{2}&=1\oplus e^B_\a \oplus m^A_\p m^A_\a \oplus m^A_\p e^B_\a m^A_\a \oplus e^A_\p e^A_\a \oplus e^A_\p e^A_\a e^B_\a \oplus e^A_\p m^A_\p e^A_\a m^A_\a \oplus e^A_\p m^A_\p e^A_\a e^B_\a m^A_\a \oplus \nn\\
    &\oplus e^B_\p \oplus e^B_\p e^B_\a \oplus e^B_\p m^A_\p m^A_\a \oplus e^B_\p m^A_\p e^B_\a m^A_\a \oplus e^A_\p e^B_\p e^A_\a \oplus e^A_\p e^B_\p e^A_\a e^B_\a \oplus \nn\\
    &\oplus e^A_\p e^B_\p m^A_\p e^A_\a m^A_\a \oplus e^A_\p e^B_\p m^A_\p e^A_\a e^B_\a m^A_\a  \\[1mm]
    \cL_{3}&=1\oplus e^A_\a e^B_\a \oplus m^A_\p m^B_\p m^A_\a m^B_\a \oplus m^A_\p m^B_\p e^A_\a e^B_\a m^A_\a m^B_\a \oplus e^A_\p e^A_\a \oplus e^A_\p e^B_\a \oplus e^A_\p m^A_\p m^B_\p e^A_\a m^A_\a m^B_\a \oplus \nn\\
    &\oplus e^A_\p m^A_\p m^B_\p e^B_\a m^A_\a m^B_\a \oplus e^B_\p e^A_\a \oplus e^B_\p e^B_\a \oplus e^B_\p m^A_\p m^B_\p e^A_\a m^A_\a m^B_\a \oplus e^B_\p m^A_\p m^B_\p e^B_\a m^A_\a m^B_\a \oplus e^A_\p e^B_\p \oplus \nn\\
    &\oplus e^A_\p e^B_\p e^A_\a e^B_\a \oplus e^A_\p e^B_\p m^A_\p m^B_\p m^A_\a m^B_\a \oplus e^A_\p e^B_\p m^A_\p m^B_\p e^A_\a e^B_\a m^A_\a m^B_\a  \\[1mm]
    \cL_{4}&=1\oplus m^B_\p m^B_\a \oplus m^A_\p m^A_\a \oplus m^A_\p m^B_\p m^A_\a m^B_\a \oplus e^A_\p e^A_\a \oplus e^A_\p m^B_\p e^A_\a m^B_\a \oplus e^A_\p m^A_\p e^A_\a m^A_\a \oplus \nn\\
    &\oplus e^A_\p m^A_\p m^B_\p e^A_\a m^A_\a m^B_\a \oplus e^B_\p e^B_\a \oplus e^B_\p m^B_\p e^B_\a m^B_\a \oplus e^B_\p m^A_\p e^B_\a m^A_\a \oplus e^B_\p m^A_\p m^B_\p e^B_\a m^A_\a m^B_\a \oplus \nn\\
    &\oplus e^A_\p e^B_\p e^A_\a e^B_\a \oplus e^A_\p e^B_\p m^B_\p e^A_\a e^B_\a m^B_\a \oplus e^A_\p e^B_\p m^A_\p e^A_\a e^B_\a m^A_\a \oplus e^A_\p e^B_\p m^A_\p m^B_\p e^A_\a e^B_\a m^A_\a m^B_\a  \\[1mm]
    \cL_{5}&=1\oplus m^A_\a m^B_\a \oplus m^B_\p m^B_\a \oplus m^B_\p m^A_\a \oplus m^A_\p m^B_\a \oplus m^A_\p m^A_\a \oplus m^A_\p m^B_\p \oplus m^A_\p m^B_\p m^A_\a m^B_\a \oplus \nn\\
    &\oplus e^A_\p e^B_\p e^A_\a e^B_\a \oplus e^A_\p e^B_\p e^A_\a e^B_\a m^A_\a m^B_\a \oplus e^A_\p e^B_\p m^B_\p e^A_\a e^B_\a m^B_\a \oplus e^A_\p e^B_\p m^B_\p e^A_\a e^B_\a m^A_\a \oplus e^A_\p e^B_\p m^A_\p e^A_\a e^B_\a m^B_\a \oplus \nn\\
    &\oplus e^A_\p e^B_\p m^A_\p e^A_\a e^B_\a m^A_\a \oplus e^A_\p e^B_\p m^A_\p m^B_\p e^A_\a e^B_\a \oplus e^A_\p e^B_\p m^A_\p m^B_\p e^A_\a e^B_\a m^A_\a m^B_\a  \\[1mm]
    \cL_{6}&=1\oplus e^A_\a e^B_\a m^A_\a m^B_\a \oplus m^A_\p m^B_\p m^A_\a m^B_\a \oplus m^A_\p m^B_\p e^A_\a e^B_\a \oplus e^A_\p m^B_\p e^A_\a m^B_\a \oplus e^A_\p m^A_\p e^B_\a m^A_\a \oplus \nn\\
    &\oplus e^A_\p m^A_\p e^A_\a m^A_\a \oplus e^A_\p m^A_\p e^B_\a m^B_\a \oplus e^B_\p m^B_\p e^A_\a m^A_\a \oplus e^B_\p m^B_\p e^B_\a m^B_\a \oplus e^B_\p m^A_\p e^A_\a m^B_\a \oplus e^B_\p m^A_\p e^B_\a m^A_\a \oplus \nn\\
    &\oplus e^A_\p e^B_\p m^A_\a m^B_\a \oplus e^A_\p e^B_\p e^A_\a e^B_\a \oplus e^A_\p e^B_\p m^A_\p m^B_\p \oplus e^A_\p e^B_\p m^A_\p m^B_\p e^A_\a e^B_\a m^A_\a m^B_\a  \\[1mm]
    \cL_{7}&=1\oplus m^B_\a \oplus m^B_\p \oplus m^B_\p m^B_\a \oplus m^A_\p m^A_\a \oplus m^A_\p m^A_\a m^B_\a \oplus m^A_\p m^B_\p m^A_\a \oplus m^A_\p m^B_\p m^A_\a m^B_\a \oplus \nn\\
    &\oplus e^A_\p e^A_\a \oplus e^A_\p e^A_\a m^B_\a \oplus e^A_\p m^B_\p e^A_\a \oplus e^A_\p m^B_\p e^A_\a m^B_\a \oplus e^A_\p m^A_\p e^A_\a m^A_\a \oplus e^A_\p m^A_\p e^A_\a m^A_\a m^B_\a \oplus \nn\\
    &\oplus e^A_\p m^A_\p m^B_\p e^A_\a m^A_\a \oplus e^A_\p m^A_\p m^B_\p e^A_\a m^A_\a m^B_\a  \\[1mm]
    \cL_{8}&=1\oplus e^A_\a m^B_\a \oplus m^B_\p m^B_\a \oplus m^B_\p e^A_\a \oplus e^A_\p m^B_\a \oplus e^A_\p e^A_\a \oplus e^A_\p m^B_\p \oplus e^A_\p m^B_\p e^A_\a m^B_\a \oplus e^B_\p m^A_\p e^B_\a m^A_\a \oplus \nn\\
    &\oplus e^B_\p m^A_\p e^A_\a e^B_\a m^A_\a m^B_\a \oplus e^B_\p m^A_\p m^B_\p e^B_\a m^A_\a m^B_\a \oplus e^B_\p m^A_\p m^B_\p e^A_\a e^B_\a m^A_\a \oplus e^A_\p e^B_\p m^A_\p e^B_\a m^A_\a m^B_\a \oplus \nn\\
    &\oplus e^A_\p e^B_\p m^A_\p e^A_\a e^B_\a m^A_\a \oplus e^A_\p e^B_\p m^A_\p m^B_\p e^B_\a m^A_\a \oplus e^A_\p e^B_\p m^A_\p m^B_\p e^A_\a e^B_\a m^A_\a m^B_\a \\[1mm]
    \cL_{9}&=1\oplus m^A_\a \oplus m^B_\p m^B_\a \oplus m^B_\p m^A_\a m^B_\a \oplus m^A_\p \oplus m^A_\p m^A_\a \oplus m^A_\p m^B_\p m^B_\a \oplus m^A_\p m^B_\p m^A_\a m^B_\a \oplus \nn\\
    &\oplus e^B_\p e^B_\a \oplus e^B_\p e^B_\a m^A_\a \oplus e^B_\p m^B_\p e^B_\a m^B_\a \oplus e^B_\p m^B_\p e^B_\a m^A_\a m^B_\a \oplus e^B_\p m^A_\p e^B_\a \oplus e^B_\p m^A_\p e^B_\a m^A_\a \oplus \nn\\
    &\oplus e^B_\p m^A_\p m^B_\p e^B_\a m^B_\a \oplus e^B_\p m^A_\p m^B_\p e^B_\a m^A_\a m^B_\a  \\[1mm]
    \cL_{10}&=1\oplus e^B_\a m^A_\a \oplus m^A_\p m^A_\a \oplus m^A_\p e^B_\a \oplus e^A_\p m^B_\p e^A_\a m^B_\a \oplus e^A_\p m^B_\p e^A_\a e^B_\a m^A_\a m^B_\a \oplus e^A_\p m^A_\p m^B_\p e^A_\a m^A_\a m^B_\a \oplus \nn\\
    &\oplus e^A_\p m^A_\p m^B_\p e^A_\a e^B_\a m^B_\a \oplus e^B_\p m^A_\a \oplus e^B_\p e^B_\a \oplus e^B_\p m^A_\p \oplus e^B_\p m^A_\p e^B_\a m^A_\a \oplus e^A_\p e^B_\p m^B_\p e^A_\a m^A_\a m^B_\a \oplus \nn\\
    &\oplus e^A_\p e^B_\p m^B_\p e^A_\a e^B_\a m^B_\a \oplus e^A_\p e^B_\p m^A_\p m^B_\p e^A_\a m^B_\a \oplus e^A_\p e^B_\p m^A_\p m^B_\p e^A_\a e^B_\a m^A_\a m^B_\a  
\end{align}

\subsection{Condensable Algebras for Ising}

Table \ref{tab:Ising_CondAlgs} summarizes the non-maximal mixed condensable algebras for strong $\Ising$ symmetry. \\
The Lagrangian algebras are written in equation \eqref{LIsingIsing}.

\begin{table}[H]
    \centering
    ${\begin{array}{|c|c|c|}
        \hline
       \#& \text{dim} & \text{Possible Condensable Algebra} \\
        \hline
        1&1 & 1 \\
        \hline \\[-3mm]        
        2& 2 & 1\oplus \psi_L \ol{\psi}_L \ol{\psi}_R \psi_R \\
        3&2 & 1\oplus \psi_L \ol{\psi}_R \\
        4& 2 & 1\oplus \ol{\psi}_L \psi_R \\
        \hline \\[-3mm]
        5& 4 & 1\oplus \psi_L \ol{\psi}_R\oplus \sigma_L \ol{\psi}_L \ol{\sigma}_R \psi_R \\
        6& 4 & 1\oplus \psi_L \ol{\psi}_R\oplus \sigma_L \ol{\sigma}_R \\
        7& 4 & 1\oplus \ol{\psi}_L \ol{\psi}_R\oplus \psi_L \psi_R\oplus \psi_L \ol{\psi}_L \ol{\psi}_R \psi_R \\
        7& 4 & 1\oplus \ol{\psi}_L \psi_R\oplus \psi_L \ol{\sigma}_L \ol{\psi}_R \sigma_R \\
        8& 4 &  1\oplus \ol{\psi}_L \psi_R\oplus \psi_L \ol{\psi}_R\oplus \psi_L \ol{\psi}_L \ol{\psi}_R \psi_R \\
        9&4 & 1\oplus \ol{\psi}_L \psi_R\oplus \ol{\sigma}_L \sigma_R \\
        10& 4 & 1\oplus \ol{\psi}_R \psi_R\oplus \psi_L \ol{\psi}_L\oplus \psi_L \ol{\psi}_L \ol{\psi}_R \psi_R \\
        \hline \\[-3mm]
       11& 8 & 1\oplus \ol{\psi}_L \ol{\psi}_R\oplus \psi_L \psi_R\oplus \psi_L \ol{\psi}_L \ol{\psi}_R \psi_R\oplus \sigma_L \ol{\sigma}_L \ol{\sigma}_R \sigma_R \\
       12& 8 & 1\oplus \ol{\psi}_L \psi_R\oplus \psi_L \ol{\psi}_R\oplus \psi_L \ol{\psi}_L \ol{\psi}_R \psi_R\oplus \sigma_L \ol{\sigma}_L \ol{\sigma}_R \sigma_R \\
        13& 8 & 1\oplus \ol{\psi}_L \psi_R\oplus \psi_L \ol{\psi}_R\oplus \psi_L \ol{\psi}_L \ol{\psi}_R \psi_R\oplus \sigma_L \ol{\sigma}_R\oplus \sigma_L \ol{\psi}_L \ol{\sigma}_R \psi_R \\
       14& 8 & 1\oplus \ol{\psi}_L \psi_R\oplus \ol{\sigma}_L \sigma_R\oplus \psi_L \ol{\psi}_R\oplus \psi_L \ol{\psi}_L \ol{\psi}_R \psi_R\oplus \psi_L \ol{\sigma}_L \ol{\psi}_R \sigma_R \\
       15&  8 & 1\oplus \ol{\psi}_R \psi_R\oplus \psi_L \ol{\psi}_L\oplus \psi_L \ol{\psi}_L \ol{\psi}_R \psi_R\oplus \sigma_L \ol{\sigma}_L \ol{\sigma}_R \sigma_R \\
        16 & 8 & 1\oplus \ol{\psi}_R \psi_R\oplus \ol{\psi}_L \psi_R\oplus \ol{\psi}_L \ol{\psi}_R\oplus \psi_L \psi_R\oplus \psi_L \ol{\psi}_R\oplus \psi_L \ol{\psi}_L\oplus \psi_L \ol{\psi}_L \ol{\psi}_R \psi_R \\
        \hline
    \end{array}}$
    \caption{possible non-maximal condensable algebras in $\cZ(\Ising)_L\boxtimes\ol{\cZ(\Ising})_R$ invariant under $L\leftrightarrow R$ exchange.}
    \label{tab:Ising_CondAlgs}
\end{table}

\twocolumngrid

%%%%%%%%%%%%%%%%%%%%%%%%%%%%%%%%%%%%%

\bibliographystyle{ytphys}
\small 
\baselineskip=.94\baselineskip
\let\bbb\bibitem\def\bibitem{\itemsep4pt\bbb}
\bibliography{GenSym}

\end{document}